\documentclass[twocolumn]{aastex631}

\usepackage{amsmath}
\usepackage{appendix}

\usepackage{booktabs}
\usepackage{textgreek}
\usepackage{threeparttable}
\def    \apjl  		{\rm {ApJL}}
\def    \apj  		{\rm {ApJ}}
\def    \mnras  	{\rm {MNRAS}}
\def    \araa  		{\rm {ARA\& A}}
\def    \apjl  		{\rm {ApJL}}

\def	\cm		{\,{\rm {cm}}}
\def	\K		{\,{\rm K}}
\def	\g		{\,{\rm {g}}}

\def \bea {\begin{eqnarray}}
\def \ena {\end{eqnarray}}

\def	\B	{{\rm B}}

\def	\C	{{\rm C}}

\def	\cm	{\,{\rm cm}}
\def	\d	{{\rm d}}

\def	\D	{{\rm D}}

\def	\erg	{\,{\rm erg}}

\def	\g	{\,{\rm g}}

\def	\G	{{\rm G}}

\def	\H	{{\rm H}}

\def	\K	{{\rm K}}

\def	\N	{{\rm N}}

\def	\s	{\,{\rm s}}

\def	\V	{{\rm V}}

\DeclareUnicodeCharacter{0301}{\'{e}}

\begin{document}
\shorttitle{Betelgeuse envelope}
\shortauthors{Truong et al.}
\title{Modeling extinction and reddening effects by circumstellar dust in the Betelgeuse envelope in the presence of radiative torque disruption}

\author{Bao Truong}
\affiliation{Department of Physics, International University, Ho Chi Minh City, Vietnam}
\affiliation{Vietnam National University, Ho Chi Minh City, Vietnam}
\email{truonglegiabao@gmail.com}

\author{Le Ngoc Tram}
\affiliation{Stratospheric Observatory for Infrared Astronomy, Universities Space Research Association, NASA Ames Research Center, MS 232-11, Moffett Field, 94035 CA, USA}
\affiliation{Max-Planck-Institut für Radioastronomie, Auf dem Hügel 69, D-53121, Bonn, Germany}

\author{Thiem Hoang}
\affiliation{Korea Astronomy and Space Science Institute, 776 Daedeokdae-ro, Yuseong-gu, Daejeon 34055, Republic of Korea}
\affiliation{Korea University of Science and Technology, 217 Gajeong-ro, Yuseong-gu, Daejeon 34113, Republic of Korea}

\author{Nguyen Chau Giang}
\affiliation{Korea Astronomy and Space Science Institute, 776 Daedeokdae-ro, Yuseong-gu, Daejeon 34055, Republic of Korea}
\affiliation{Korea University of Science and Technology, 217 Gajeong-ro, Yuseong-gu, Daejeon 34113, Republic of Korea}

\author{Pham Ngoc Diep}
\affiliation{Department of Astrophysics, Vietnam National Space Center, Vietnam Academy of Science and Technology, 18 Hoang Quoc Viet, Hanoi, Vietnam}

\author{Dieu D. Nguyen}
\affiliation{Université de Lyon 1, Ens de Lyon, CNRS, Centre de Recherche Astrophysique de Lyon (CRAL) UMR5574, F-69230 Saint-Genis-Laval, France}

\author{Nguyen Thi Phuong}
\affiliation{Korea Astronomy and Space Science Institute, 776 Daedeokdae-ro, Yuseong-gu, Daejeon 34055, Republic of Korea}
\affiliation{Department of Astrophysics, Vietnam National Space Center, Vietnam Academy of Science and Technology, 18 Hoang Quoc Viet, Hanoi, Vietnam}

\author{Thuong D. Hoang}
\affiliation{University of Science and Technology of Hanoi, Vietnam Academy of Science and Technology, 18 Hoang Quoc Viet, Hanoi, Vietnam}
\affiliation{Kavli Institute for the Physics and Mathematics of the Universe, The University of Tokyo, 5-1-5 Kashiwanoha, Kashiwa, Chiba, 277-8583, Japan}

\author{Nguyen Bich Ngoc}
\affiliation{Department of Astrophysics, Vietnam National Space Center, Vietnam Academy of Science and Technology, 18 Hoang Quoc Viet, Hanoi, Vietnam}
\affiliation{Graduate University of Science and Technology, Vietnam Academy of Science and Technology, 18 Hoang Quoc Viet, Hanoi, Vietnam}

\author{Nguyen Fuda}
\affiliation{Department of Physics, International University, Ho Chi Minh City, Vietnam}
\affiliation{Vietnam National University, Ho Chi Minh City, Vietnam}

\author{Hien Phan}
\affiliation{University of Science and Technology of Hanoi, Vietnam Academy of Science and Technology, 18 Hoang Quoc Viet, Hanoi, Vietnam}

\author{Tuan Van Bui}
\affiliation{University of Science and Technology of Hanoi, Vietnam Academy of Science and Technology, 18 Hoang Quoc Viet, Hanoi, Vietnam}

\correspondingauthor{Le Ngoc Tram, Thiem Hoang}
\email{nle@mpifr-bonn.mpg.de; thiemhoang@kasi.re.kr}

\begin{abstract}
Circumstellar dust is formed and evolved within the envelope of evolved stars, including Asymptotic Giant Branch (AGB) and Red Supergiant (RSG). The extinction of stellar light by circumstellar dust is vital for interpreting RSG/AGB observations and determining high-mass RSG progenitors of core-collapse supernovae. Nevertheless, circumstellar dust properties are not well understood. Modern understanding of dust evolution suggests that intense stellar radiation can radically change the dust properties across the circumstellar envelope through the RAdiative Torque Disruption (RAT-D) mechanism. In this paper, we study the impacts of RAT-D on the grain size distribution (GSD) of circumstellar dust and model its effects on photometric observations of $\alpha$ Orionis (Betelgeuse). Due to the RAT-D effects, large grains formed in the dust formation zone are disrupted into smaller species of size $a < 0.5\,\rm\mu m$. Using the GSD constrained by the RAT-D effects, we model the visual extinction of background stars and Betelgeuse. We find that the extinction decreases at near-UV, optical, and infrared wavelengths while increasing at far-UV wavelengths. The resulting flux potentially reproduces the observation from the near-UV to near-IR range. Our results can be used to explain dust extinction and photometric observations toward other RSG/AGB stars.

\end{abstract}

\section{Introduction} \label{sec:intro}
When helium burning in stellar cores is turned on, stars with initial masses of $12\,\rm M_{\odot} - 30\,M_{\odot}$ evolve into Red Supergiants (hereafter RSGs; see \citealt{Levesque2006} for a review). RSG stars experience significant stellar activities such as global pulsation (\citealt{Goldberg1984}; \citealt{Heger1997}; \citealt{Gou2002}; \citealt{Kiss2006}) and large convection cells (\citealt{Schwarzchild1975}; \citealt{Antia1984}; \citealt{Fadeyev2012}; \citealt{Ren2020}), leading to the loss of materials through the stellar winds. Heavy elements are synthesized in the core until they reach the Fe fusion. Due to its maximum binding energy per nucleon, the Fe fusion is impossible, and the fusion core can no longer be sufficient to support against its gravity. Hence, the RSG implodes and ends their lives as core-collapse supernovae (CCSNe). Chemical elements heavier than iron and cosmic dust are formed in the ejecta, which enriches the chemical abundance of the interstellar medium (ISM)(see \citealt{Tielens2005} and \citealt{DeBeck2019} for reviews). Therefore, RSG research is essential for understanding dust formation and stellar feedback in the ISM. Moreover, RSG studies have substantially impacted on determining the initial mass of supernovae progenitors, which relates to the physical properties of supernovae (SNe) and their remnants. 

However, observations revealed that not all CCSNe progenitors are coincident with RSGs. \cite{Smartt2009} indicated the link between RSGs and CCSNe through observations of 20 Type IIP-SNe progenitors, and showed the upper limit mass of $16.5\,\rm M_{\odot}$, which was below the maximum estimated RSG mass of $25\,\rm M_{\odot}$ for CCSNe (\citealt{Meynet2003}). Several explanations were proposed, including the loss of the H envelope through stellar pulsation (\citealt{Yoon2010}), creating "failed SNe" that directly collapse to stellar-mass black holes (\citealt{Woosley2012}; \citealt{Lovegrove2013}) or the limit mass of RSG of $20\,\rm M_{\odot}$ in stellar evolution (\citealt{Groh2013}). 

Recent studies have shown that circumstellar dust may be responsible for the missing of high mass RSG progenitors. Circumstellar dust obscures stellar photons and produces an additional extinction, causing the underestimation of stellar luminosity and linking to failed searches for high mass RSG progenitors (\citealt{Smartt2009}; \citealt{Walmswell2012}; \citealt{Kochanek2012}; \citealt{Kilpatrick2018}). Another significant impact of circumstellar dust is the reddening effects at short wavelengths. Circumstellar dust absorbs and scatters incoming stellar photons in UV-optical and near-IR wavelengths, resulting in a strong continuum absorption in the RSG/AGB spectra (\citealt{Massey2005}; \citealt{Fonfria2020}; \citealt{Fonfria2021}). Thus, studying circumstellar dust in RSGs and their extinction properties may help resolve the RSG problems and identify high mass RSG progenitors. 

Among dust properties, the grain size distribution (GSD) and its optical properties are essential parameters for determining dust extinction and stellar reddening. The first study of GSD through UV extinction by \cite{Seab1989} reported a lower cutoff of the GSD at $0.08\,\rm\mu m$ due to grain shattering. The study of dust condensation in RSGs by \textcite{Velhoelst2009} and in the RSG progenitor of SN 2012aw by \cite{Kochanek2012} adopted a standard Mathis-Rumpl-Nordsieck (MRN) size distribution of the ISM with $dn/da \varpropto a^{-3.5}$ for $0.005\,\mu$m $< a < 0.25\,\rm\mu m$ (\citealt{Mathis1977}). \cite{vanLoon2005} adopted a single-size of $0.1\,\rm\mu m$ for their study. From RSG dusty envelopes observed in the IR spectra, several modelings proposed the presence of larger grains $a > 0.1\,\rm\mu m$. \cite{Groenewegen2009} found grains with $1\,\rm\mu m$ in O-rich RSGs, while \cite{Scicluna2015} investigated the maximum grain size of $0.5\,\rm\mu m$ in the RSG stellar winds.

Dust properties are modified during their formation and evolution within the circumstellar envelope (CSE). Dust grains are first formed by gas nucleation in the formation zone of $\sim 3 - 10\,\rm R_{\ast}$ (\citealt{Cherchneff2013}). Subsequently, they can grow by gas accretion and grain-grain collisions in the inner envelope (\citealt{Velhoelst2006}). When being expelled by radiation pressure into the ISM, dust properties can also be affected by radiative and mechanical feedback. In general, dust grains can be destroyed by thermal sublimation, grain shattering, and sputtering (\citealt{Dominik1989}; \citealt{Biscaro2016}). Nevertheless, thermal sublimation is inefficient in the AGB/RSG cases, because dust temperature is below the sublimation threshold of $1500$ K starting from the formation zone. Moreover, shattering and sputtering are only efficient in shocks when dust is injected into the ISM.

Recently, \cite{Hoang2019} introduced a new mechanism of grain disruption based on radiative torques, which is termed RAdiative Torque Disruption (RAT-D). The RAT-D mechanism can be effective in radiation fields in which thermal sublimation is insufficient. The mechanism implies that large grains spun-up by radiative torques (RATs) to fast rotation can be fragmented by centrifugal stress into smaller particles, including nanoparticles (i.e., $a < 100\,\rm\AA$). \cite{Tram2020} first studied the RAT-D effect for AGB envelopes and modeled electric dipole emission from rapidly spinning nanoparticles using the GSD constrained by RAT-D. The authors demonstrated that the model within a proper GSD can explain the observations of both carbon-rich and oxygen-rich AGB stars (see their Figure 10). Therefore, the approach of the RAT-D mechanism is plausible in determining proper GSD and materials in the CSE of the evolved stars.  

In this work, we will apply the RAT-D for the RSG envelopes and discuss its implications for a particular star, the $\alpha$ Orionis (Betelgeuse) - a M2 supergiant star in the end phase of the evolution for high-mass stars. The star is a luminous source with the bolometric luminosity of $\sim 1.26\times 10^{5}L_{\odot}$ and a distance of $\rm 197\,pc$ (see in Table \ref{tab:Betelgeuse}). Hence, the CSE dust grains are expected to be affected by the intense radiation field and their gaseous environment.  

Using the GSD constrained by RAT-D, we determine starlight absorption and scattering characteristics. We will examine the extinction properties of dust grains through modeling extinction curves along the sightline. We will demonstrate the dependency of extinction features on grain properties and circumstellar environment. As a result, we will test grain effects by reproducing the reddened spectrum in Betelgeuse envelope. 

This paper is organized as follows. Section \ref{section: Gas profile} describes the physical properties of the Betelgeuse CSE. In Section \ref{section: Modeling} and \ref{section: Procedure}, we describe the principle of the RAT-D mechanism along with procedures of modeling extinction by circumstellar dust. The modeling results, including GSD, extinction features, and reddening effects in the stellar spectrum, are presented in Section \ref{section:Result}. We discuss the applications of the RAT-D mechanism on studying grain size distribution in AGB/RSG envelopes and the reddening in AGB/RSG spectra, and analyzing the Betelgeuse CSE grain properties in Section \ref{section:Discussion}. Finally, our main findings are summarized in Section \ref{section: Summary}.

\section{Physics model of Betelgeuse envelope}
\label{section: Gas profile}

\subsection{Circumstellar envelope structure}
\label{section:circumstellar}

\begin{table}
    \centering
    \caption{Physical parameters of the Betelgeuse circumstellar envelope}
    \label{tab:Betelgeuse}
    \begin{tabular}{l c c}
         \toprule
         Parameters & Values & References \\
         \midrule
         Distance ($\rm pc$) & $197\pm 45$ & (1) \\
         Stellar luminosity ($L_{\odot}$) &  $1.26\times 10^{5}$ & (1) \\
         Stellar mass ($M_{\odot}$) & $19.7$ & (1)\\
         Stellar radius ($R_{\odot}$) & $887\pm 203$ & (1) \\
         Stellar diameter ($\rm mas$) & $55.64\pm 0.04$ & (1) \\
         Effective temperature ($\K$) & $3500\pm 200$ & (1); (2) \\
         Mass-loss rate ($M_{\odot} \, \rm yr^{-1}$) & $2\pm 1\times 10^{-6}$ & (1); (3) \\
         Expansion velocity ($\rm km \s^{-1}$) & $15$ & (3)\\
         Inner envelope radius ($R_{\ast}$) & $3$ & (3); (4) \\
         Outer envelope radius ($R_{\ast}$) & $17,000$ & (5); (6) \\
         \bottomrule
    \end{tabular}
    \begin{tablenotes}
      \small
      \item \textbf{References:} (1) \cite{Dolan2016}; (2) \cite{Fay1973}; (3) \cite{Velhoelst2006}; (4) \cite{Kervella2016}; (5) \cite{Noriega1997}; (6) \cite{Decin2013}
    \end{tablenotes}
\end{table}

The characteristics of the Betelgeuse envelope have been studied through near-IR (NIR) and mid-IR (MIR) observations. \cite{Velhoelst2006} revealed an excess astrosilicate emission at an inner dust shell of $20\,\rm R_{\ast}$ from the mid-IR observations. The authors also showed a thin $\rm Al_2O_3$ dust shell at $2 - 3\,\rm R_{\ast}$ above Betelgeuse, proving the scenarios of first dust condensation in the hot chromosphere. \cite{Kervella2016} resolved a dust shell at $3\,\rm R_{\ast}$ from the light scattering of $\rm Al_2O_3$ dust by VLT/SPHERE/ZIMPOL. For an outer envelope, \cite{Noriega1997} observed an arc structure at $7' \sim 17000\,\rm R_{\ast}$ by IRAS, which was confirmed by Herschel PACS and SPIRE images (\citealt{Decin2013}). \cite{Decin2013} showed this outer layer is a boundary region of the Betelgeuse envelope where the circumstellar winds collide with the ISM. 

Our model adopts a spherical symmetry of the dusty circumstellar envelope, characterised by the observations of \cite{Noriega1997} and \cite{Decin2013}. We choose an innermost dust shell at $r_{\rm in} = 3\,\rm R_{\ast}$ followed by polarimetric results of \cite{Kervella2016}, in which dust grains formed. An outermost layer is chosen at $r_{\rm out} = 17000\,\rm R_{\ast}$ from \cite{Noriega1997} as we mentioned earlier. Our modelling parameters are listed in Table \ref{tab:Betelgeuse}.

\subsection{Gas density profile}

Stellar activities in the photosphere produce strong shocks and eject most of the stellar materials above an escape velocity (see \citealt{Bennett2010} for a review). The outflows accelerate until reaching a certain distance $R_0$, where materials are able to condense and form the first solids. With the high flux opacity of grains, radiation pressure is sufficient to accelerate the newly formed grains and produce the drift of dust relative to the gas. This mechanism is known as the radiation-driven winds (see e.g., \citealt{Hofner2008}; \citealt{Hofner2009}; \citealt{Hofner2016}; \citealt{Hofner2018}). 

\begin{figure}
    \centering
    \includegraphics[width = 0.48\textwidth]{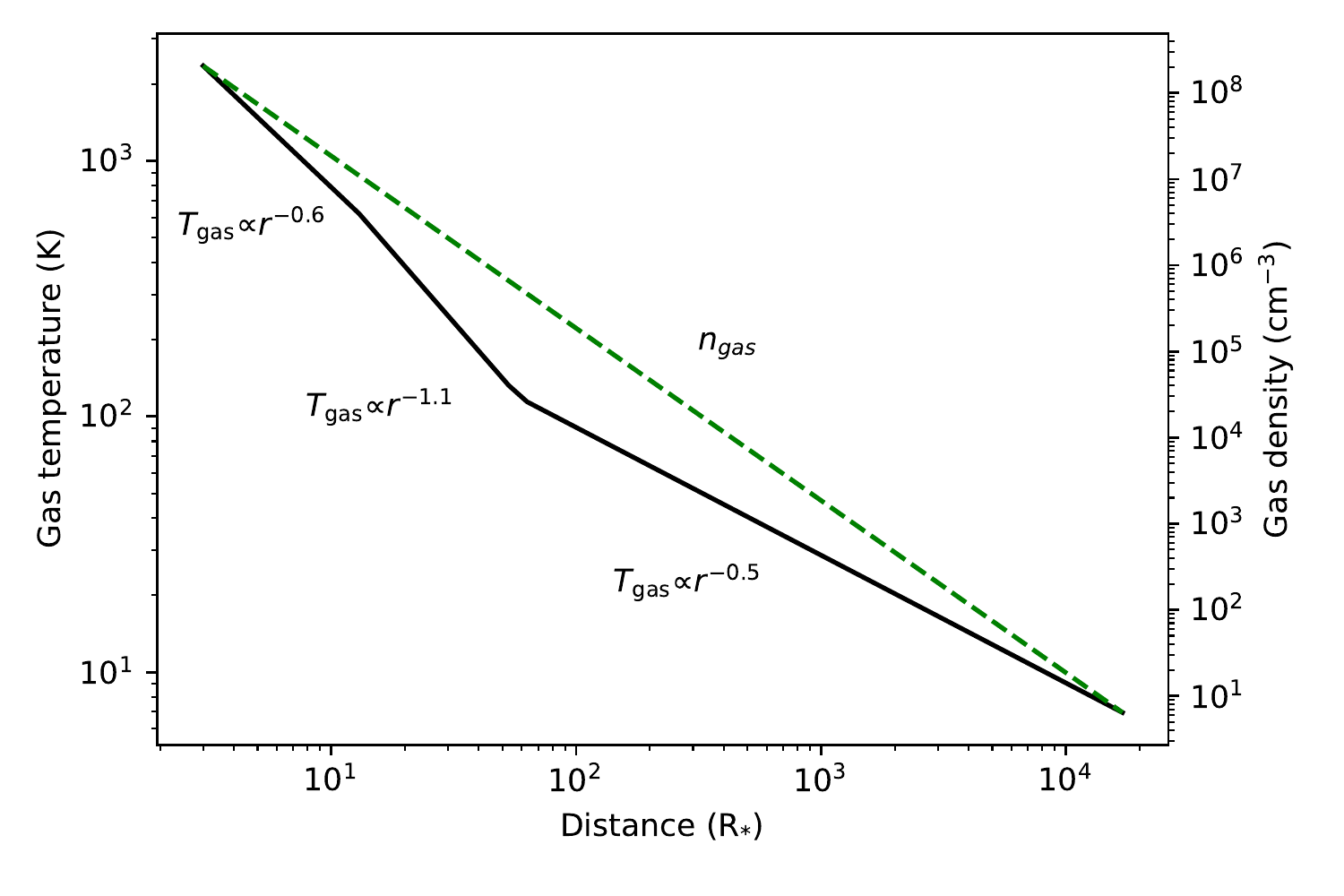}
    \caption{The mean gas density (dashed green line) and gas temperature profile (solid black line) of the Betelgeuse circumstellar envelope.}
    \label{fig:gas properties}
\end{figure}

In this work, we assume that the CSE expands spherically within a constant terminal velocity $v_{\rm exp}$ (\citealt{Tielens1983}). Hence, the gas density at a distance $r$ is given by

\begin{equation}
\label{eq:gas_density}
\begin{split}
     n_{\rm gas}(r) &= \frac{\dot{M}}{4 \pi r^2 m_{\rm gas} v_{\rm exp} } \\
&\simeq 10^{7}  \, \left(\frac{\dot{M}}{10^{-5} M_{\odot}}\right) \, \left(\frac{10^{15} \, \rm cm}{r}\right)^2 \, \left(\frac{10 \rm\,km \,s^{-1}}{v_{\rm exp}}\right) \, \rm cm^{-3}   
\end{split}
\end{equation} 
with $\dot{M}$ the mass-loss rate and $m_{\rm gas} = 1\,\rm amu$ for the hydrogen atom mass (see e.g., \citealt{Tram2020}\footnote{ Note that the pre-factor is $10^{7}\,\rm cm^{-3}$ instead of $10^{6}\,\rm cm^{-3}$ as a typo in Equation 1 of \cite{Tram2020}, but we checked that their gas density profile (their Figure 1) is correct.}). For Betelgeuse, we adopt $\dot{M}\sim 2\, \times \, 10^{-6} \, \rm M_{\odot} \, \rm yr^{-1}$ and $v_{\rm exp} = 15\,\rm km\,s^{-1}$(\citealt{Dolan2016}; \citealt{Velhoelst2006}). The gas density profile in Betelgeuse envelope is shown by the dashed line in Figure \ref{fig:gas properties}.

\subsection{Gas temperature profile}
The pressure winds transfer the gas kinetic energy into thermal energy, caused by the Alfvén wave flux, linear acoustic waves, and global pulsation (\citealt{Hartmann1984}; \citealt{Pijpers1989}; \citealt{Lim1998}). As a result, it heats surrounding gas to reach above an effective surface temperature of Betelgeuse ($T_{\rm eff}$, see, e.g., \citealt{Gillard1996} and \citealt{Lim1998}). When the gas expands spherically by radiation-driven winds, the temperature is balanced by the drag heating of circumstellar dust and adiabatic cooling, and drops significantly to below $100\,\rm K$ at the outer CSE (see e.g., \citealt{Glassgold1986}; \citealt{Rodger1991}).
Therefore, we construct a power-law mean gas temperature profile as 
\begin{equation}
    T_{\rm gas} = T_{0}\,\left(\frac{r}{r_{0}}\right)^{-\alpha},
\end{equation}
where $T_{0}$ is the initial temperature; $\alpha$ and $r_{0}$ vary in each certain distance of the extended atmosphere. Observations and semi-empirical models (\citealt{Harper2001}; \citealt{Gorman2015}) showed that the temperature profile of the chromosphere at $2\,\rm R_{\ast} < r < \rm 6\,R_{\ast}$ drops from $3000\,\rm K$ to $1800\,\rm K$ as a form of $T_{\rm gas} \varpropto r^{-0.6}$. At $6\,\rm R_{\ast} < r < \rm 60\,R_{\ast}$ , the temperature decreases as $T_{\rm gas} \varpropto r^{-1.1}$. At a larger distance of $r > \rm 60\,R_{\ast}$, the gas temperature decreases dramatically to $\sim 6\,\rm K$ at $r \sim 11000\,\rm R_{\ast}$ (\citealt{Rodger1991}). 
In this stage, the temperature can be approximately of $T_{\rm gas} \varpropto r^{-0.5}$. The radial profile of the gas temperature is shown by the solid line in Figure \ref{fig:gas properties}.  

\section{Radiative Torque Disruption Mechanism and grain-size distribution}
\label{section: Modeling}

\subsection{Radiation field from a RSG star}
\label{section: Radiation_RSG}
For a spherical envelope, the specific energy density at wavelength $\lambda$ at a distance $r$ from the central star is given by
\begin{equation}
\label{eq:u_star}
    u_{\lambda}(T_{\ast}) = \frac{\pi B(\lambda,T_{\ast})}{c} \, \left(\frac{R_{\ast}}{r}\right)^2
\end{equation}
where $T_{\ast}$ is the star effective temperature. For the sake of convenience, we consider the radiation of a central star as a black-body, and $B(\lambda,T_{\ast})$ follows the Planck function.

Photons emitting from the central source will be attenuated by dust extinction in the CSE. Consequently, the specific energy density is reduced to
\begin{equation}
\label{eq:u_lambda}
    u_{\lambda} = u_{\lambda,0}\,e^{-\tau_{\lambda}}
\end{equation}
where $u_{\lambda,0}$ is the intrinsic energy density as shown in Equation \ref{eq:u_star}, and $\tau_{\lambda}$ is the optical depth at wavelength $\lambda$ measured from the central star.

The total energy density of radiation field in the range of wavelengths from $\lambda_{\rm min}$ to $\lambda_{\rm max}$
\begin{equation}
\label{eq:u_rad}
    u_{\rm rad} = \int_{\lambda_{\rm min}}^{\lambda_{\rm max}} u_{\lambda}\, d\lambda.
\end{equation}

We adopt $\lambda_{\rm min} = 0.1\,\rm \mu m$, and $\lambda_{\rm max} = 20\,\rm\mu m$, which covers the peak of the black-body radiation of a RSG star with the lower limit being close to the Lyman cutoff. For convenience, we use a dimensionless parameter $U = u_{\rm rad}/u_{\rm ISRF}$ to refer to the strength of radiation fields with $u_{\rm ISRF}$ the energy density of the average interstellar radiation field in the solar neighborhood ($u_{\rm ISRF} = 8.64\,\times\,10^{-13}\,\rm erg\,\rm cm^{-3}$) (\citealt{Mathis1983}).

Finally, the mean wavelength of the radiation field is given by
\begin{equation}
\label{eq:mean_wave}
    \bar{\lambda} = \frac{\int_{\lambda_{\rm min}}^{\lambda_{\rm max}} \lambda u_{\lambda}d\lambda}{\int_{\lambda_{\rm min}}^{\lambda_{\rm max}} u_{\lambda}d\lambda}.
\end{equation}

For Betelgeuse, $\bar{\lambda} = 1.45\,\rm \mu m$ for unreddened radiation field with $u_{\lambda}$ given by Equation (\ref{eq:u_star}). This value is smaller for the case of AGB stars, e.g., $\bar{\lambda} = 2.42\,\rm \mu m$ for IRC+10216 and $\bar{\lambda} = 2.53\,\rm \mu m$ for IK Tau (see \citealt{Tram2020}).

\subsection{Rotational Disruption by Radiative Torques}
\label{section: RATD}
The intense radiation field of Betelgeuse can spin up the large irregular grain shapes to an extremely fast rotation. If the induced centrifugal stress can exceed the binding energy that holds the grain constituents, the grain is spontaneously disrupted to smaller species. This destruction is called the RAT-D mechanism (see more in \citealt{Hoang2019} and \citealt{Lazarian2021} for details).

Here we recall the principle of the RAT-D mechanism. The maximum angular velocity that a dust grain of size $a$ can achieve by RATs is given by
\begin{equation}
\label{eq:angular_RAT}
    \omega_{\rm RAT} = \frac{\Gamma_{\rm RAT}\,\tau_{\rm damp}}{I},
\end{equation}
where $I = 8\pi \rho a^{5} / 15 $ is the inertia moment of the spherical dust grain, and $\tau_{\rm damp}$ is the total rotational damping timescale. Above, $\Gamma_{\rm RAT}$ is the radiative torque, which is given by (see Equation 6 in \citealt{Giang2020})

\begin{equation}
\label{eq:radiative_torque}
    \Gamma_{\rm RAT} = \int_{\lambda_{\rm min}}^{\lambda_{\rm max}} \pi\,a^2\,\gamma_{\rm rad}\,u_{\lambda,0}\,e^{-\tau_{\lambda}}\,\left(\frac{\lambda}{2\pi}\right)\,Q_{\lambda}\,d\lambda
\end{equation}
where $u_{\lambda,0}$ is given by Equation \ref{eq:u_star}, $\gamma_{\rm rad}$ is the degree of anisotropy of the radiation field. For the unidirectional radiation field from a central star, one has $\gamma_{\rm rad} = 1$. Here, $Q_{\lambda} = \alpha\,\left(\lambda/a\right)^{-\eta}$ is the RAT efficiency. From numerical calculations of RATs for different shapes and orientations of dust grains (\citealt{Lazarian2007} and \citealt{Hoang2021}), the RAT efficiency is approximately equal to $2.33\,\left(\lambda/a\right)^{3}$ for $a < a_{\rm trans}$ and $0.4$ for $a_{\rm trans} < a < \lambda/0.1$, where $a_{\rm trans} = \lambda/1.8$ is the transition grain size from the flat to power-law RAT efficiency. 

The grain rotation is slowed down by a combination of collisional damping by gas collision and the IR re-emission of grains. The characteristic damping timescale is
\begin{equation}
\label{eq:tau_damp}
    \tau_{\rm damp} = \frac{\tau_{\rm gas}}{1+F_{\rm IR}}.
\end{equation}

The corresponding characteristic timescale of collisional damping is
\begin{equation}
\label{eq:tau_gas}
    \tau_{\rm gas} \simeq 8.74 \times 10^{4}\,a_{-5}\hat{\rho}\left(\frac{30\,\rm cm^{-3}}{n_{\rm H}}\right)\left(\frac{100\,\rm K}{T_{\rm gas}}\right)^{1/2}\,\rm yr,
\end{equation} 
and the dimensionless IR damping coefficient is
\begin{equation}
\label{eq:F_IR}
    F_{\rm IR} \simeq \left(\frac{0.4U^{2/3}}{a_{-5}}\right)\left(\frac{30\,\rm cm^{-3}}{n_{\rm H}}\right)\left(\frac{100\,\rm K}{T_{\rm gas}}\right)^{1/2},
\end{equation}
where $a_{-5} = a/(10^{-5} \rm cm)$ and $\hat{\rho} = \rho/(3\,\rm g\,\rm cm^{-3})$ (see Equations 12 and 13 in \citealt{Tram2020}). One can see from Equations \ref{eq:angular_RAT} - \ref{eq:F_IR} that the angular velocity depends on the local gas density $n_{\rm H}$, gas temperature $T_{\rm gas}$, and radiation strength $U$. 

The critical angular velocity at which the grain is fragmented due to the centrifugal force is given as (see Equation 3 in \citealt{Hoang2019}):

\begin{equation}
\label{eq:omega_crit}
    \omega_{\rm disr} = \frac{2}{a}\,\left(\frac{S_{\rm max}}{\rho}\right)^{1/2} \simeq 3.65\times10^{9}a_{-5}^{-1}\hat{\rho}^{-1/2}S_{\rm max,9}^{1/2}\,\rm rad\,s^{-1},
\end{equation}
where $S_{\rm max,9} = S_{\rm max}/(10^{9}\,\rm erg\,\rm cm^{-3})$ is the maximum tensile strength of grains (\citealt{Hoang2019}). The tensile strength is characterized by the grain internal structure. As discussed in \cite{Hoang2019}, porous/composite grains have low values of $S_{\rm max} = 10^{6} - 10^{9}\,\rm erg\,\rm cm^{-3}$, whereas compact grains have higher values as $S_{\rm max} \gtrsim 10^{10}\,\rm erg\,\rm cm^{-3}$. 

Consequently, the maximum grain size determined by the RAT-D mechanism ($a_{\rm disr}$) is calculated numerically by setting $\omega_{\rm RAT} \equiv \omega_{\rm disr}$, and grains larger than $a_{\rm disr}$ would be disrupted.

Note that the condition of the RAT-D mechanism is valid for grains aligned at attractor points of high angular momentum (a.k.a high-J attractors) due to the spin-up effects (see in \citealt{Hoang2021}). These grains can experience the disruption with the timescale less than the gas damping. For grains at low-J attractors, their angular momentum is, otherwise, smaller and coupled by thermal rotation of gas (denoted by $\rm J_{\rm thermal}$), thus they are inefficiently disrupted by the RAT-D mechanism (see in \citealt{Hoang2021} for further explanation). The rotational disruption properties of grains have a significant impact on our calculation of the GSD, and will be discussed in the next Section.

\subsection{Grain size distribution (GSD)}
\label{section: grain_disrtribution}

Our model assumes that the Betelgeuse envelope contains mostly O-rich dust (e.g., crystalline olivine and amorphous silicate, see \citealt{Draine1984} and \citealt{Greenberg1996}), which was presented in spectroscopic observations taken by \cite{Velhoelst2009} and \cite{Kervella2011}. The grain size distribution follows a power-law as
 \begin{equation}
\label{eq:grain_size_distribution}
    \frac{dn_{\rm sil}}{da} = C_{\rm sil}n_{\rm H}a^{-\alpha},
\end{equation}
where $C_{\rm sil}$ is the corresponding normalization constant of astrosilicate dust. The GSD has a lower and upper cutoffs denoted by $a_{\rm min}$ and $a_{\rm max}$, respectively.

Original dust in the dust formation zone is expected to have large size with $a_{\rm max}>0.25\,\mu$m (\citealt{Scicluna2015}). While expanding outward by radiation pressure, a fraction $f_{\rm high-J}$ of original large grains at high-J attractors can achieve suprathermal rotation (i.e., $\rm J > J_{\rm thermal}$) and be effectively disrupted by RATs, and thus, the disruption size $a_{\rm disr}$ represents the upper cutoff of GSD. The depletion of large grains and enhancement of smaller grains also modify the slope of the GSD. Meanwhile, the remained portion $1 - f_{\rm high-J}$ of original grains at low-J attractors rotate thermally (i.e., $\rm J \sim J_{\rm thermal}$). With decreased angular momentum, the RAT-D mechanism is unable to occur and results in the remaining of the original GSD with a fixed $a_{\rm max}$ and $\alpha = 3.5$. As shown in \cite{Lazarian2021}, the value of $f_{\rm high-J}$ depends on the grains properties and magnetic susceptibility. From the RAT calculations by \cite{Herranen2021} for an ensemble of irregular grain shapes, $f_{\rm high-J}$ is about $10\% - 70\%$ for the ordinary paramagnetic material as astrosilicate.

For the case of grains at high-J attractors, we consider the modified GSD with a fixed normalization constant and a varying power-law index, according to model 1 in \cite{GiangSuper2020}. Hence, the fixed normalization constant $C_{\rm sil}$ is computed through the dust-to-gas mass ratio $\eta$ as
\begin{equation}
\label{eq:dust_to_gas}
     \eta = \frac{4\pi}{m_{\rm H}}\frac{a_{\rm max}^{0.5} - a_{\rm min}^{0.5}}{0.5}\,C_{\rm sil}\rho_{\rm sil}.
\end{equation}
where $\rho_{\rm sil} = 3.5\rm\,g\,cm^{-3}$ is the mass density of silicate grains. Eventually, the modification of the power-law index is determined through the mass conservation as (see Equation 14 in \citealt{GiangSuper2020})
\begin{equation}
    \frac{a_{\rm max}^{0.5} - a_{\rm min}^{0.5}}{0.5} = \frac{a_{\rm disr}^{4 + \alpha}+a_{\rm min}^{4 + \alpha}}{4 + \alpha},
\end{equation}
where $a_{\rm min}=3.5\,\rm\AA$ is fixed, which is determined by thermal sublimation of astrosilicate dust (see \citealt{Draine1979}; \citealt{GiangSuper2020}). We assume that the original dust has a maximum grain size of $0.5\,\rm\mu m$. However, we also consider the effect of the variation of $a_{\rm max}$ from $0.05\,\rm\mu m$ to $0.5\,\rm\mu m$.

\section{Extinction curve modeling}
\label{section: Procedure}
\subsection{Model set-up}
\begin{figure}
    \centering
    \includegraphics[width = 0.6\textwidth]{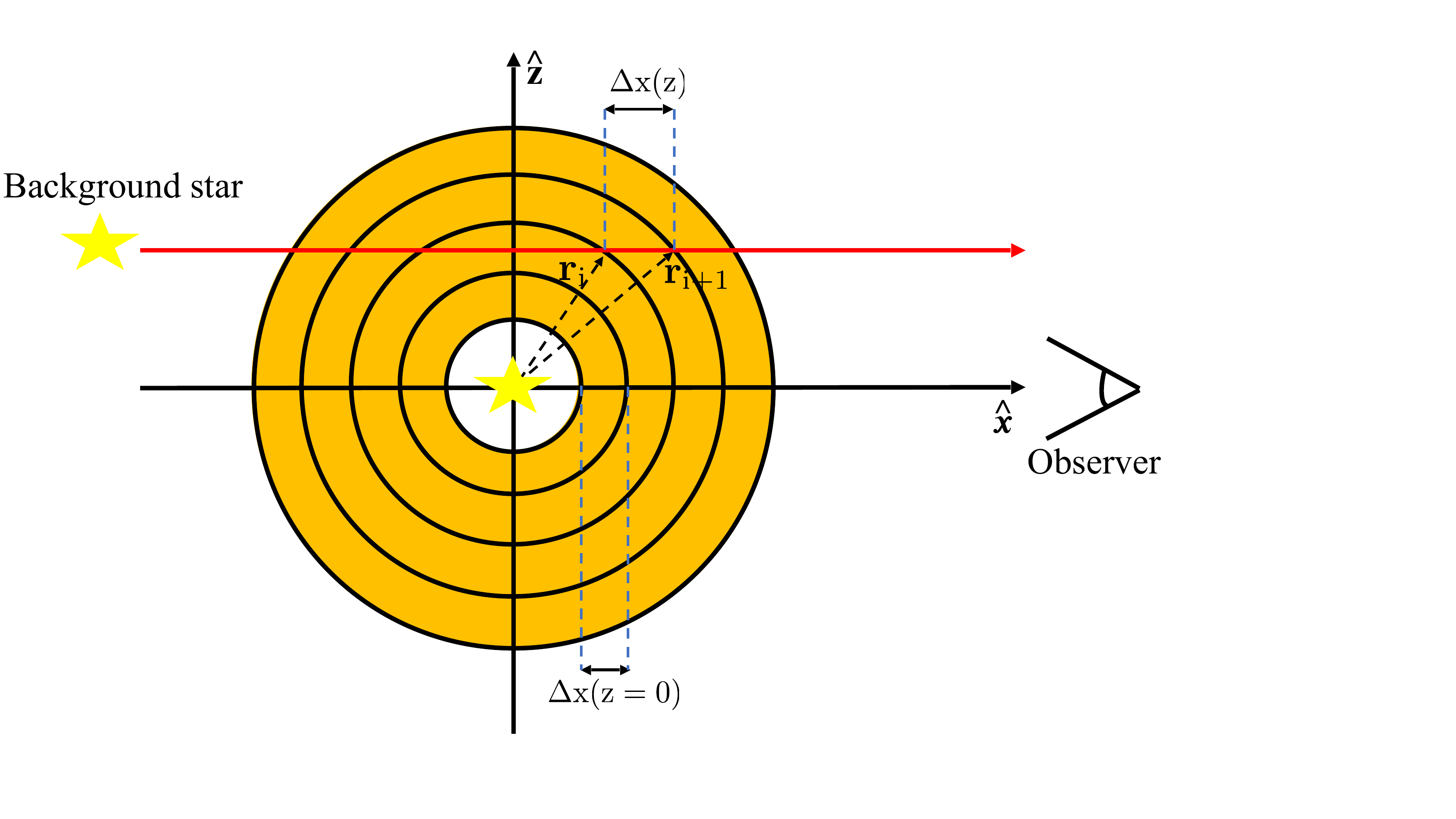}
    \caption{Schematic of radiative-transfer modeling by dust grains along a line-of-sight across the circumstellar envelope in the $\bf \hat{x}\hat{z}$ plane (middle slab with $y = 0$). A central star is considered as an origin. The positive direction of $x$ is chosen directly toward the observer on Earth.}
    \label{fig:Betel_model}
\end{figure}

Our work adopts a spherical geometry of the CSE within a coordinate system $\bf \hat{x}\hat{y}\hat{z}$ having the origin at the center of Betelgeuse in a unit of $1\,\rm R_{\ast}$ as seen in Figure \ref{fig:Betel_model}. The circumstellar envelope is assumed to have an inner radius of $3\,\rm R_{\ast}$ and an outer radius of $17000\,\rm R_{\ast}$. The line-of-sight (LOS) toward the central star is along $x$ with $z=0$. For numerical modeling, the envelope is divided into 1700 sub-layers, and each layer has the same thickness of $\Delta x(z=0) = 10\,\rm R_{\ast} \simeq 42.4\,\rm AU$. Within each layer, the gas and dust physical properties are uniform. We also consider the extinction of background stars behind Betelgeuse. Since the distance to Betelgeuse ($d \sim 197\,\rm pc$, see in Table \ref{tab:Betelgeuse}) is much larger than its envelope, the LOS of a background star is approximately parallel to the Betelgeuse LOS with $z\neq 0$. In this case, the thickness of each slab is $\Delta x(z)$, identified by the intersection of the parallel-LOS and sub-layers of the envelope. 

\subsection{Optical depth}
\label{section: optical depth}

For each slab $i$ along the LOS, $x$, toward an observer, the optical depth $\Delta \tau_{i}$ at wavelength $\lambda$ with thickness $\Delta x$ is calculated as
\begin{equation} \label{eq:optical_depth}
\begin{split}
    \Delta \tau_{\lambda,i} &= \int C_{\rm ext,sil}^{i}(a)\frac{dn_{i}}{da}da\Delta x \\
    &= C_{\rm sil}\,n_{\textrm{H},i}\left[f_{\rm high-J}\int_{a_{\rm min}}^{a_{\rm disr,i}} C_{\rm ext,sil}^{i}(a)a^{-\alpha}da \right.\\
    &\left.+ (1 - f_{\rm high-J})\int_{a_{\rm min}}^{a_{\rm max}} C_{\rm ext,sil}^{i}(a)a^{-3.5}da \right]\,\Delta s,
\end{split}
\end{equation}
where $C_{\rm ext,sil}^{i}$ is the extinction cross-section of astrosilicate grain of size $a$ and $dn_{i}/da$ is the GSD extracted from Equation \ref{eq:grain_size_distribution}. The optical depth is produced by two seperated GSDs: the fraction $f_{\rm high-J}$ of disrupted grains at high-J attractors in the range of $a_{\rm min} - a_{\rm disr}$, and $1 - f_{\rm high-J}$ of non-disrupted grains at low-J attractors in the range of $a_{\rm min} - a_{\rm max}$. The value of $C_{\rm ext,sil}^{i}$ is taken from \cite{Hoang2013} from photon absorption and scattering in grain surfaces, which is derived from DDSCAT code for various grain shapes and orientations (\citealt{Draine2012}). We assume that dust grains have oblate spheroidal shape with an axial ratio of 2.

Finally, the total optical depth at wavelength $\lambda$ along the LOS is the sum of the optical depths of each slab $i$ as
\begin{equation}
    \tau_{\lambda} = \sum_{i = 1}^{n} \Delta \tau_{\lambda,i},
\end{equation}
where $i$ spans from $i = 1$ to $i = n$ (with $n = 1700$), corresponding to the farthest and closest slab of a LOS across the Betelgeuse envelope toward the observer, respectively. 

\subsection{Dust extinction}
\label{section: Extinction curve formula}
The extinction at wavelength $\lambda$ along the LOS is computed as (\citealt{Hoang2013})
\begin{equation}
\label{eq: dust extinction}
    A_{\lambda} = 1.086\tau_{\lambda} = \sum_{i = 1}^{n} 1.086\Delta \tau_{\lambda,i}
\end{equation}
with the optical depth given in Equation \ref{eq:optical_depth}. To compare with observations, we normalize the extinction curve to the visual extinction at $\lambda=0.5448\,\mu$m in UBVRI passbands (see Table \ref{tab:UBVRI} in Appendix) (\citealt{Bessell2005}).  

Next, we compute the total-to-selective extinction ratio, $R_{\rm V}$, as
\begin{equation}
\label{eq:Rv}
    R_{\rm V} = \frac{A_{\rm V}}{E_{\rm B-V}} = \frac{A_{\rm V}}{A_{\rm B} - A_{\rm V}},
\end{equation}
where $E_{\rm B-V}=A_{B}-A_{V}$ is the color excess; $A_{\rm B}$ and $A_{\rm V}$ are the extinction at blue wavelength ($\lambda_{\rm eff} = 0.4363\,\rm\mu m$) and visible wavelength. The value of $R_{\rm V}$ determine the steepness of the extinction curve. 

\begin{figure}
    \centering
    \includegraphics[width = 0.48\textwidth]{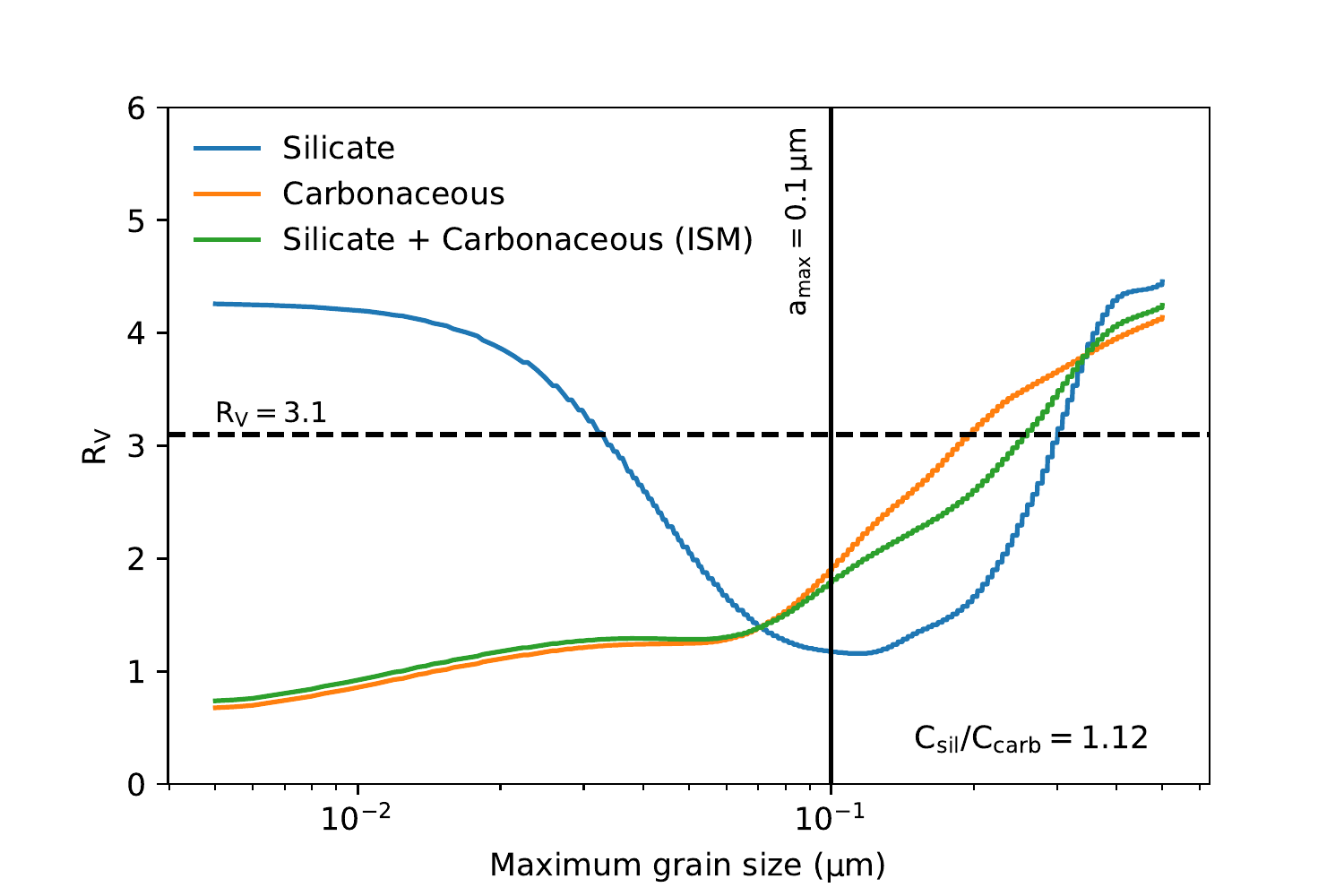}
    \caption{The calculation of the total-to-selective ratio $R_{\rm V}$ in a range of $a_{\rm max} = [5\,\rm nm - 0.5\,\rm\mu m]$ of three configurations of dust: astrosilicate (blue); carbonaceous (orange); and a mixture of both dust materials (green). In the latter case, we assume a ratio of astrosilicate and carbonaceous components of $C_{\rm sil}/C_{\rm carb} = 1.12$.}
    \label{fig:Rv_ISM}
\end{figure}

Figure \ref{fig:Rv_ISM} shows the variation of $R_{\rm V}$ with respect to maximum grain size $a_{\rm max}$, assuming $a_{\rm min} = 3.5\,\rm\AA, \alpha = 3.5$ and $C_{\rm sil}/C_{\rm carb} = 1.12$ (\citealt{Draine1984}; \citealt{Laor1993}). With a mixture of astrosilicate and carbonaceous components in ISM dust (solid green line), $R_{\rm V}$ is positively correlated with the grain size distribution, in which lower $R_{\rm V}$ implies the enhancement of smaller grains, whereas larger $R_{V}$ implies the presence of large grains in the environment (see \citealt{Hirashita2014}; \citealt{Nataf2016}).

The optical properties of dust materials can impact the observed $R_{\rm V}$. In the case of only O-rich dust (solid blue line), the extinction by small astrosilicate is only affected by the strong absorption at far-UV wavelengths ($\lambda < 0.2\,\rm\mu m$) while the visible extinction is negligible (see \citealt{Draine1984} and \citealt{DraineBook2011} for reviews). This leads to higher $R_{\rm V}$ for smaller grains of $a < 0.1 \,\rm\mu m$. For large grains of $a > 0.1\,\rm\mu m$, scattering with stellar photons becomes dominant and significantly contributes to the change in the extinction at optical wavelengths. The value of $R_{\rm V}$ start to increase with increasing $a_{\rm max}$. The standard $R_{\rm V} \simeq 3.1$ of the diffuse ISM (\citealt{Cardelli1989}) is possibly produced by the dominance of small astrosilicates of $a_{\rm max} = 0.02\,\rm\mu m$ or the abundance of large astrosilicate grains of $a_{\rm max} = 0.15\,\rm\mu m$, as depicted in Figure \ref{fig:Rv_ISM} (see in \citealt{DraineBook2011}). Hence, the major effects of optical properties of O-rich compositions should be taken in consideration in the calculation of $R_{\rm V}$. 

\subsection{Dust reddening}
Finally, we study the effect of dust reddening on the stellar radiation spectrum. The specific intensity of stellar radiation after being attenuated by circumstellar dust is calculated as 

\begin{equation}
\label{eq:Ilambda}
    I_{\lambda} = I_{\lambda, 0}\,e^{-\tau_{\lambda}}.
\end{equation}
where $I_{\lambda, 0} = B(\lambda , T_{\ast})$ is the intrinsic spectral intensity following the Planck function; $\tau_{\lambda} = A_{\lambda}/1.086$ is the optical depth derived from the dust extinction toward the central star. We study the reddened spectrum by circumstellar dust in two configurations of GSD: the fixed GSD with original grain sizes $a_{\rm max}$ (i.e., without RAT-D), and the modified GSD with the disruption sizes $a_{\rm disr}$ (i.e., with RAT-D). The enhancement of smaller grain sizes in the latter is expected to increase the attenuation at the UV wavelengths.

\section{Numerical results}
\label{section:Result}

\subsection{Grain disruption size} 
\label{section:Grain disruption sizes}
\begin{figure*}
    \centering
    \includegraphics[width = 0.48\textwidth]{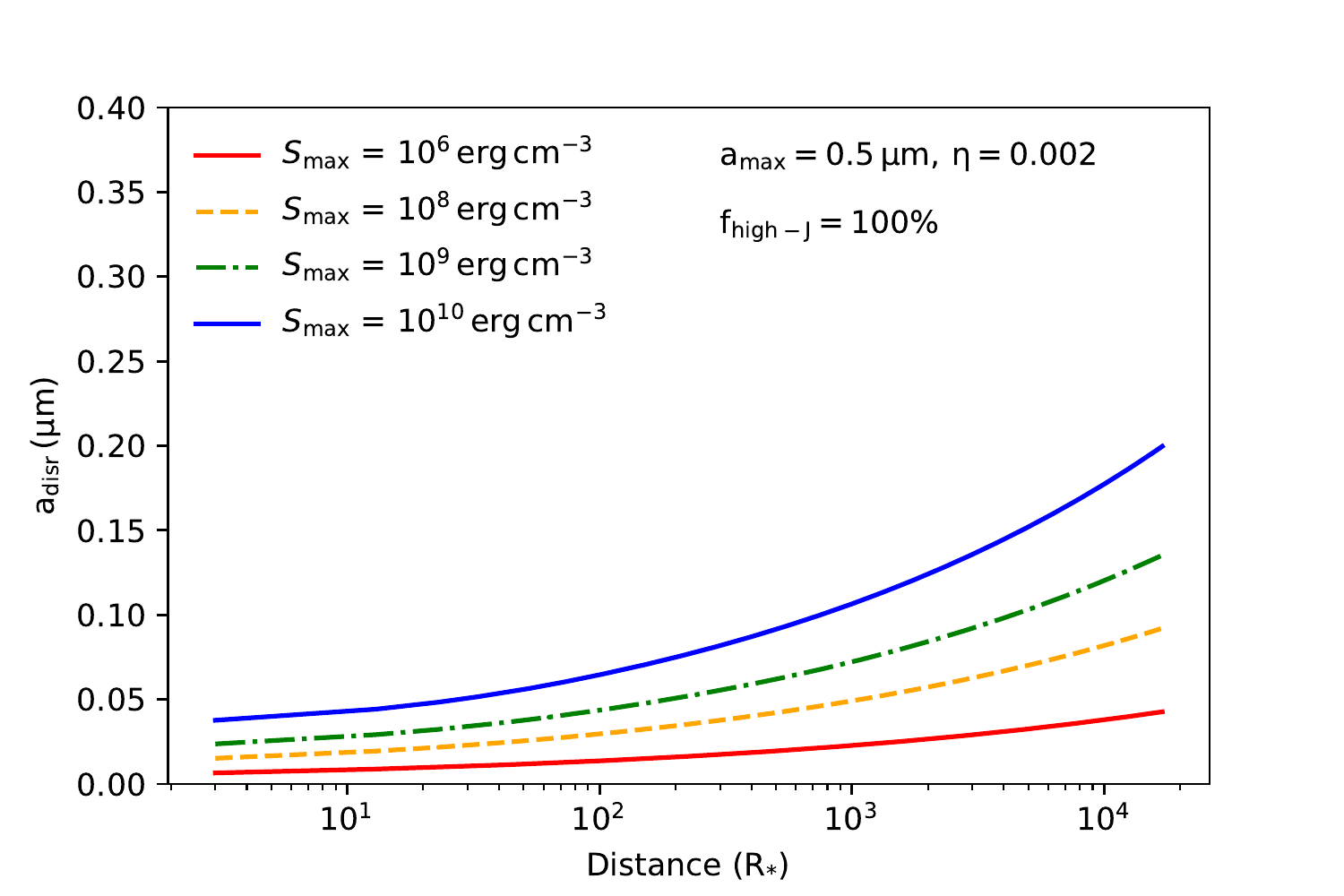} 
    \includegraphics[width = 0.48\textwidth]{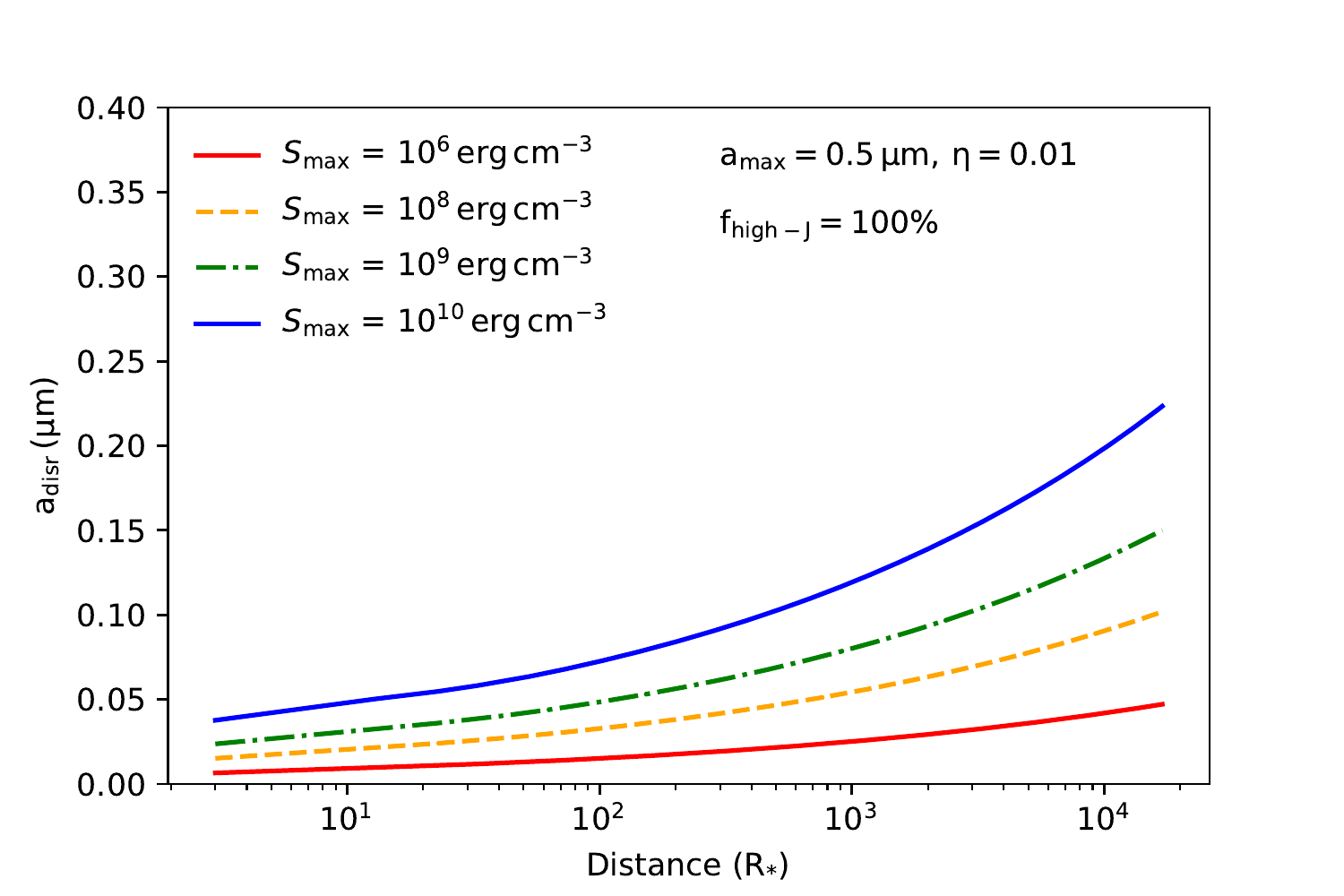} 
    \caption{Disruption grain size vs. radial distance from the central star for different values of tensile strength $S_{\rm max} = 10^{6} - 10^{10}\,\rm erg\,cm^{-3}$, and for different dust-to-gas ratios $\eta = 0.002$ (left panel) and $\eta = 0.01$ (right panel), assuming the original dust has the maximum size $a_{\rm max}=0.5\,\mu$m and is fully disrupted by RATs ($f_{\rm high-J} = 100\%$). 
    The disruption is more effective for grains with lower $S_{\rm max}$ and low dust-to-gas ratio.}
    \label{fig:Grain size_eta}
    \includegraphics[width = 0.48\textwidth]{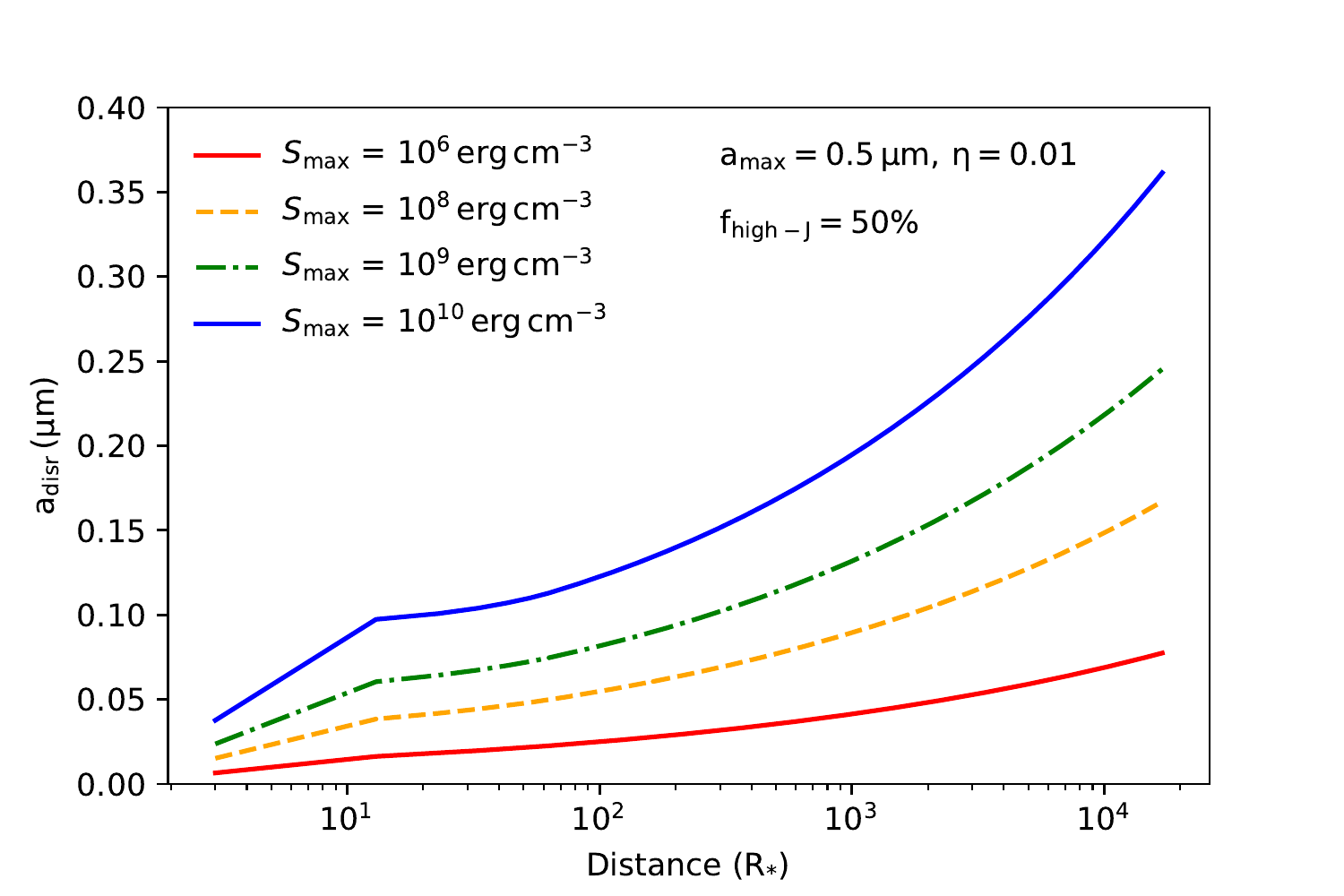} 
    \includegraphics[width = 0.48\textwidth]{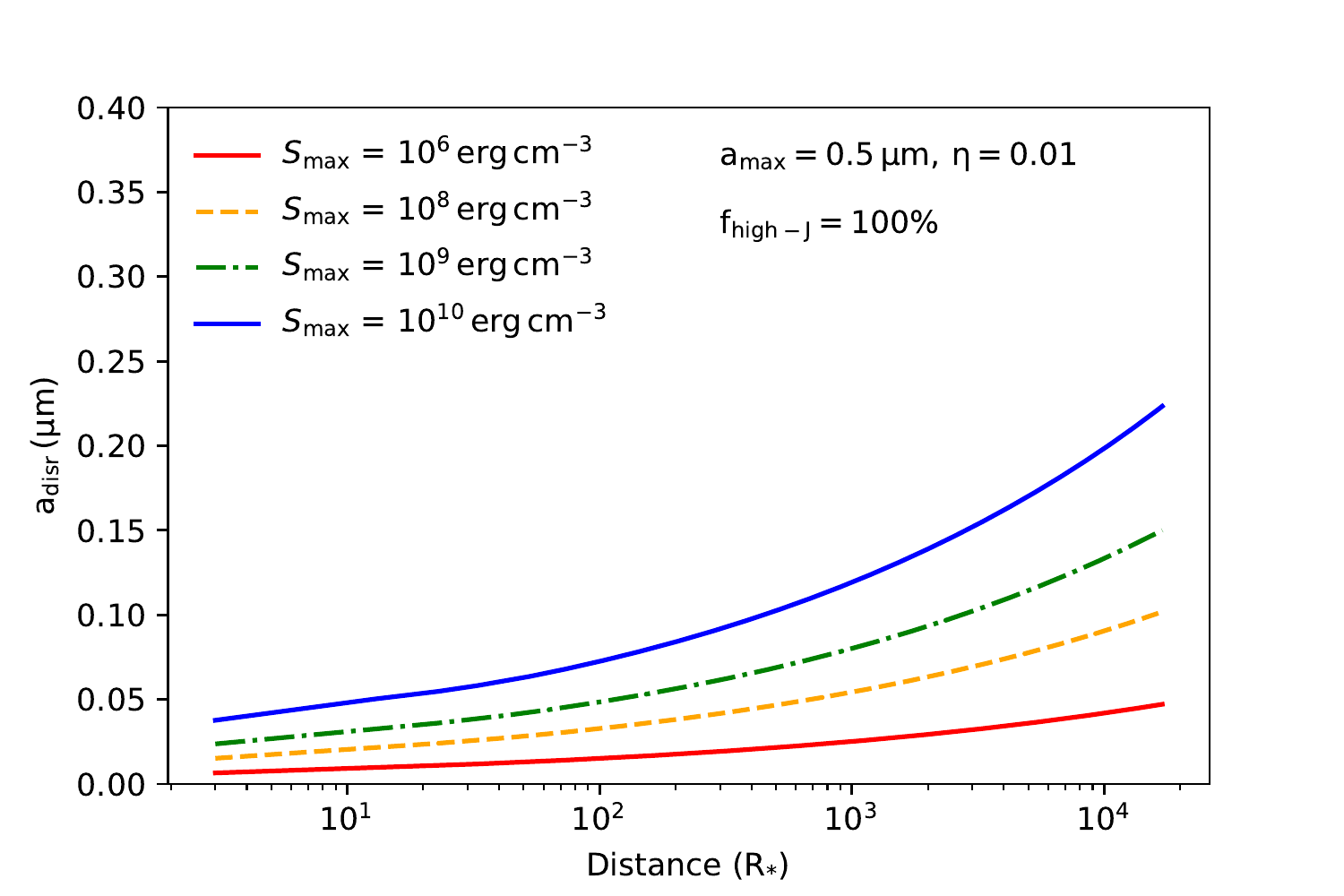} 
    \caption{Same as Figure \ref{fig:Grain size_eta} but considering the effects of $f_{\rm high-J}$ ranging from $50\%$ (left panel) to $100\%$ (right panel). A fixed dust-to-gas ratio of 0.01 is assumed. The RAT-D mechanism becomes more efficient when applying on grains at high-J attractors, resulting in the decreased disruption grain sizes with increasing $f_{\rm high-J}$. }
    \label{fig:Grain size_fraction}
\end{figure*}

Figure $\ref{fig:Grain size_eta}$ illustrates the grain disruption size as a function of the radial distance from the central star for different values of the tensile strength and the dust-to-gas mass ratio. One can see that $a_{\rm disr}$ is smaller than the upper cutoff of the original GSD $a_{\rm max}$ in the entire Betelgeuse envelope. Moreover, $a_{\rm disr}$ radially increases with the envelope distance. The disruption size decreases with decreasing $S_{\rm max}$ or dust-to-gas ratio $\eta$.  

In the framework of the RAT-D mechanism, the grain angular velocity is determined by (1) the spin-up by RATs depending on the radiation strength, and (2) the spin-down by gas collision depending on the gas density. Figure \ref{fig:gas properties} shows that the gas density decreases dramatically as $r^{-2}$. However, the decrease rate of radiation strength is faster than that of the gas density due to the optical depth effect. Thus, the maximum grain rotation rate decreases with increasing radial distance, resulting in the increase of the disruption size. For instance, $a_{\rm disr}$ increases from $\sim 0.05 \,\rm\mu m$ at $r = 3\,\rm R_{\ast}$ to $\sim 0.25\,\mu$m at $r = 17000\,\rm R_{\ast}$.

The internal structure of grains also affects the efficiency of the RAT-D mechanism. The compact grain structure with higher $S_{\rm max}$ is more difficult to disrupt, and the disruption size increases as $a_{\rm disr} \propto S_{\rm max}^{1/4}$ (see Equation 30 in \citealt{Hoang2021}).  
For example, at $r \sim 5000\,\rm R_{\ast}$ within dust-to-gas ratio of 0.002, $a_{\rm disr} \sim 0.03$ and $0.15\,\rm\mu$m for $S_{\rm max}= 10^{6}$ and $10^{10}\,\rm erg\,cm^{-3}$, respectively.

The dust-to-gas mass ratio is found to reduce the RAT-D efficiency and increase the disruption size (see Figure \ref{fig:Grain size_eta}). The reason is that, a higher dust-to-gas ratio corresponds to higher amount of dust, which increases the optical depth and increases the attenuation of stellar radiation in the envelope. Thus, it reduces the effect of RAT-D on dust grains. As an example, for grains with $S_{\rm max} = 10^{9}\erg\cm^{-3}$ located at $r \sim 5000\,\rm R_{\ast}$, $a_{\rm disr}\sim 0.06\,\rm\mu m$ with $\eta = 0.002$, whereas this value $\sim 0.08\,\rm\mu m$ with $\eta = 0.01$.

\begin{figure*}
    \centering
    \includegraphics[width = 0.48\textwidth]{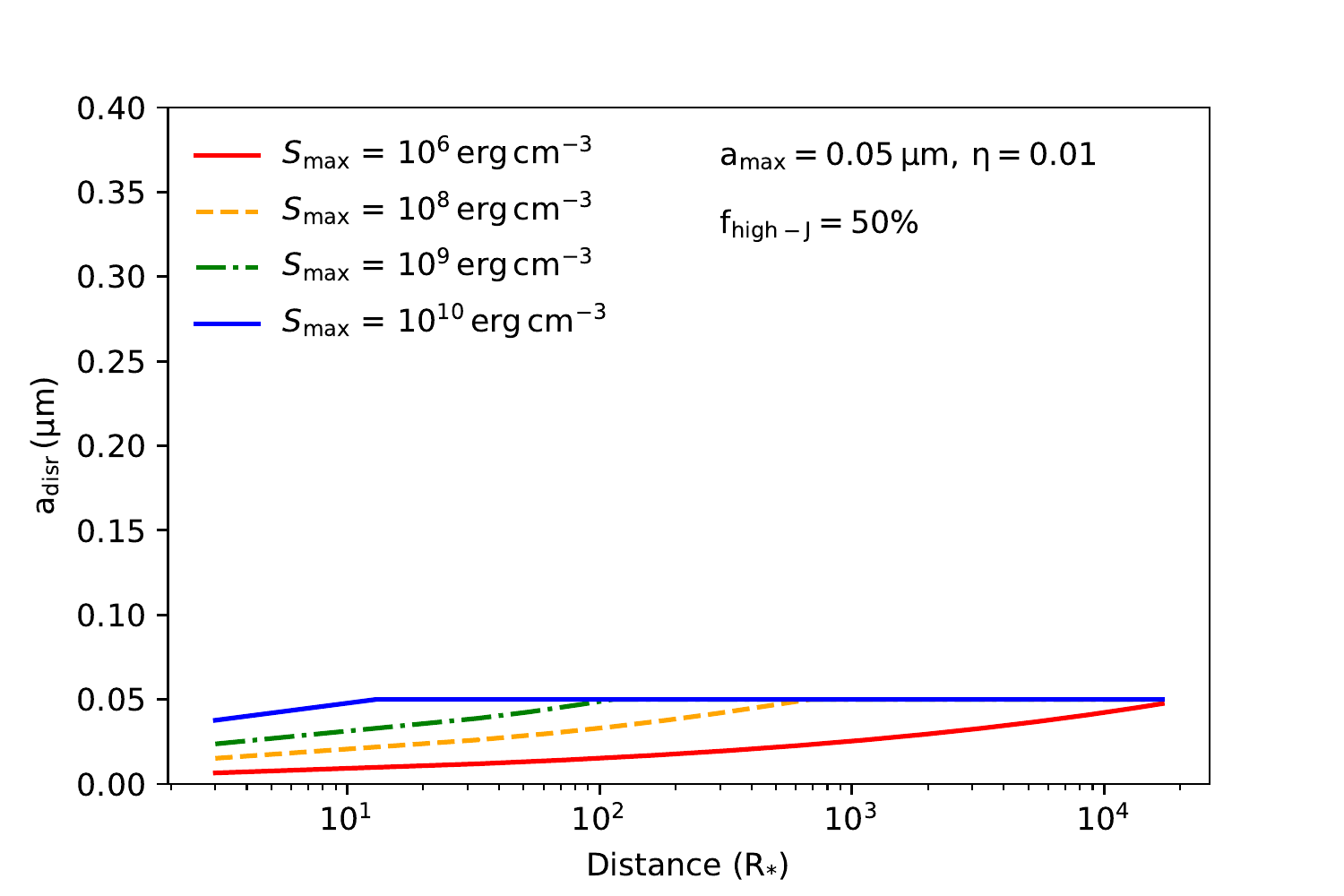}
    \includegraphics[width = 0.48\textwidth]{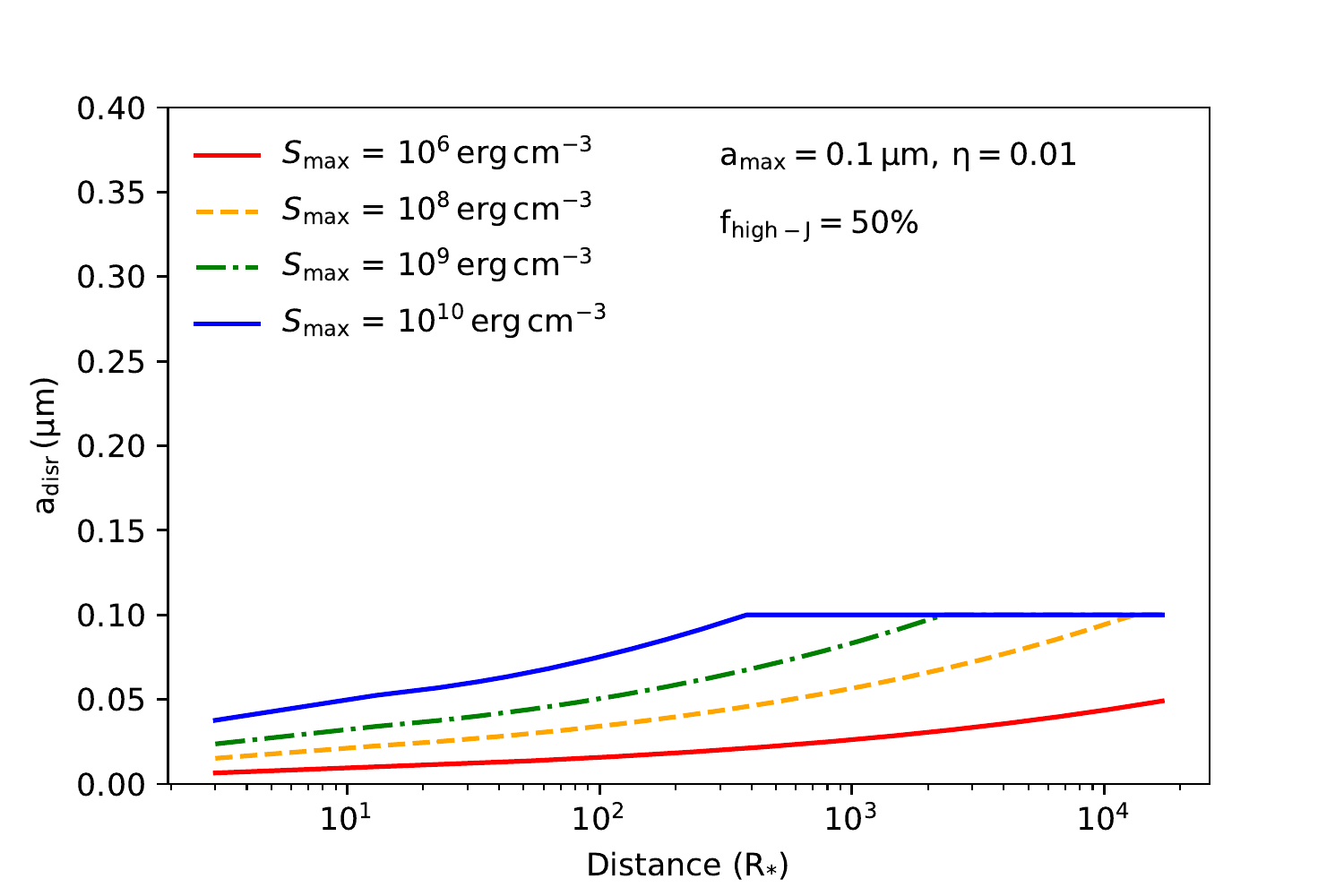}
    \includegraphics[width = 0.48\textwidth]{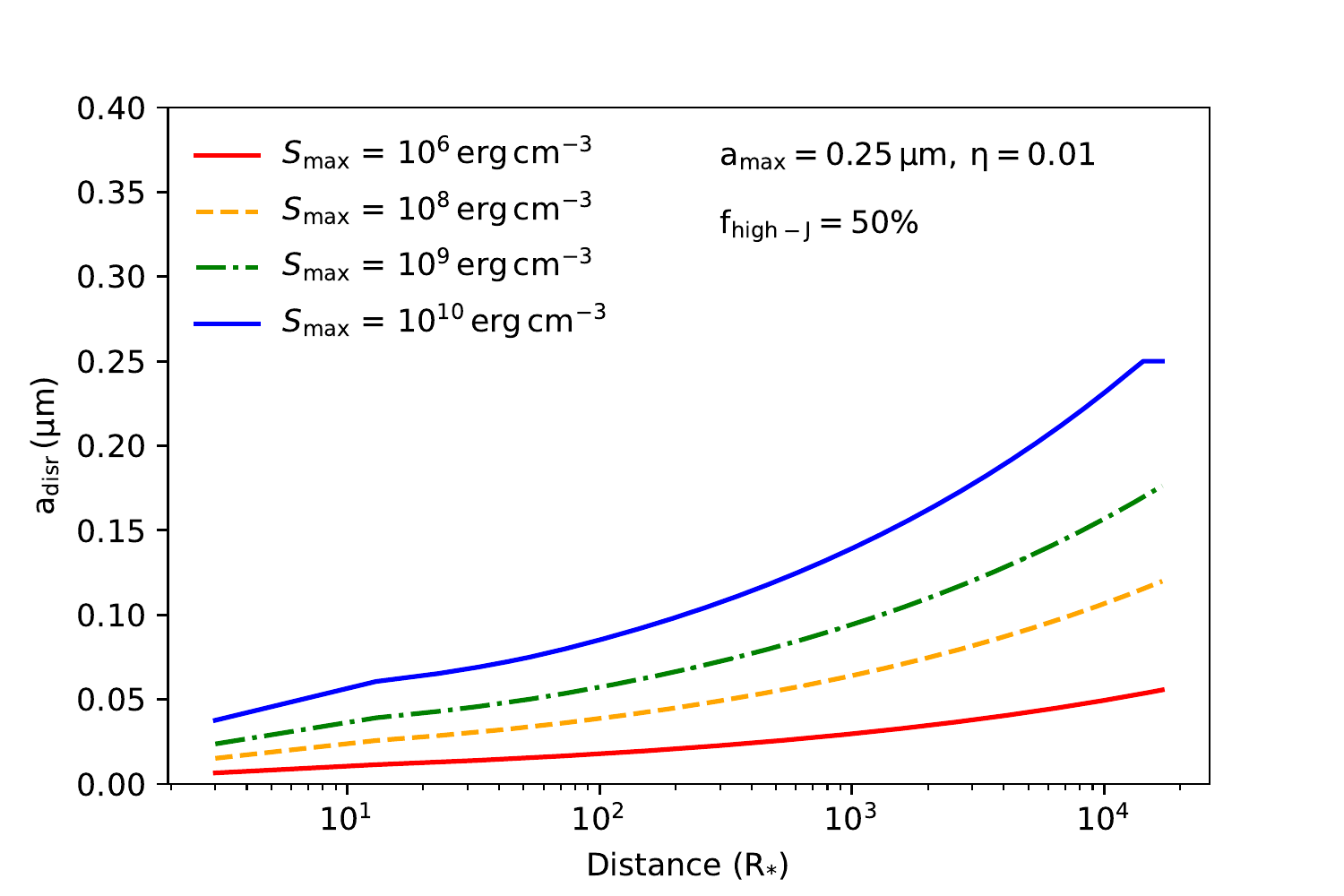}
    \includegraphics[width = 0.48\textwidth]{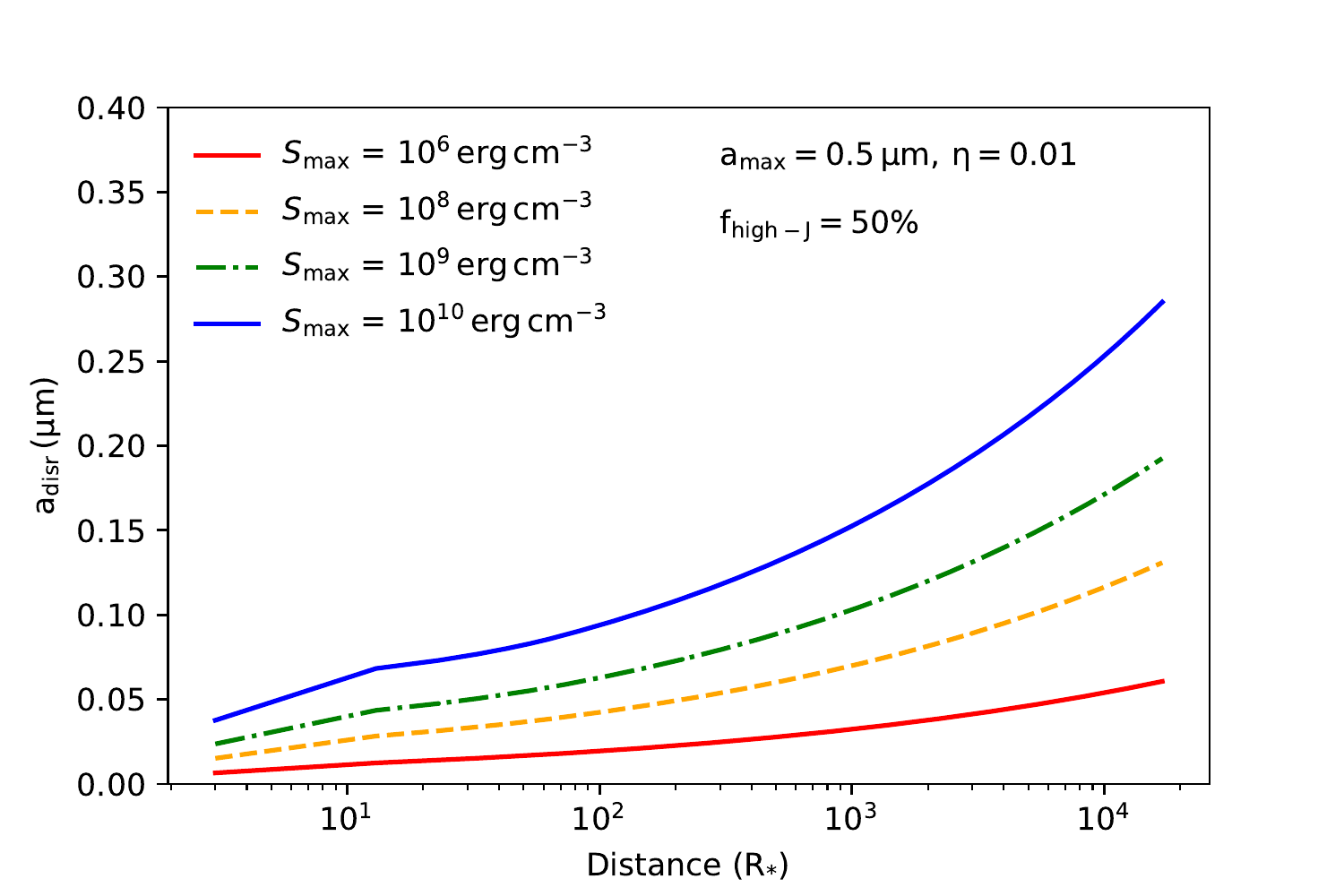}
    \caption{Similar to Figure \ref{fig:Grain size_fraction} but for a different initial maximum grain size $a_{\rm max}$ from 0.05 to 0.5$\,\mu$m. Assuming a fixed $f_{\rm high-J} = 50\%$ and $\eta = 0.01$. The disruption zone is more extended to the entire envelopes as $a_{\rm max}$ increases and $S_{\rm max}$ decreases.}
    \label{fig:Grain size_amax}
\end{figure*} 

Figure \ref{fig:Grain size_fraction} depicts the dependency of the RAT-D efficiency on the fraction of grains at high-J attractors, denoted by $f_{\rm high-J}$, assuming a fixed dust-to-gas-ratio of $\eta=0.01$. A higher fraction of grains at high-J attractors (i.e., higher $f_{\rm high-J}$) are being spun up and undergo a fast disruption, resulting in a lower grain disruption size $a_{\rm disr}$. Meanwhile, the higher abundance of grains at low-J attractors (i.e., lower $f_{\rm high-J}$) requires a larger timescale to be disrupted by RATs than the gas damping time, which reduces the RAT-D efficiency and leads to a higher $a_{\rm disr}$. Taking grains with $S_{\rm max} = 10^{9}\erg\cm^{-3}$ at $r \sim 5000\,\rm R_{\ast}$ as an example, $a_{\rm disr}$ decreases from $\sim 0.3$ to $0.2\,\rm\mu m$ with increasing $f_{\rm high-J}$ from $50\%$ to $100\%$.

Figure \ref{fig:Grain size_amax} shows the similar profiles of $a_{\rm disr}$, but for different values of $a_{\rm max}$, considering $50\%$ of grains at high-J attractors. The disruption zone (in which RAT-D is effective) increases with decreasing tensile strength if $a_{\rm max} < 0.25\,\rm\mu m$ (top panel). For $a_{\rm max}=0.1\,\mu$m, as an example, the disruption zone is $r=300\,\rm R_{\ast}$ for $S_{\rm max}\geq10^{10}\erg\cm^{-3}$ to $r=10000\,\rm R_{\ast}$ for $S_{\rm max} = 10^{8}\erg\cm^{-3}$. And for the initial $a_{\rm max} > 0.25\,\rm\mu m$ (bottom panel), the disruption occurs in the entire envelope.    
   
\subsection{Extinction curve along a line-of-sight (LOS)}
\label{section: Extinction curve results}
We calculate the extinction curve along a LOS toward the central star of Betelgeuse ($z=0$) and the extinction curve toward a background star ($z\neq 0$) (see Figure \ref{fig:Betel_model}).

\begin{figure*}
    \centering
    \includegraphics[width = 0.48\textwidth]{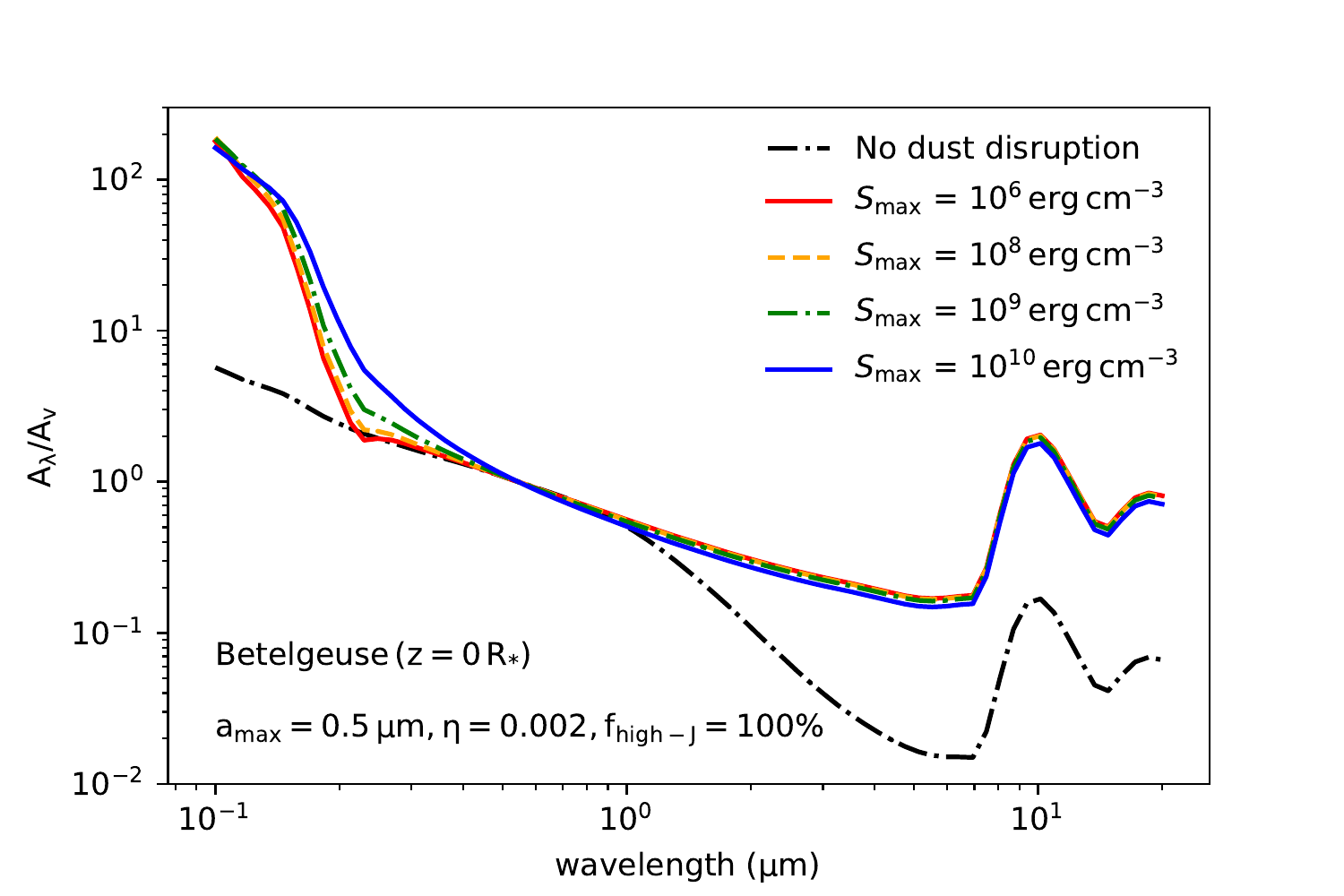}
    \includegraphics[width = 0.48\textwidth]{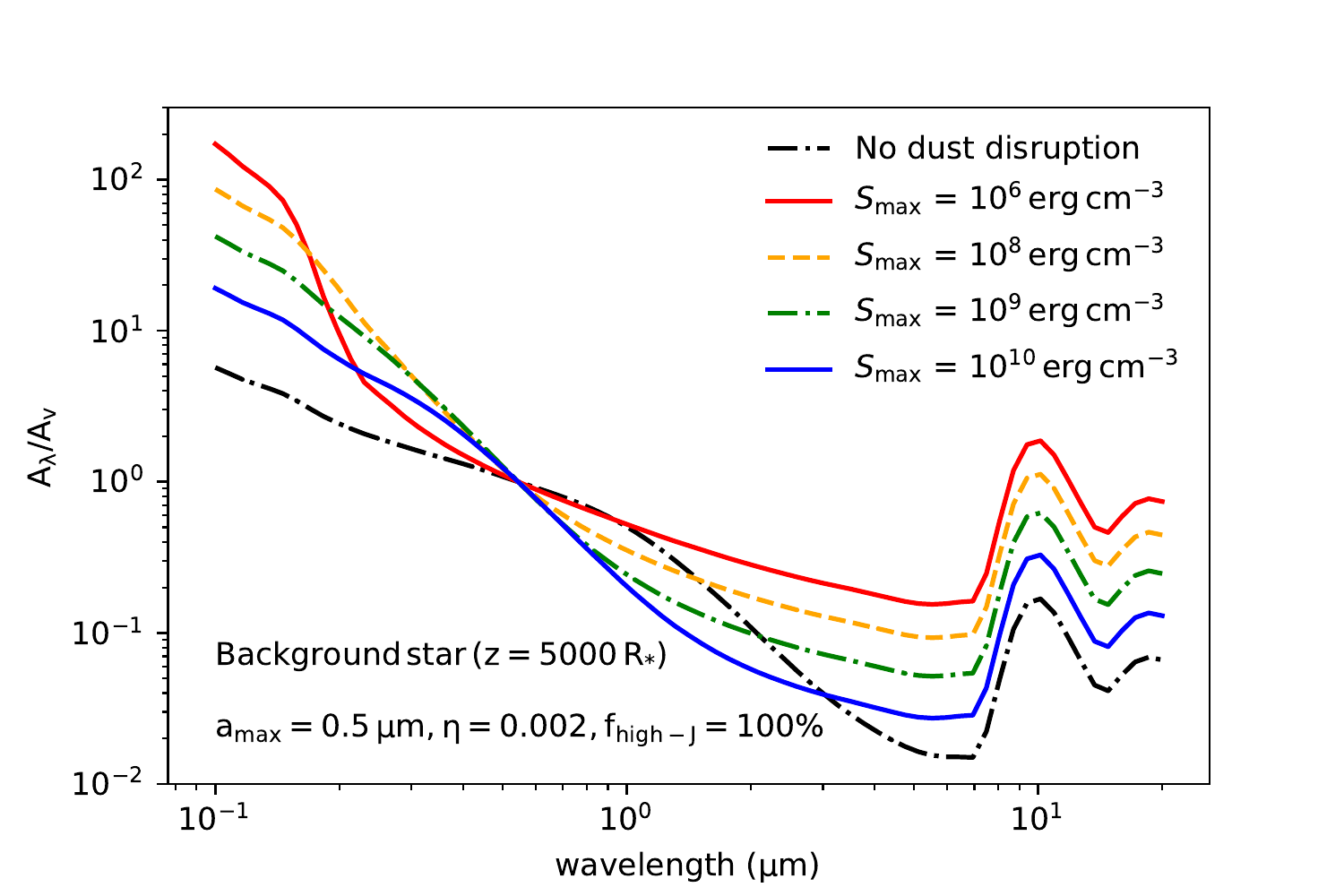}
        \caption{Extinction curve normalized by the visible extinction $A_{\lambda}/A_{\rm V}$ on the LOS toward the central star, Betelgeuse ($z = 0\,\rm R_{*}$, left panel) and toward a background star ($z = 5000\,\rm R_{*}$, right panel). The original dust with $a_{\rm max}=0.5\,\mu$m and $\eta=0.002$ are adopted. Compared with the case of no dust disruption (dash-dotted black line), the RAT-D effect results in a decrease in optical-IR extinction and an increase in far-UV extinction produced by porous grains (lower $S_{\rm max}$).}
        \label{fig:Extinction_Smax_Av}
    \includegraphics[width = 0.48\textwidth]{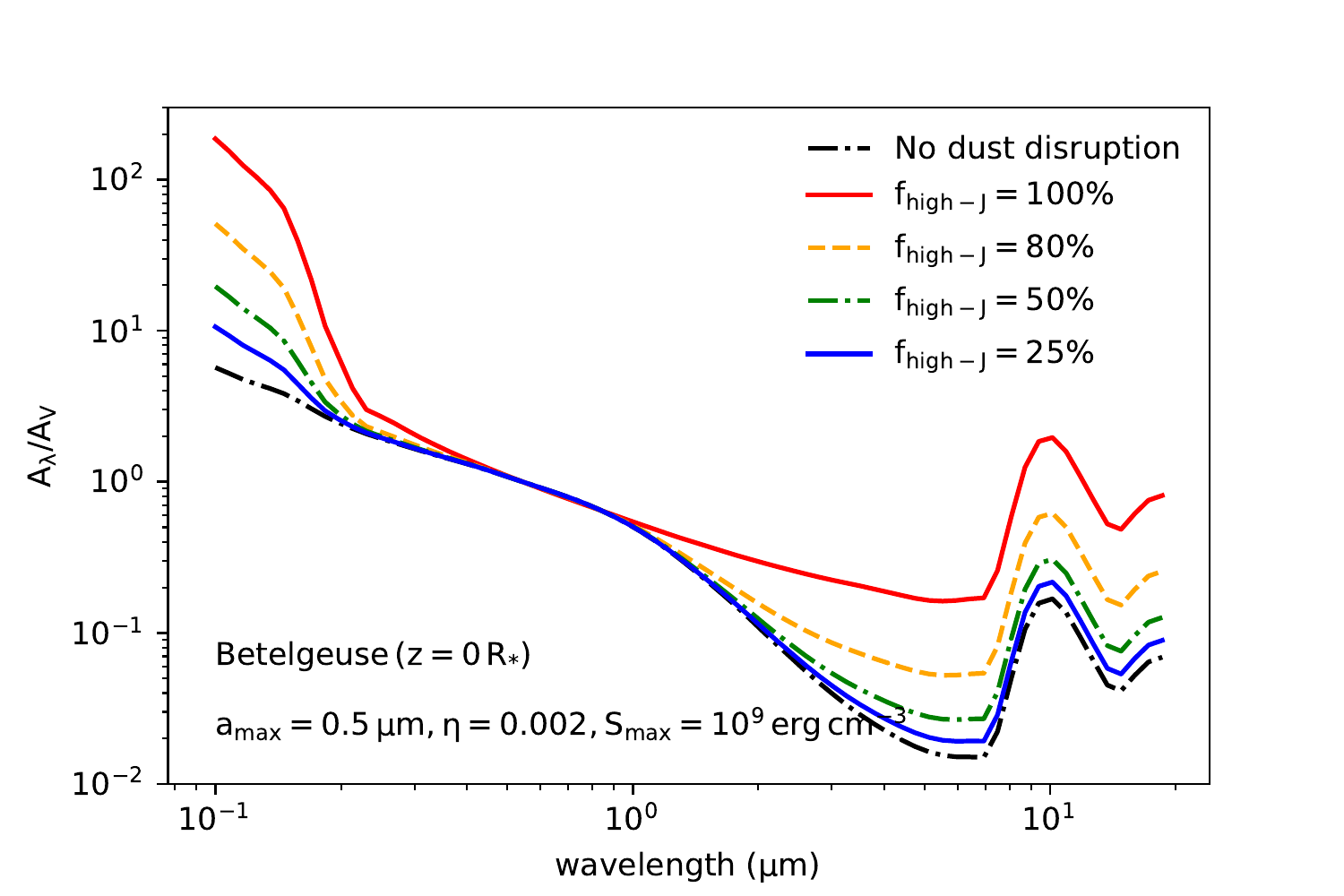}
    \includegraphics[width = 0.48\textwidth]{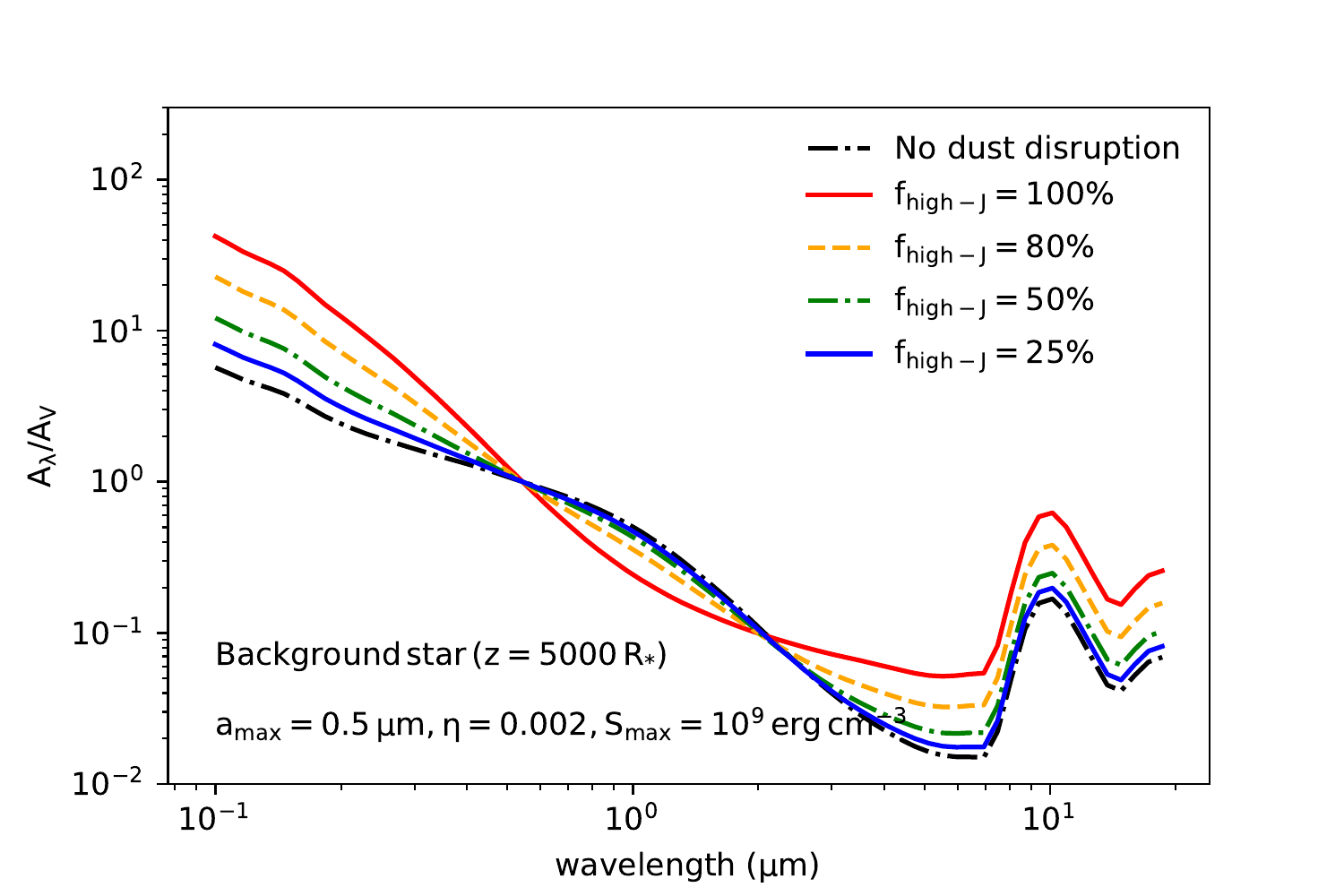}
        \caption{Same as Figure \ref{fig:Extinction_Smax_Av} but with $f_{\rm high-J}$ ranging from $25\%$ to $100\%$, assuming compact grains with $S_{\rm max} = 10^{9}\,\rm erg\,cm^{-3}$. The enhancement of grains at high-J attractors being disrupted by RATs causes the increased far-UV extinction and the decreased optical-IR extinction.}
        \label{fig:Extinction_fraction_Av}
    \includegraphics[width = 0.48\textwidth]{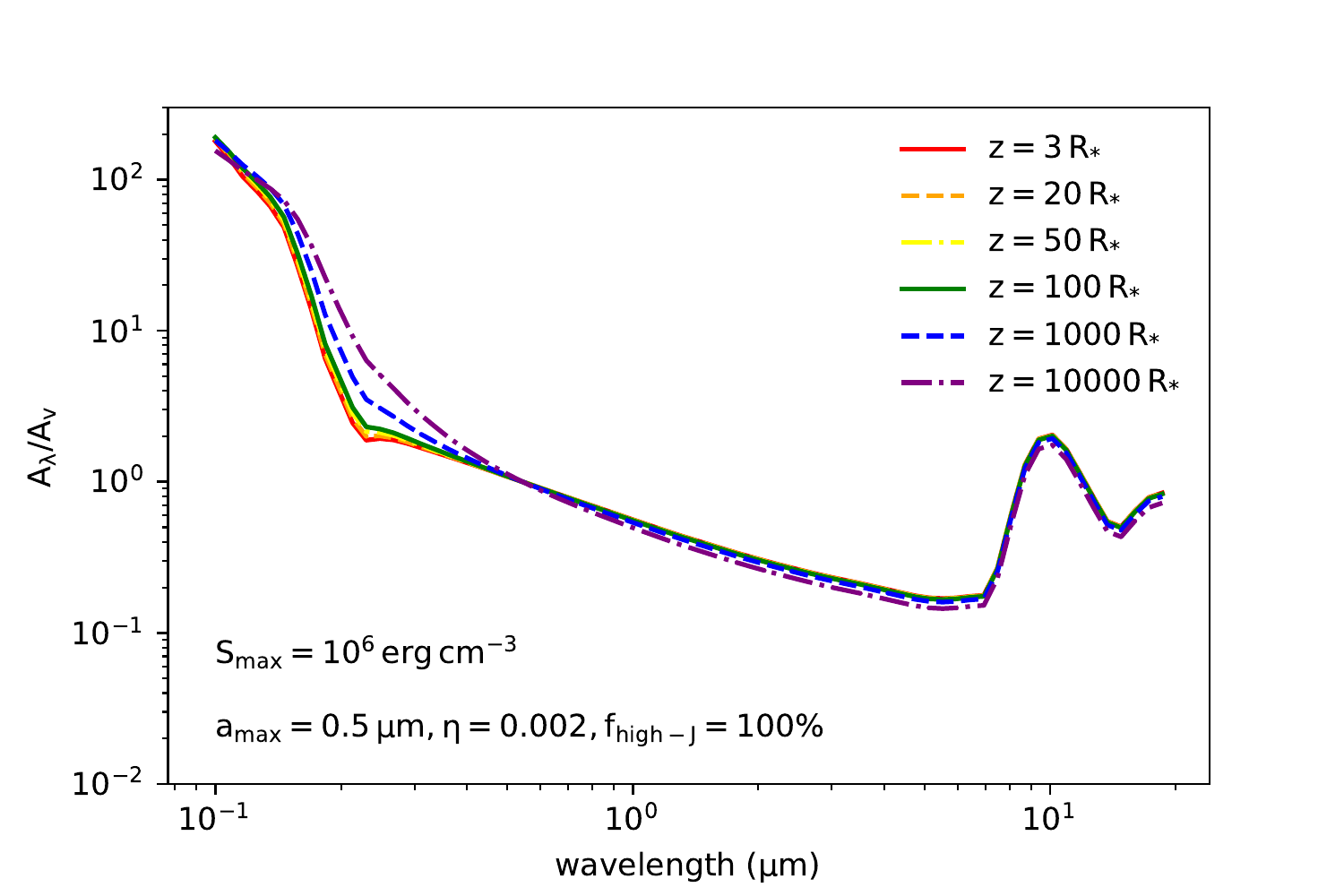}
    \includegraphics[width = 0.48\textwidth]{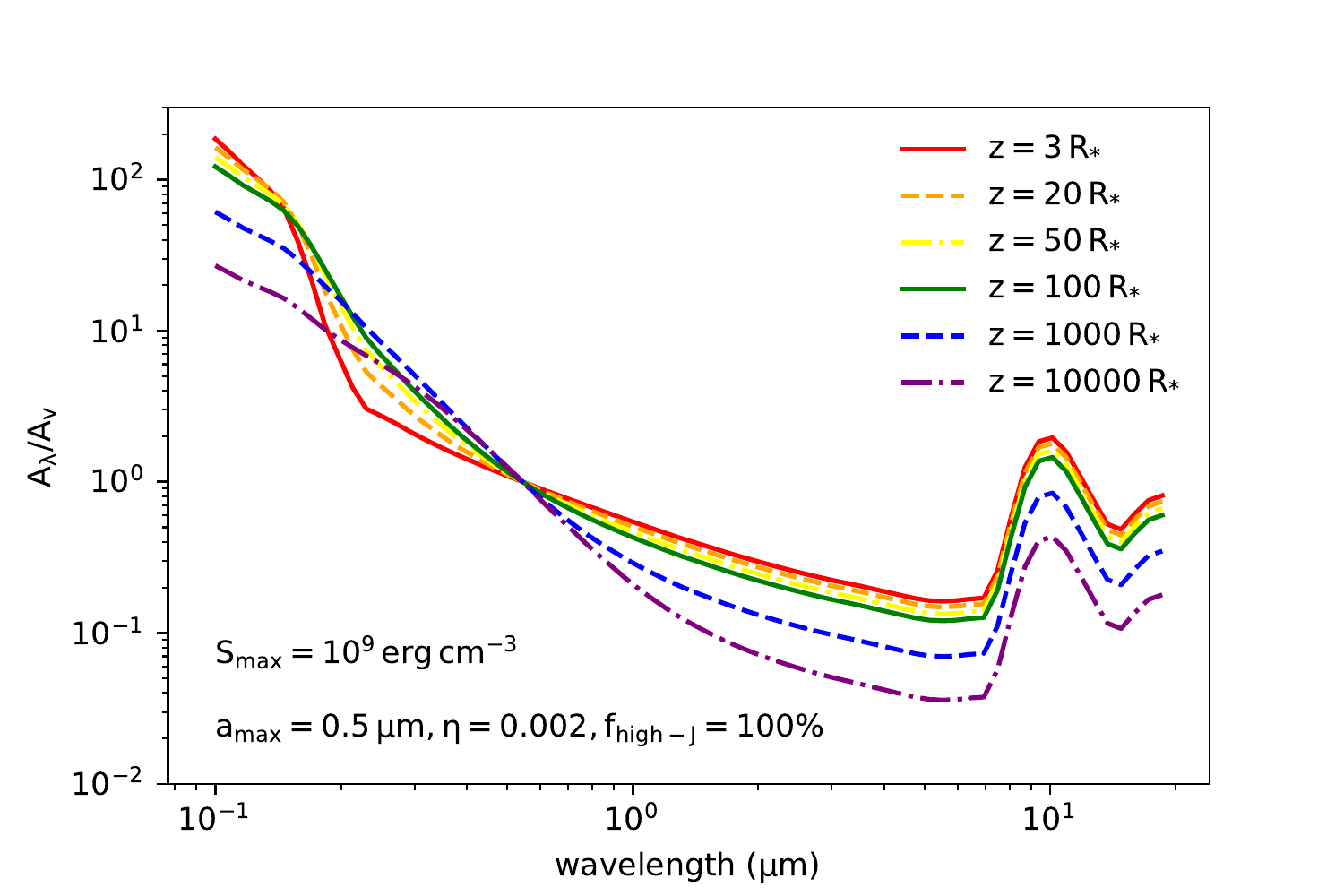}
        \caption{The normalized extinction curve $A_{\lambda}/A_{\rm V}$ for dust grains at $S_{\rm max} = 10^{6}\,\rm erg\,cm^{-3}$ (left panel) and $S_{\rm max} = 10^{9}\,\rm erg\,cm^{-3}$ (right panel) located along the LOS at different positions $z$, considering $f_{\rm high-J} = 100\%$. The original dust with $a_{\rm max}=0.5\,\mu$m and $\eta=0.002$ are adopted. The extinction curve is steeper when observing at lower $z$.}
        \label{fig:Extinction_z_Av}
\end{figure*}

Figure \ref{fig:Extinction_Smax_Av} shows the normalized extinction curve, $A_{\lambda}/A_{\rm V}$, at two comparable projected heights of LOS: $z = 0$ and $z =  5000\,\rm R_{\ast}$. There is a strong extinction bump at 9.7$\,\mu$m - the absorption feature produced by the Si-O stretching mode of astrosilicate grains. Due to the strong enhancement of smaller grains by RAT-D, the extinction curves are significantly higher at far-UV wavelengths, producing steeper slopes than that in the case of no dust disruption (dash-dotted black line). Meanwhile, the extinction increases at IR wavelengths, which is a result of the normalization effects at a visible band ($\lambda=0.5448\,\mu$m) (see Figure \ref{fig:Extinction_Smax_NH} in Appendix).

The left panel of Figure \ref{fig:Extinction_Smax_Av} shows the normalized extinction curve along the LOS of $z = 0$. The slope in the UV-optical ($\lambda\sim 0.2 - 0.6\,\mu$m) is steeper for lower value of $S_{\rm max}$. It is associated with the decrease in $a_{\rm disr}$ with decreasing $S_{\rm max}$ as discussed in Section \ref{section:Grain disruption sizes}. Grains with porous structure (lower $S_{\rm max}$) are easier to be disrupted (lower $a_{\rm disr}$) by the RAT-D mechanism, which results in the high abundance of smaller grains. As a consequence, the extinction gets decreased at UV-optical. On the contrary, the disruption of grains with compact structure (higher $S_{\rm max}$) is harder to occur, which shows a higher extinction at UV-optical wavelengths.

The right panel of Figure \ref{fig:Extinction_Smax_Av} shows the normalized extinction curves toward a background star at projected distance $z = 5000\,\rm R_{\ast}$. The effect of small porous astrosilicate grains is more significant at the far-UV regimes, with higher extinction at lower $S_{\rm max}$. Meanwhile, the IR extinction increases with decreasing $S_{\rm max}$. The reason is that the disruption size in all slabs along the LOS at $z=5000\,\rm R_{\ast}$ (see Figure \ref{fig:Betel_model}) is larger and significantly different with varying $S_{\rm max}$ (see Figure \ref{fig:Grain size_eta}). The discrepancy of extinction curves consequently extends from UV-optical to IR regimes (see Figure \ref{fig:Extinction_Smax_NH} in Appendix). The weaker disruption effect at lower $S_{\rm max}$ is unable to increase the extinction at IR wavelengths, such that the normalization at a visible band causes an opposite order as seen in Figure \ref{fig:Extinction_Smax_Av} (right panel).

Figure \ref{fig:Extinction_fraction_Av} shows the effects of $f_{\rm high-J}$ on the normalized extinction curves, assuming compact grains with $S_{\rm max}=10^{9}\erg\cm^{-3}$. One can see that for grains located at $z = 0$ (left panel), with lower $f_{\rm high-J}$, fewer grains at the high-J attractor points are rapidly fragmented by RAT-D, leading to the increased population of non-disrupted large grains in the envelope and causing a decrease in far-UV extinction. Meanwhile, the normalized extinction gets increased at optical-IR wavelengths, which is a result of the increase in optical extinction for smaller fraction $f_{\rm high-J}$ (see Figure \ref{fig:Extinction_fraction_NH} in Appendix). The right panel shows the similar patterns for grains located along the LOS at $z = 5000\,\rm R_{\ast}$, but with flatter curves caused by higher abundance of large grains in this region.

Figure \ref{fig:Extinction_z_Av} shows the variation of extinction curves with $z$ for grains that are fully disrupted ($f_{\rm high-J} = 100\%$), with a fixed tensile strength of $S_{\rm max}=10^{6}\erg\cm^{-3}$ (left panel) and $S_{\rm max}=10^{9}\erg\cm^{-3}$ (right panel). The extinction is higher at far-UV regimes for lower $z$, which results from the enhancement of the small astrosilicate grains caused by the disruption. Notably, the extinction curve at IR wavelengths is higher for smaller $z$ with $S_{\rm max}=10^{9}\erg\cm^{-3}$, which is opposite the case of $S_{\rm max}=10^{6}\erg\cm^{-3}$ at UV-optical wavelengths. At the same height $z$, a porous grain is much easier to be disrupted than the compact one. The extinction at optical wavelength is less affected for grains with higher $S_{\rm max}$ (see Figure \ref{fig:Extinction_z_NH} in Appendix). Thus, the normalization of the extinction curve at the optical wavelength leads to the difference in Figure \ref{fig:Extinction_z_Av}.

Another notable difference between the left and right panels in Figures \ref{fig:Extinction_Smax_Av}, \ref{fig:Extinction_fraction_Av}, and \ref{fig:Extinction_z_Av} is that the steepness in the right panel is flatter than that in the left panel. The explanation is due to the total-to-selective extinction $R_{\rm V}$, which will be described in the next section.

\subsection{Total-to-selective extinction, $R_{\rm V}$}

\begin{figure*}
    \centering
    \includegraphics[width = 0.48\textwidth]{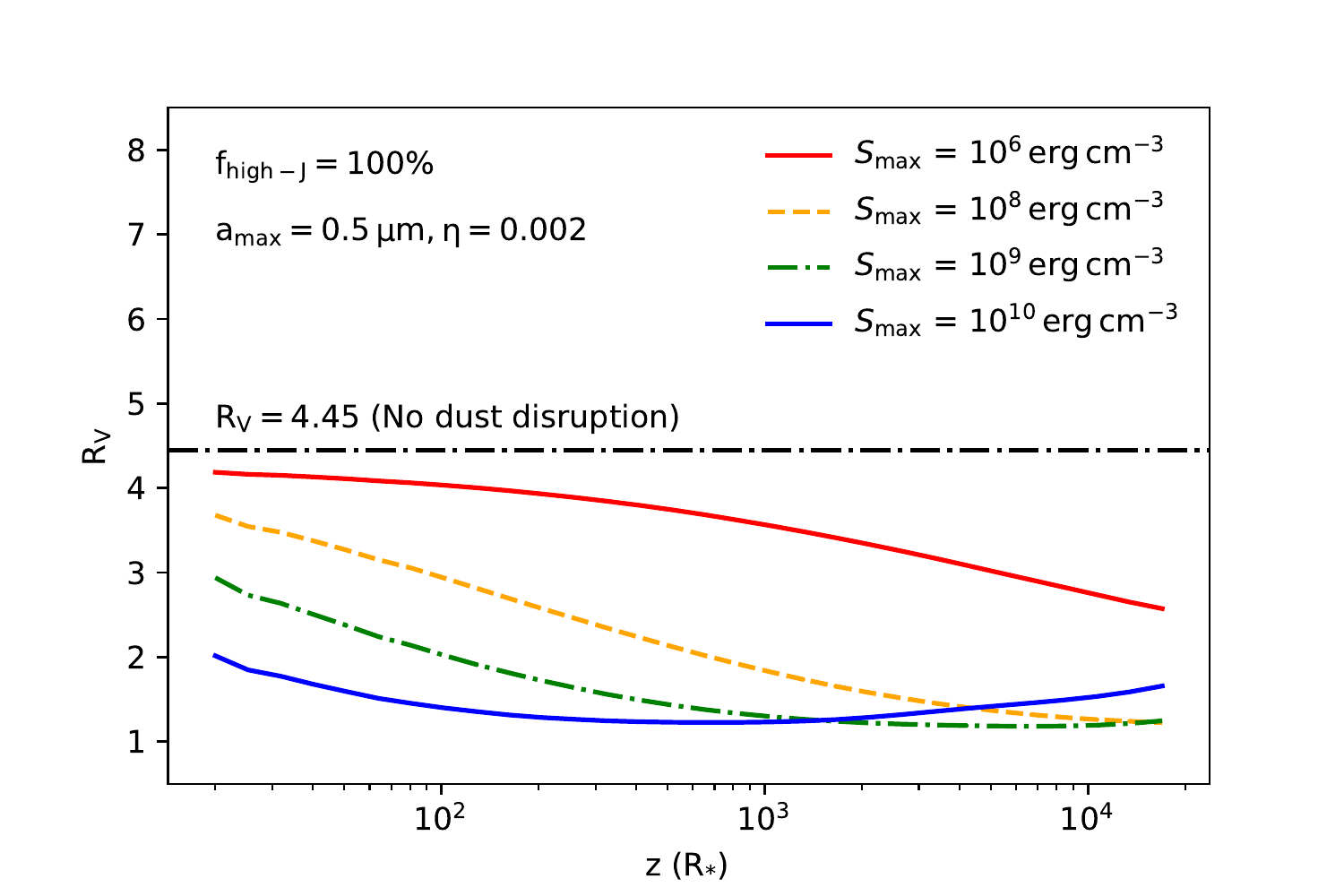}
    \includegraphics[width = 0.48\textwidth]{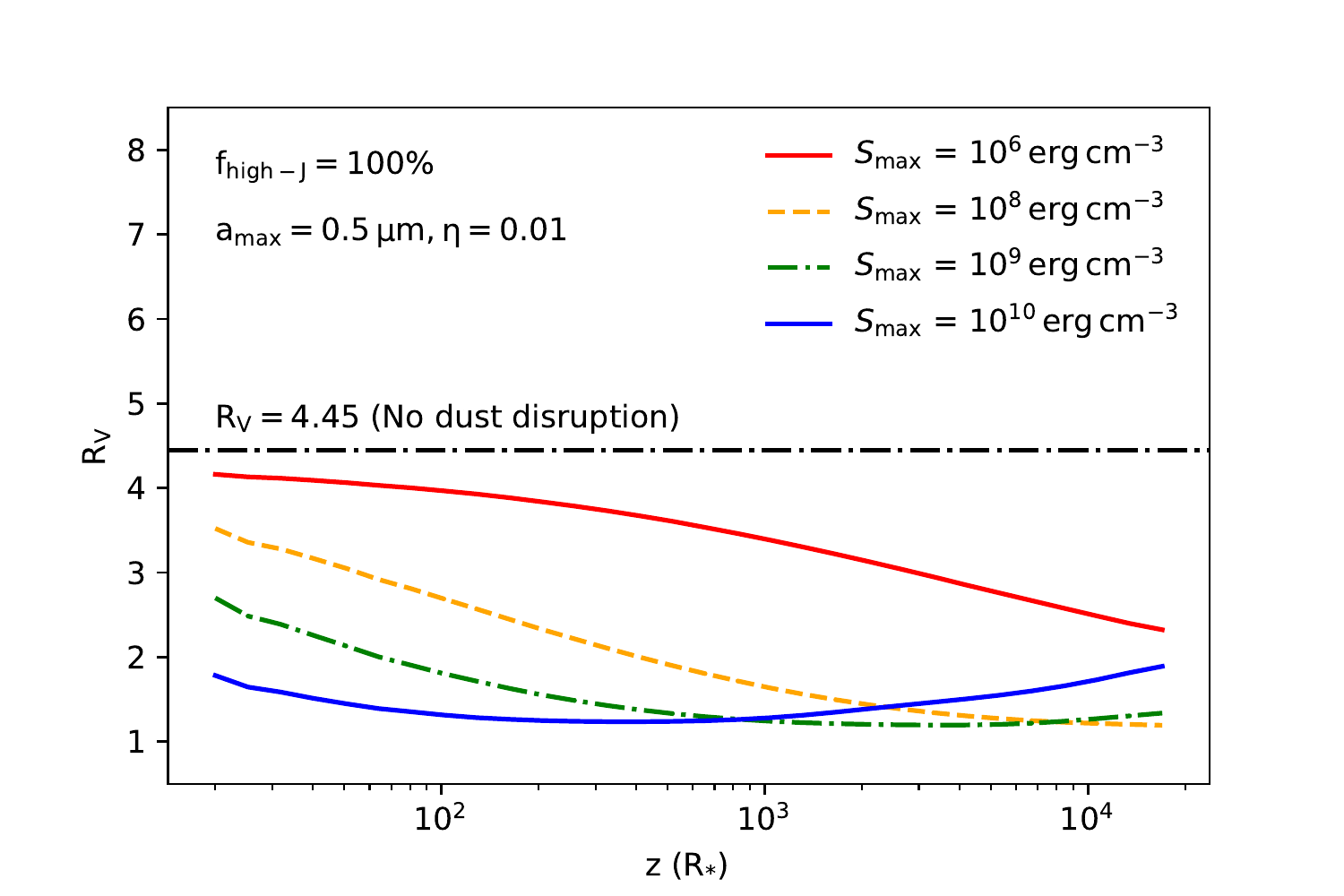}
    \caption{The change in total-to-selective ratio, $R_{\rm V}$, over the projected height $z$ of the LOS toward a background star, assuming at different values of $S_{\rm max}$ and dust-to-gas ratio $\eta$. Under the effect of RAT-D, the slope of extinction curve is steeper, resulting in lower $R_{\rm V}$ than a fixed value produced by the original dust (dash-dotted black line). The value of $R_{\rm V}$ varies with increasing the height $z$ and the tensile strength.}
    \label{fig:Rv_Smax}
    \includegraphics[width = 0.48\textwidth]{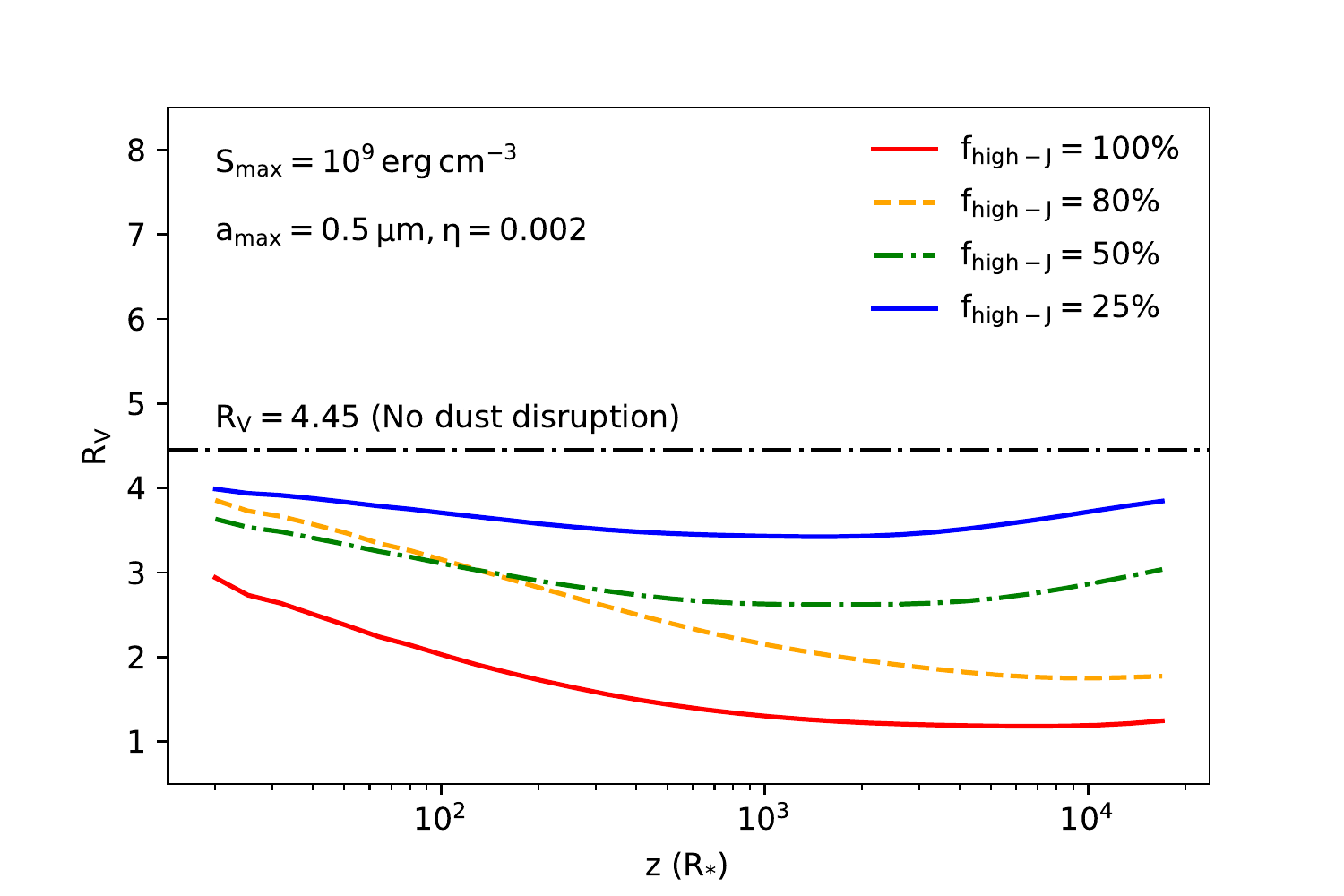}
    \includegraphics[width = 0.48\textwidth]{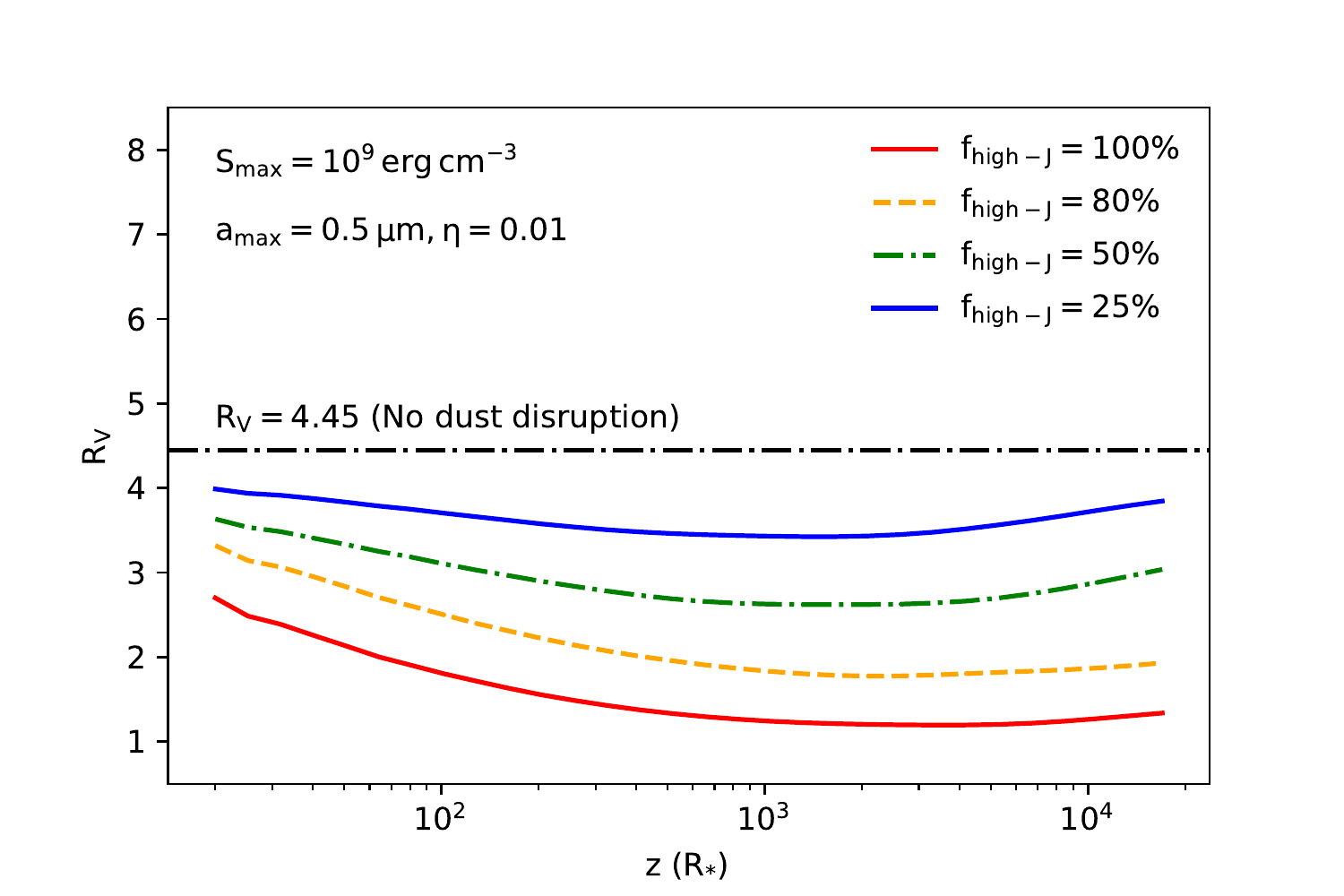}
    \caption{Similar to Figure \ref{fig:Rv_Smax} but for compact grains (i.e., $S_{\rm max} = 10^{9}\erg\cm^{-3}$) with $f_{\rm high-J}$ ranging from $25\%$ to $100\%$. The total-to-selective ratio $R_{\rm V}$ tends to rise with lower fraction of grains at high-J attractors in the CSE.}
    \label{fig:Rv_fraction}
\end{figure*}

 As shown in the previous section, the enhancement of smaller grains decreases the dust extinction in the optical-IR wavelengths, resulting in the steeper extinction curve (see Section \ref{section: Extinction curve results}). Additionally, the extinction patterns are affected by the optical properties of O-rich materials, resulting in different values of $R_{\rm V}$ with respect to grain sizes as illustrated in Figure \ref{fig:Rv_ISM}. In this section, we calculate the total-to-selective extinction, $R_{\rm V}$, and show how $R_{V}$ changes concerning the height $z$ of the LOS.

Figure \ref{fig:Rv_Smax} shows the variation of $R_{\rm V}$ with the projected height $z$ for different values of tensile strength and dust-to-gas ratio. Small disrupted grains of $a < a_{\rm disr}$ cause the increase in far-UV extinction and the decrease in optical-IR extinction, leading to the observed $R_{\rm V} < 4.45$ - lower than the standard value of $R_{\rm V}$ produced by the original GSD of $a < a_{\rm max} = 0.5\,\rm\mu m$ (dash-dotted black line).

One can see from Figure \ref{fig:Rv_Smax} that $R_{\rm V}$ increases when observing at lower position $z$. This is because of the increasing effects of RAT-D toward the central star, which reduces the abundance of large grains, as shown in Figure \ref{fig:Grain size_eta}. Since the major contribution of small astrosilicates of $a < 0.05 \,\rm\mu m$ is only in the far-UV extinction, the extinction curve is flatter at optical-IR regimes and results in higher $R_{\rm V}$. Meanwhile, the increased disruption size over the envelope distance (see Figure \ref{fig:Grain size_eta}) causes the enhancement of optical-IR extinction, leading to a decrease in $R_{\rm V}$ with increasing height $z$. The results are consistent with the features of the extinction curve along the LOS at lower $z$ as depicted in Figure \ref{fig:Extinction_z_Av}.

The grain internal structure plays a crucial role on the value of $R_{\rm V}$. Close to the central star, large astrosilicate grains with lower $S_{\rm max}$ are fragmented into smaller sizes of $a < 0.05 \,\rm\mu m$, resulting in higher $R_{\rm V}$. Meanwhile, compact grains are difficult to be disrupted by RATs, leading to higher grain disruption size and lower value of $R_{\rm V}$. For instance, at the height $z = 20\,\rm R_{\ast}$ within $\eta = 0.002$, $R_{\rm V}$ decreases considerably from 4.1 to 2 for grains with $S_{\rm max} = 10^{6}\,\rm erg\,cm^{-3}$ and $S_{\rm max} = 10^{10}\,\rm erg\,cm^{-3}$. However, at the outer CSE, astrosilicate grains of $a > 0.1\,\rm\mu m$ with high tensile strength of $S_{\rm max} > 10^{9}\,\rm erg\,cm^{-3}$ can survive against the RAT-D mechanism and enhance the optical-IR extinction. The value of $R_{\rm V}$ increases slightly, as a example at $z = 10000\,\rm R_{\ast}$, from 1.3 for grains with $S_{\rm max} = 10^{9}\,\rm erg\,cm^{-3}$ to 1.8 for these with $S_{\rm max} = 10^{10}\,\rm erg\,cm^{-3}$. The results are linked to the change in the slope of the observed extinction curve $A_{\lambda}/A_{\rm V}$ in Figure \ref{fig:Extinction_Smax_Av}.

One contributing factor that affects $R_{\rm V}$ is the dust-to-gas mass ratio. In a low dust-to-gas ratio environment, dust reddening effect is weaker and the RAT-D mechanism becomes more efficient. Due to stronger disruption, the effects of grain internal structure is more significant. As can be seen from Figure \ref{fig:Rv_Smax}, the envelope with a lower dust-to-gas ratio has a higher $R_{\rm V}$ for porous grains (i.e., lower $S_{\rm max}$), while this value is lower for compact grains (i.e., higher $S_{\rm max}$). As an example of grains with $S_{\rm max} = 10^{6}\erg\cm^{-3}$, $R_{\rm V} \sim 2.8$ as maximum within $\eta=0.002$ and decreases to $\sim 2.4$ with $\eta=0.01$, while it can increase from $\sim 1.7$ to $\sim 2$ for grains with $S_{\rm max} = 10^{10}\erg\cm^{-3}$.

Figure \ref{fig:Rv_fraction} shows the effects of grains at low-J and high-J attractors with $f_{\rm high-J} = 25\% - 100\%$. We consider these grains have a compact structure with $S_{\rm max} = 10^{9}\erg\cm^{-3}$ located in the envelope with $\eta = 0.002 - 0.01$. The RAT-D efficiency increases with increasing $f_{\rm high-J}$, causing smaller $R_{\rm V} < 3.1$ when $f_{\rm high-J} > 50\%$. By constrast, the GSD of non-disrupted grains at low-J attractors enhances the extinction at optical-IR wavelengths, as shown in Figure \ref{fig:Extinction_fraction_Av}. Consequently, when $f_{\rm high-J} < 25\%$, $R_{\rm V}$ becomes higher and can achieve its maximum up to 4.2 (above the standard value of $R_{\rm V} \sim 3.1$ in the ISM, see \citealt{Cardelli1989}).

\subsection{Reddening effects on the stellar radiation spectrum}
\begin{figure*}
    \centering
    \includegraphics[width = 0.48\textwidth]{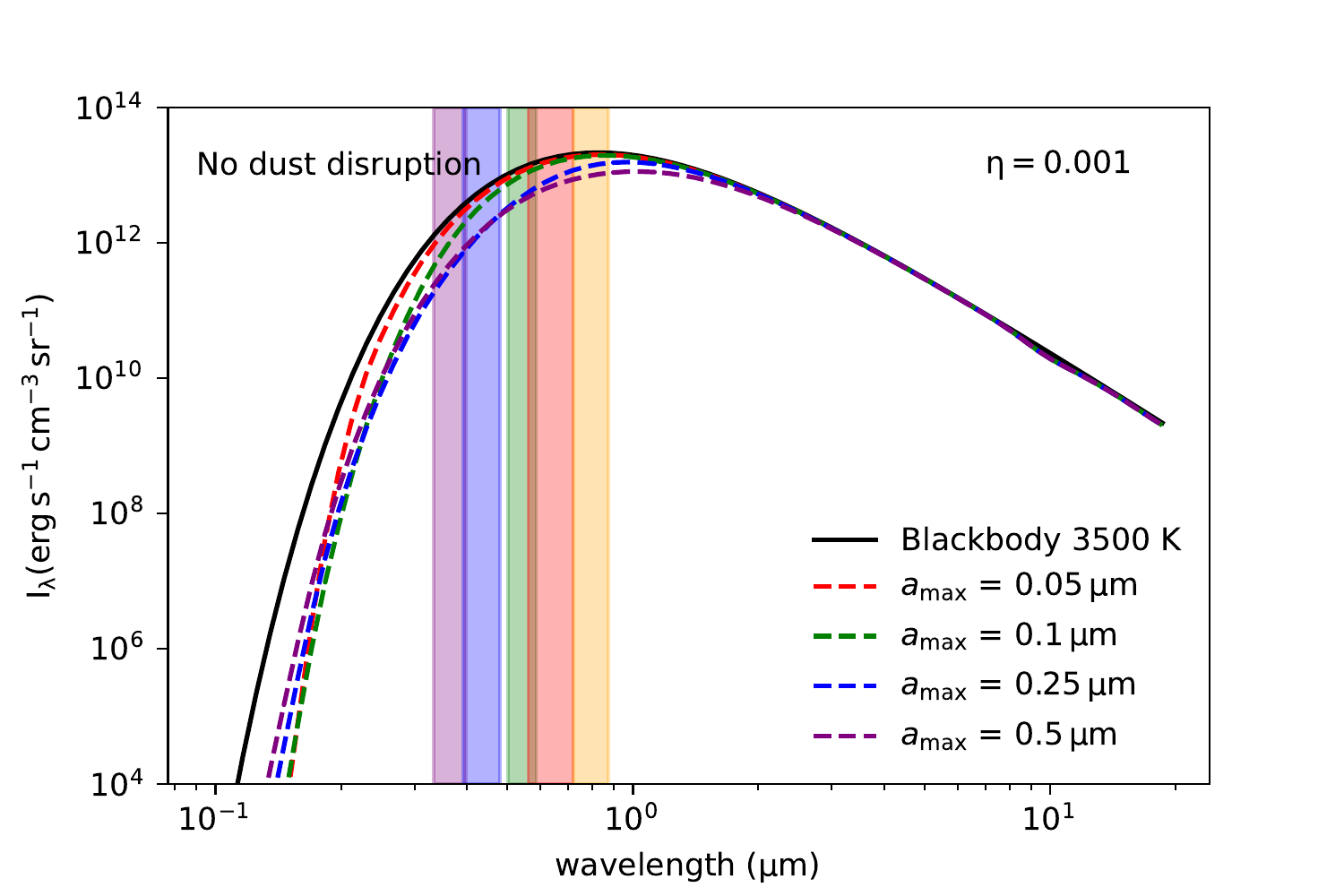}
    \includegraphics[width = 0.48\textwidth]{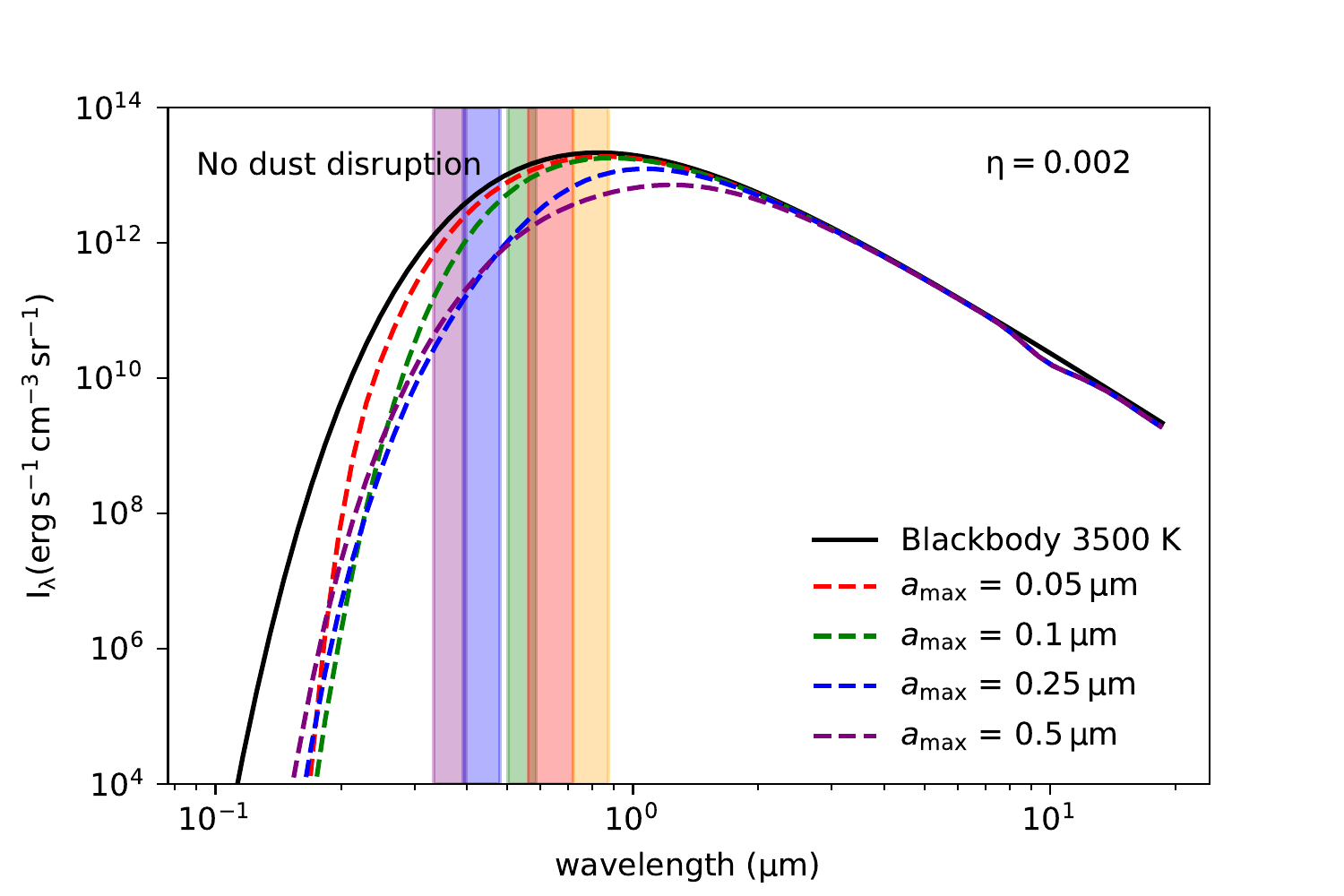}
    \includegraphics[width = 0.48\textwidth]{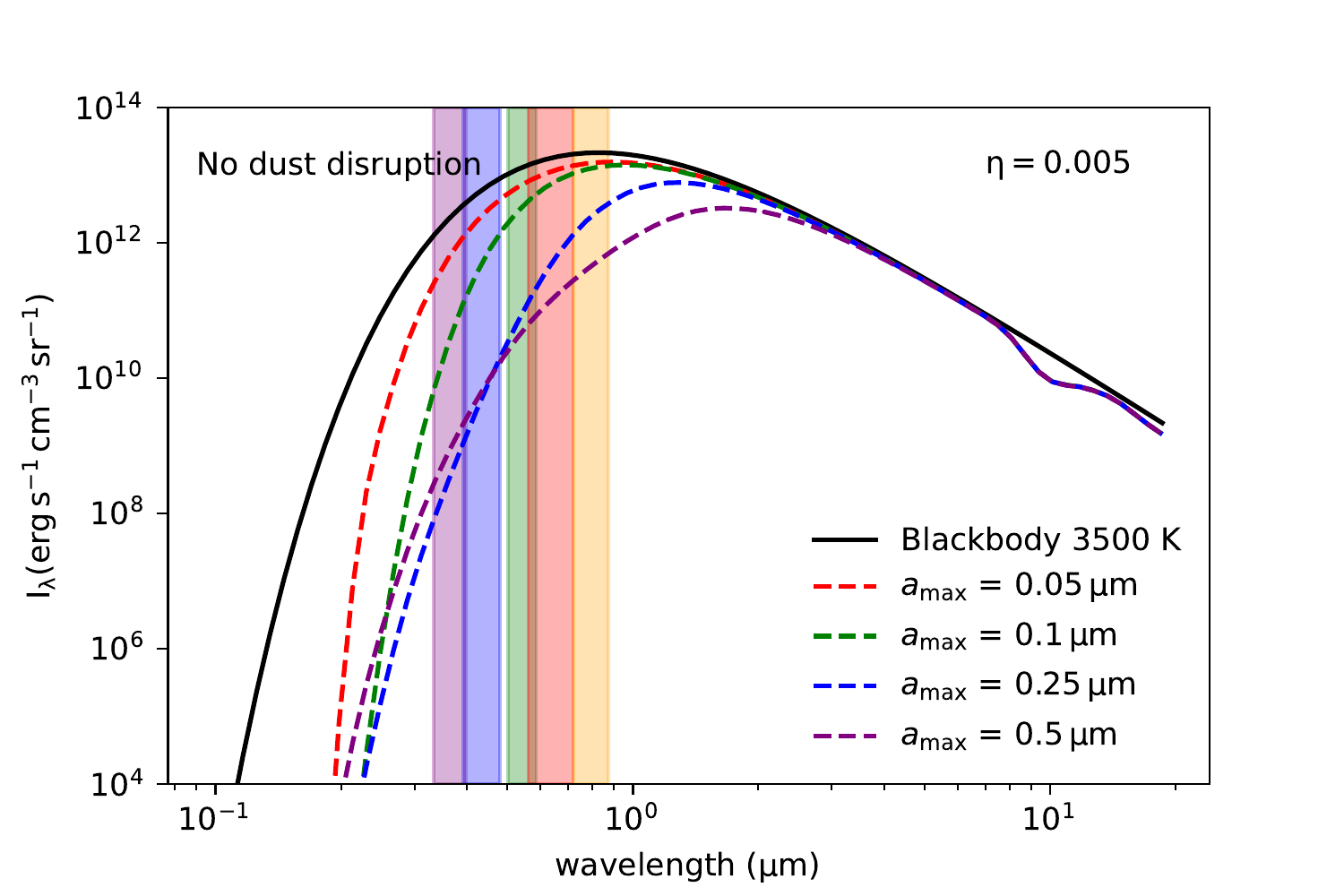}
    \includegraphics[width = 0.48\textwidth]{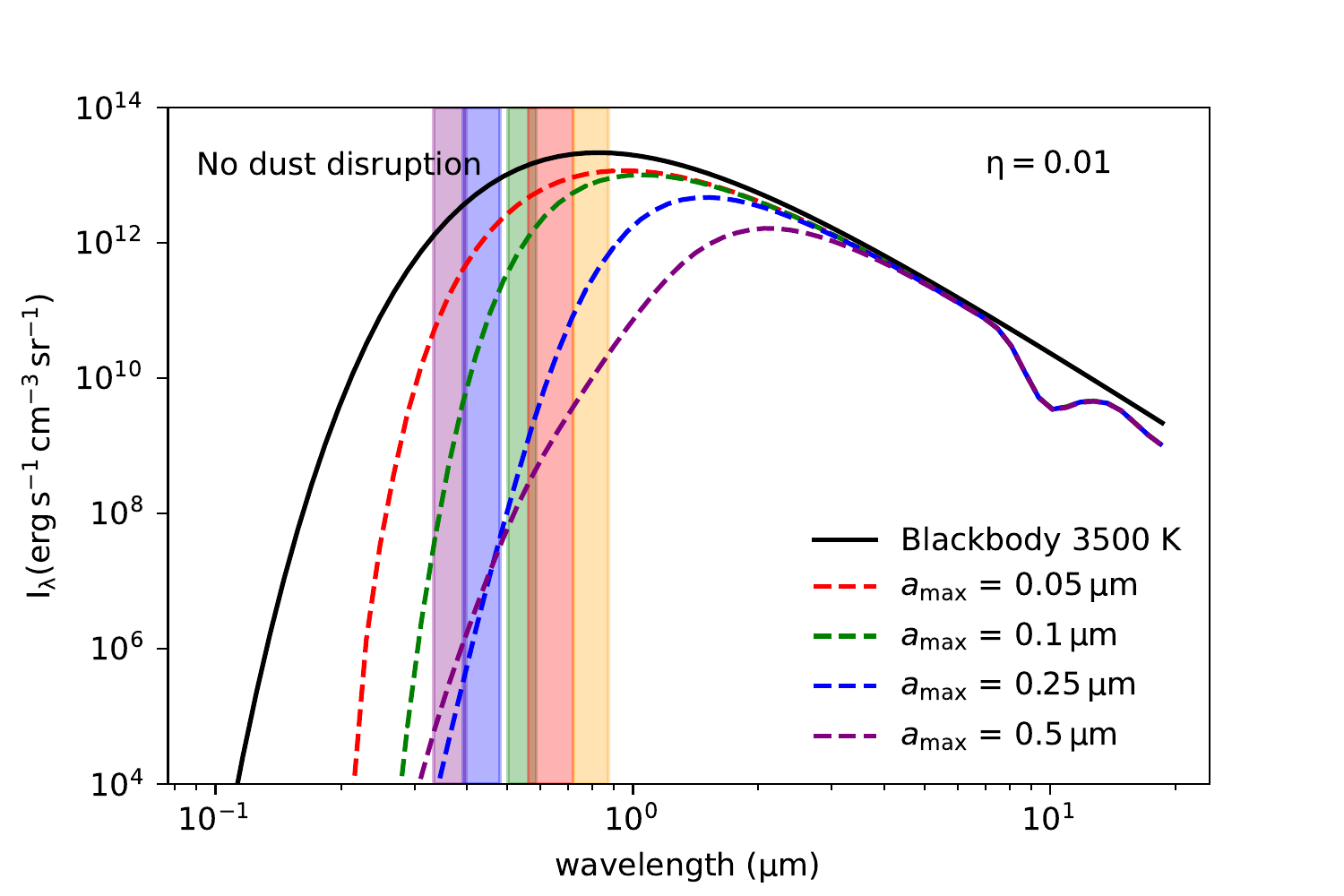}
    \caption{The reddened spectra by CSE dust grains with the variation of $a_{\rm max}$ without RAT-D (dashed color lines). Compared with the intrinsic intensity (solid black line), the spectra are more attenuated by large grains from UV to near-IR wavelengths. Rectangle boundaries illustrate photometric bands (see Table \ref{tab:UBVRI} in Appendix).}
    \label{fig:Radiation_amax}
\end{figure*}

Figure \ref{fig:Radiation_amax} displays the intrinsic spectrum of Betelgeuse at 3500 K (solid black line). The dashed lines are the radiation spectra accounting for the extinction effect for different values of $a_{\rm max}$ and without the disruption effect. Since the extinction is higher at shorter wavelengths (i.e., UV-optical, see Figures \ref{fig:Extinction_Smax_Av}, \ref{fig:Extinction_fraction_Av} and \ref{fig:Extinction_z_Av}), the stellar radiation spectrum is more reddened at the short wavelength range. In this case, the grain sizes follow a fixed power-law distribution, as described by Equation \ref{eq:grain_size_distribution}. The level of reddening depends on the GSD from $a_{\min}$ to $a_{\max}$. Larger grains are presented in the envelope for the broader GSD (i.e., larger maximum grain size $a_{\rm max}$), thus the extinction is strong at UV-NIR wavelengths. Therefore, it significantly reduces the brightness in UV-NIR photometric bands.

\begin{figure*}
    \centering
    \includegraphics[width = 0.48\textwidth]{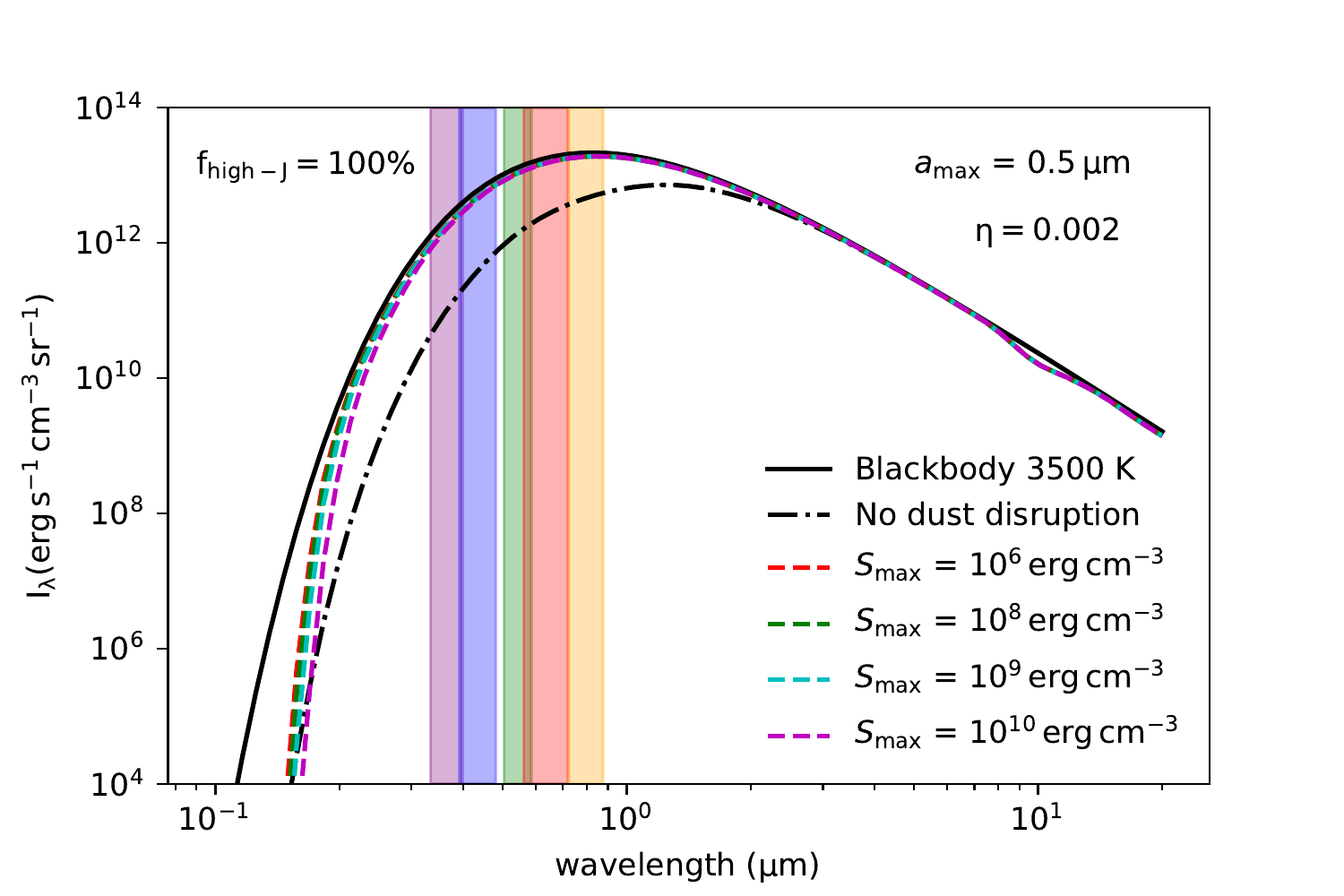}
    \includegraphics[width = 0.48\textwidth]{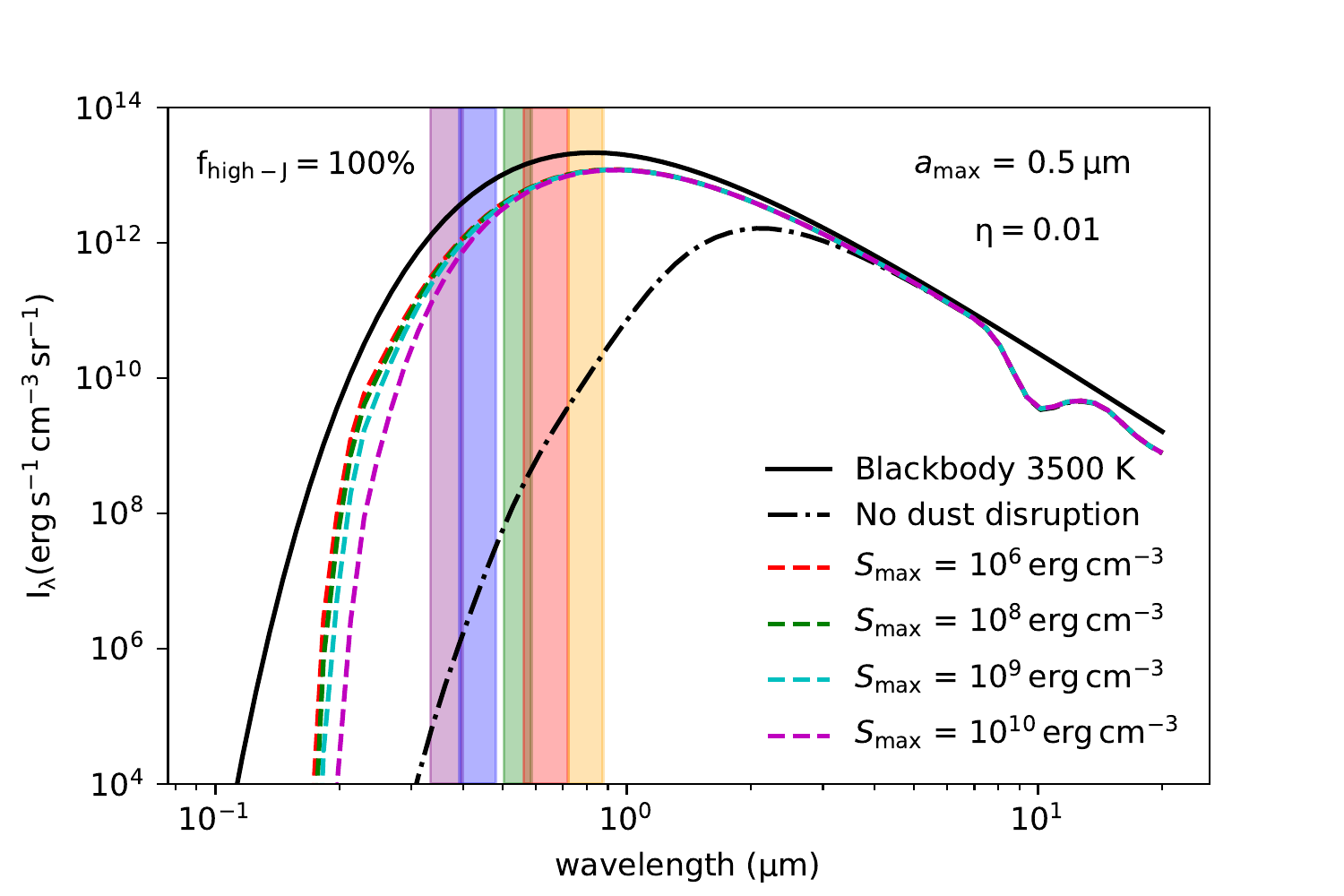}
    \caption{The reddening effects by disrupted grains at various tensile strength $S_{\rm max}$ (dashed color lines). The RAT-D effect can enhance the spectra from UV to near-IR wavelengths, compared with the case of no dust disruption with a fixed $a_{\rm max} = 0.5\,\rm\mu m$ (dash-dotted black line). The spectra produced by porous grains (i.e., lower $S_{\rm max}$) are more enhanced.}
    \label{fig:Radiation_Smax}
    \includegraphics[width = 0.48\textwidth]{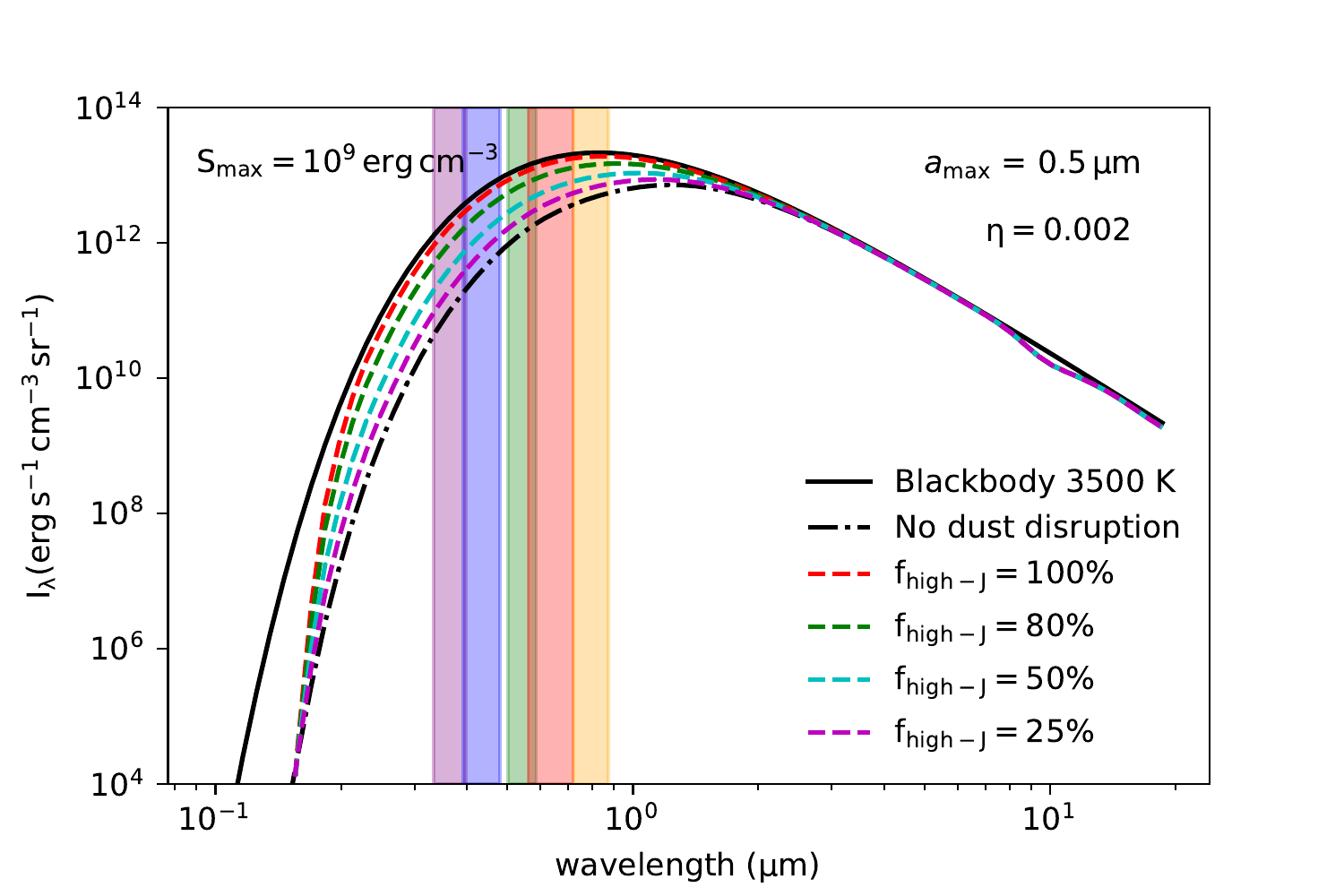}
    \includegraphics[width = 0.48\textwidth]{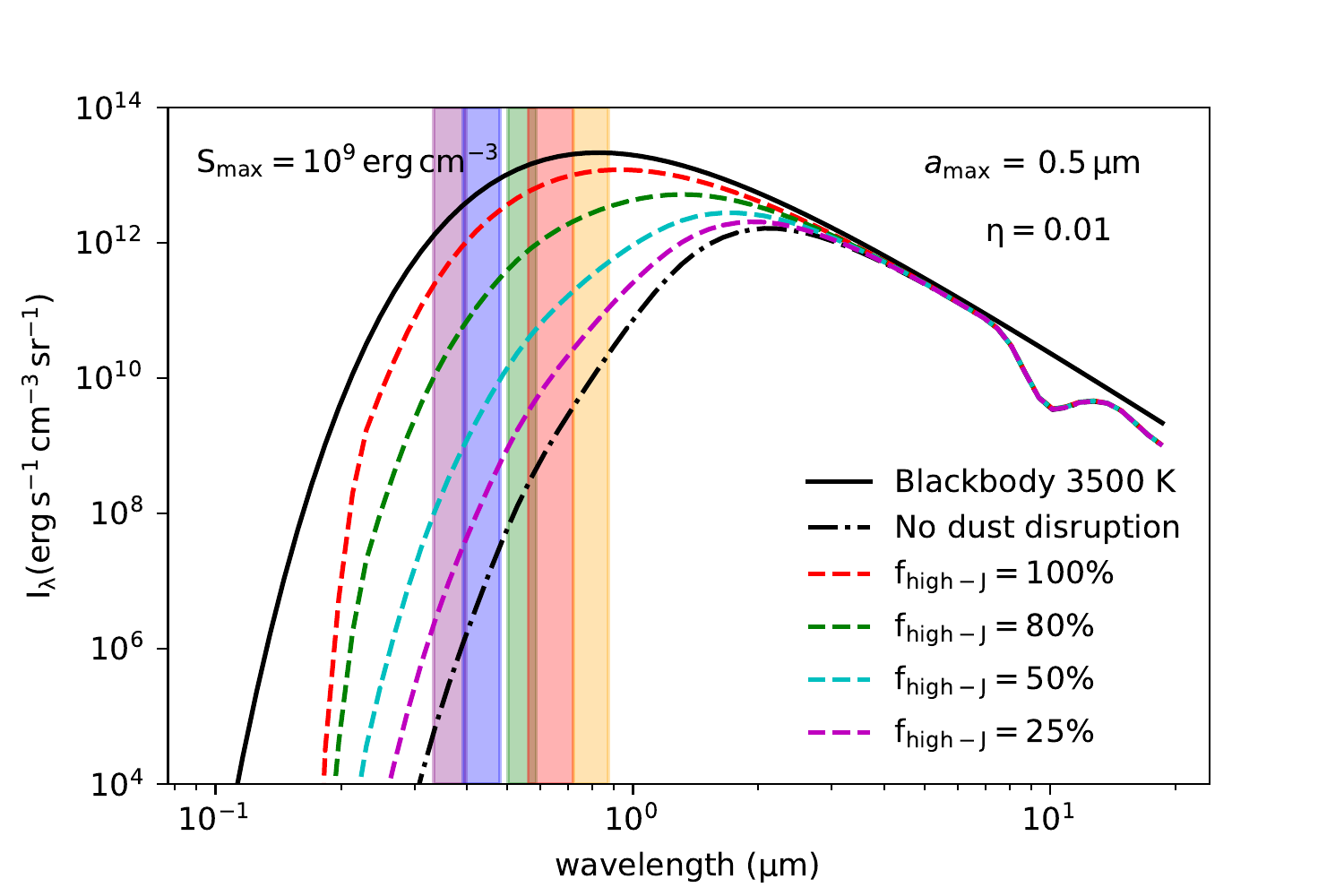}
    \caption{Similar to Figure \ref{fig:Radiation_Smax} but with fraction of grains at high-J attractors of $f_{\rm high-J} = 25\% - 100\%$ (dashed color lines). The spectra becomes redder as the abundance of grains at low-J attractors is significantly higher (i.e., lower $f_{\rm high-J}$).}
    \label{fig:Radiation_fraction}
\end{figure*}

Figure \ref{fig:Radiation_Smax} shows the radiation spectra when the RAT-D effect is taken into account (colored lines). Compared with no disruption (dash-dotted black line), the spectra are more enhanced at UV-optical wavelengths.

One important point that we emphasize in Figure \ref{fig:Radiation_Smax} is the effect of $S_{\rm max}$ on the reddened spectra. Compact grains (higher $S_{\rm max}$) retain larger sizes in the CSE, while the RAT-D disrupts porous grains (lower $S_{\rm max}$). As a consequence, grains with higher $S_{\rm max}$ attenuate more UV-NIR radiation, while this radiation is less extincted by grains with lower $S_{\rm max}$.

Another crucial point is the major impact of $f_{\rm high-J}$ on the reddened spectra, as shown in Figure \ref{fig:Radiation_fraction}. In the environment with higher fraction of grains at high-J attractors (i.e., high $f_{\rm high-J}$), the GSD of circumstellar dust is dominated by disrupted grains of $a < a_{\rm disr}$ (see Figure \ref{fig:Grain size_fraction}), and extincts less UV-NIR radiation. By contrast, for a high portion of non-disrupted grains at low-J attractors (i.e., low $f_{\rm high-J}$), large grains remain its original sizes of $a < a_{\rm max}$, and the stellar brightness becomes dimmer when observing at UV-NIR regimes.

Regarding the dust-to-gas ratio, it also influences the distinction of the reddened spectra. For a low dust-to-gas ratio ($\eta \lesssim 0.005$), the extinction effect is weak, thus the reddened spectra vary slightly. However, for a higher dust-to-gas ratio ($\eta \gtrsim 0.005$), a larger amount of dust reduces the strength of radiation fields, thus, diminishing the disruption of grains by RAT-D. Consequently, the discrepancy of reddened spectra is more significant and shifted from UV-optical bands (U,V and R bands) to NIR bands (I band) as indicated in Figure \ref{fig:Radiation_amax} for different $a_{\max}$, Figure \ref{fig:Radiation_Smax} for various $S_{\max}$ and Figure \ref{fig:Radiation_fraction} for different fractions $f_{\rm high-J}$.

\section{Discussion}
\label{section:Discussion}

In this section, we first mention the use of our model in interpreting grain size distribution in RSG/AGB envelopes, as well as dust extinction and reddening in RSG/AGB observations. Then, we discuss the implications of the results for the observed spectrum of Betelgeuse.

\subsection{Implications for studying grain size distribution in RSG/AGB envelopes}
\label{section:Implications_GSD}
The major process of driving evolved star outflows is radiation pressure on circumstellar dust in AGB and RSG envelopes (\citealt{Velhoelst2006}; \citealt{Hofner2008}; \citealt{Velhoelst2009}). The modeling results in Figures \ref{fig:Grain size_eta}, \ref{fig:Grain size_fraction} and \ref{fig:Grain size_amax} implied that, along with translation motion, stellar radiation can produce rotational motion of dust grains and disrupt them by RATs, leading to the abundance of smaller grains of $a < a_{\rm max} = 0.5\,\rm\mu m$. The variation of GSD can support in understanding dust formation and evolution in these environments.

The RAT-D effects not only present in massive RSGs but also in AGB envelopes, which was previously studied by \cite{Tram2020}. The results of GSD in that work showed that, stellar radiation can also disrupt larger grains into smaller species of  $a < a_{\rm max} = 0.25\,\rm\mu m$ at certain regions of both C-rich and O-rich AGB envelopes (see their Figure 3 as a reference). The authors emphasized that the RAT-D mechanism is independent of grain minerals and can happen for both C-rich dust (e.g., amorphous carbon, SiC, and PAHs) and O-rich dust (e.g., amorphous silicate and crystalline olivine). Thus, rotational disruption by RATs is mainly responsible for modifying the GSD in evolved star envelopes, and should be considered in the studies of both RSGs and AGBs.

Besides, the results of GSD taken by RAT-D are correlated with the chemical modeling results in evolved star envelopes by \citealt{Sande2019}. The abundance of small grains can result in higher gas depletion levels due to desorption/sputtering on grain surfaces. Hence, the approaches of interpreting depletion level from molecular line profiles (\citealt{Sande2019}) and spinning dust emission by RATs (\citealt{Tram2020}) can support in probing the existence of smaller grains in the CSEs of evolved stars.

\subsection{Implications for dust extinction and reddening in RSG/AGB observations}
The previous study by \cite{Levesque2005} described the physical properties of RSGs from broadband filter photometry of 74 Milky Way RSGs. The spectra of these RSGs were estimated by the MARCS models along with the ISM reddening law of \cite{Cardelli1989} with $R_{\rm V} = 3.1$. The results depicted the excess in the visual magnitude $A_{\rm V}$ of RSGs due to the occurrence of circumstellar dust. This dust extinction was also estimated in Small Magellanic Cloud (SMC), Large Magellanic Cloud (LMC) (\citealt{Levesque2006}), and M31 red supergiants (\citealt{Massey2009}). 

The study of \cite{Massey2005} highlighted in more detail the reddening effects of circumstellar dust. Several RSGs demonstrated that the observed fluxes mismatched with the reddened stellar atmospheric models at near-UV (NUV) wavelengths from $0.35\,\rm\mu m$ to $0.39\,\rm\mu m$ (\citealt{Massey2005}). There are some explanations for this phenomenon. One possibility is the presence of a hot chromosphere above a stellar photosphere. The heating gas profile in the extended atmosphere and strong emissions such as in the MgII, $\rm H\alpha$, and K lines are responsible for the rise in UV images and spectra (\citealt{Gillard1996}; \citealt{Carpenter1994}; \citealt{Uitenbroek1998}; \citealt{Harper2001}).  Another explanation is the scattering of blue light from light beams along a line-of-sight by larger grains, as proposed by \cite{Massey2005}. These possibilities require high spectral resolution observations. However, it is challenging to extract CSE components for further reddening studies due to high optical depth. 

The modeled GSD considering the RAT-D mechanism opens up a new approach in studying dust extinction. The results from Figure \ref{fig:Rv_Smax} and \ref{fig:Rv_fraction} reveal the variation of the observed $R_{\rm V}$ along the LOS toward background stars as a result of the change in the GSD within the envelope. The measured $R_{\rm V}$ can achieve up to 4.2, possibly induced by small porous O-rich grains along the LOS across the inner envelope (i.e., $a < 0.05\,\rm\mu m$ and $S_{\rm max} \sim 10^{6} \,\rm erg\,cm^{-3}$) or the increased population of non-disrupted grains at low-J attractors (i.e., $f_{\rm high-J} < 25\%$) with larger sizes. This is consistent with $R_{\rm V} > 3.1$ derived from RSG photometry (\citealt{Lee1970}; \citealt{Humphreys1978}). Consequently, the dust extinction determined by RAT-D contributes to correcting photometric observations of RSGs with the modified $R_{\rm V}$.

For the NUV problem, there are other explanations with the implication of the RAT-D mechanism. According to reddening properties in Figures \ref{fig:Radiation_amax}, \ref{fig:Radiation_Smax} and \ref{fig:Radiation_fraction}, the NUV excess can be produced by the survival of large grains in stellar radiation fields under the effects of RAT-D. This can be derived from grain properties, both grain internal structures and rotational properties. One possibility is that grains may have a strong internal structure with $S_{\rm max} > 10^{9}\,\rm erg\,cm^{-3}$, which are difficult to be disrupted by RATs. This explanation can be reasonable since grains are likely to have a compact structure (e.g., crystalline olivine) at the condensation radius where first solids are formed by dust nucleation (see \citealt{Tielens1998}; \citealt{Cherchneff2013}; \citealt{Velhoelst2009}). Another possibility is that grains at low-J attractors may be abundant in the RSG envelopes with $f_{\rm high-J} < 25\%$, requiring a larger disruption timescale (see Figure \ref{fig:Grain size_fraction}). These suggestions raise the requirements of studying radiative feedback in RSG envelopes to precisely evaluate grain properties from stellar observations.

Circumstellar dust also makes an important impact on interpreting the SED of AGB stars. \cite{Fonfria2020} and \cite{Fonfria2021} showed that the modeling can reproduce the observed SED at the range of wavelengths $\lambda \sim 2.5 - 26\,\rm\mu m$. However, the model mismatched and can not reproduce the spectra for shorter wavelengths at $\lambda < 2.5\,\rm\mu m$ for carbon-rich and oxygen-rich AGB stars. Figures \ref{fig:Radiation_amax}, \ref{fig:Radiation_Smax} and \ref{fig:Radiation_fraction} suggested that our model can be a good explanation for these observational puzzles. As we mentioned in Section \ref{section:Implications_GSD}, the RAT-D mechanism applies in both C-rich and O-rich dust, and the reddening effects can be presented in both cases of AGB stars (see \citealt{Tram2020}). Hence, the effects of RAT-D on circumstellar dust are essential to explain the extinction at UV-optical wavelengths of AGB envelopes.

\subsection{Implications for the dimming of Betelgeuse}
\label{section: Fitting observation}

\begin{figure*}
    \centering
    \includegraphics[width = 0.48\textwidth]{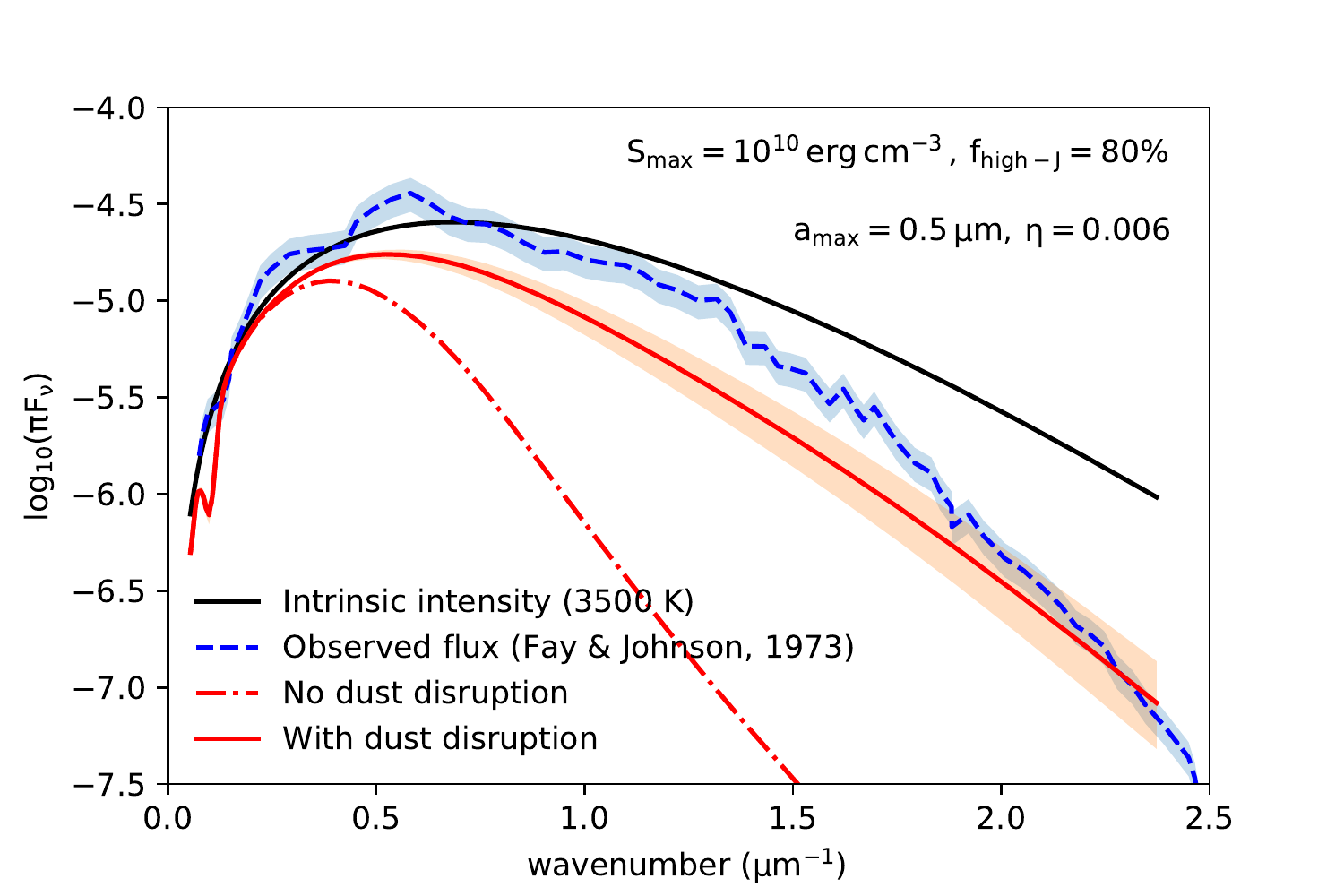}
    \includegraphics[width = 0.48\textwidth]{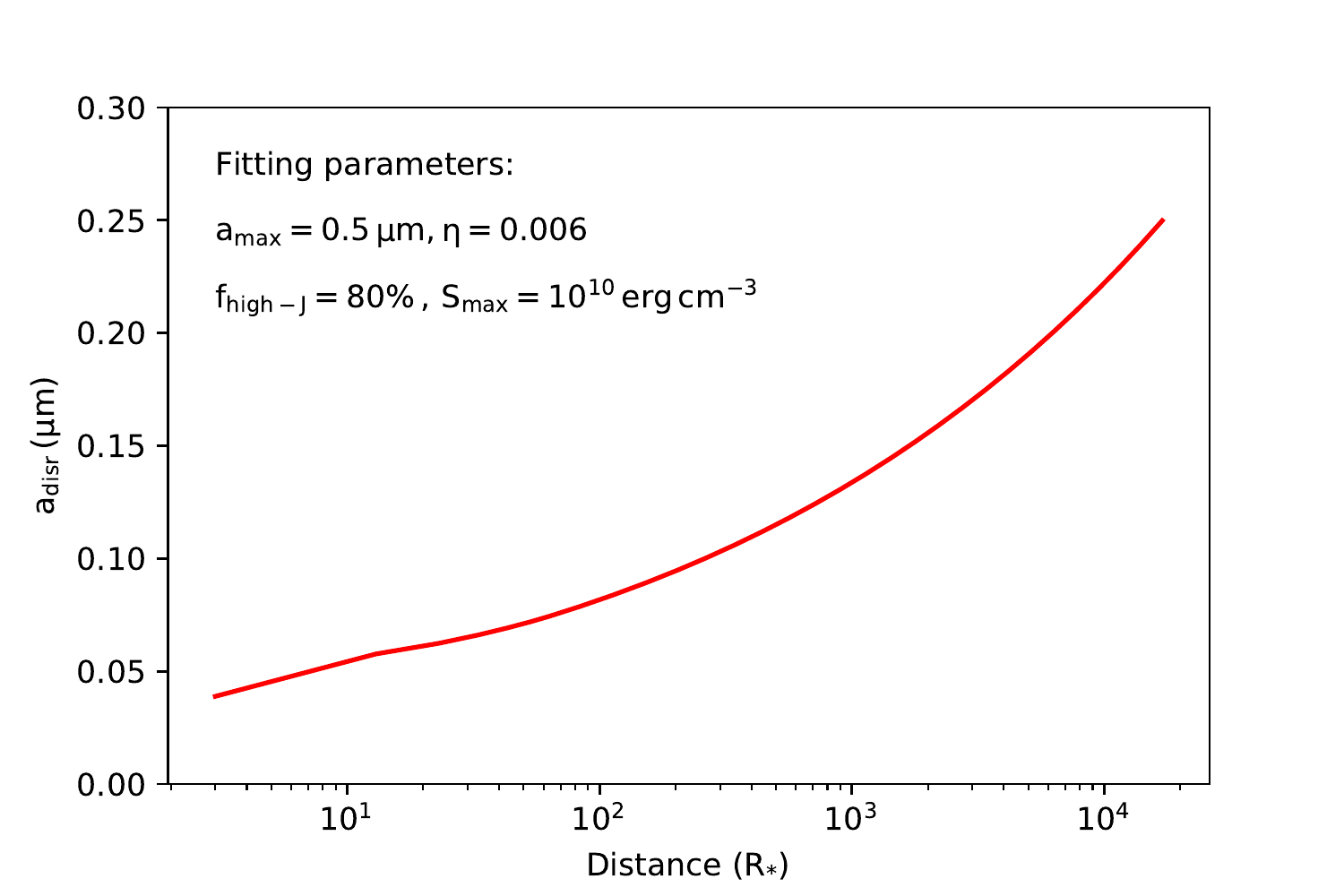}
    \caption{The absorption in the observed flux of Betelgeuse by \cite{Fay1973} (dashed blue line) slightly matched with the fitting reddened spectrum considering the RAT-D mechanism (solid red line) within the confidence intervals of $1\sigma$ for each parameter (red shaded region) in the left panel. The right panel is grain disruption size given by grain parameters from the best-fit model.}
    \label{fig:Fitting observation}
\end{figure*}

Circumstellar dust affects the observational properties of Betelgeuse, especially its observed flux. The left panel of Figure \ref{fig:Fitting observation} shows the spectral energy distribution of Betelgeuse observed by \cite{Fay1973} (dashed blue line), in comparison with the black-body radiation at 3500 K (solid black line). The observed flux reveals a strong absorption from the UV to NIR regions at $\gtrsim 0.8\,\mu$m$^{-1}$. We interpret the circumstellar reddening effects in the observed Betelgeuse spectrum with the applications of the RAT-D mechanism.

We first assume that the maximum grain size is equal to $0.5\,\rm\mu m$ (\citealt{Scicluna2015}). Thus, three main free parameters are (1) the maximum tensile strength $S_{\rm max}$, (2) the fraction of grains at high-J attractors $f_{\rm high-J}$ and (3) dust-to-gas ratio $\eta$, which are determined through the reduced $\chi^2$ minimization as
\begin{equation}
     \chi^2_{\rm red} = \frac{1}{L}\sum_{i = 1}^n \frac{(y_i - \overline{y}_i)^2}{\sigma_i^2}
\end{equation}
where $L$ is the number of freedom degrees, $y_i$ and $\overline{y}_i$ are observed and modeling values, respectively, and $\sigma_i$ is the measurement error of the observed flux. We test the effect of assuming flux errors on the minimum $\chi^2_{\rm red}$. 

The best model is illustrated by the solid red line in Figure $\ref{fig:Fitting observation}$ (left panel) with $S_{\rm max} = 10^{10}\,\rm erg\,cm^{-3}$, $f_{\rm high-J} = 80\%$ and the dust-to-gas ratio $\eta = 0.006$ within $1\sigma$ level of significance (red shaded region), assuming a $20\%$ error of the total flux (blue shaded region). These parameters change negligibly when we vary the error from $5\%$ to $30\%$. The constrained dust-to-gas ratio is higher than that of 0.0025 predicted by \cite{Velhoelst2006}. The fitted $S_{\rm max}$ and $f_{\rm high-J}$ imply that approximately $80\%$ of original large grains at high-J attractors may obtain compact structure before being fragmented by RAT-D. We compare our best fit with the case of no RAT-D, which is shown by a dash-dotted red line. This standard case considerably underestimates the observed flux at optical and UV wavelengths due to a lack of small grains.

From the best fit, we compute the change in the GSD of disrupted grains as a function of the envelope distance, which is shown in the right panel of Figure $\ref{fig:Fitting observation}$. The grain size varies from $\sim 0.05\,\rm\mu m$ at $r = 3\,\rm R_{\ast}$ to $\sim 0.25\,\rm\mu m$ at $r = 17000\,\rm R_{\ast}$, which is apparently higher than the mean grain size of ISM of $0.1\,\rm\mu m$ (\citealt{Cardelli1989}). 

The results found in Figure $\ref{fig:Fitting observation}$ illustrate the scenario of driving outflows by two populations of grains at high-J and low-J attractors based on the differences in dust-gas drift velocity. The former shows the dominance of smaller grains (i.e., $a  \lesssim 0.1 \,\rm\mu m$) induced by RAT-D at the dust formation zone, which are expelled easier to the radiation pressure. These grains subsequently collide with gas molecules and move at subsonic motion (see Section 2.5 in \citealt{Tram2020}). The latter shows the survival of large grains (i.e., $a \gtrsim 0.1 \,\rm\mu m$) due to a larger disruption timescale, which produces higher drift velocity and can achieve the supersonic speed (\citealt{Tram2020}). Consequently, RAT-D plays a significant role in defining grain properties and their dependency on radiation fields to push materials into the ISM environment.

Nevertheless, the left panel in Figure \ref{fig:Fitting observation} shows the slight deviation between the modeled SED and the observed spectrum at optical-NIR wavelengths. There are several factors that may contribute to this difference. One factor could be the impacts of the gaseous atmosphere, which can be caused by the strong absorption lines of heavy metal species (e.g., TiO and CN, see \citealt{Fay1973}; \citealt{Tsuji1976}; \citealt{Lobel2002}) or the emission by heating gas at the chromosphere region at optical-UV regimes (see \citealt{Lim1998}; \citealt{Harper2001}). Another factor might be the presence of dense clumps due to a local cooling, contributing to 3$\%$ to 12$\%$ of the total mass-loss of Betelgeuse (\citealt{Montarges2021}). Different from the continuous envelope as considered in this work, the clumpy structure is expected to cause less extinction, which could improve our model's interpretations.

\subsection{Model limitations}
This work is our first attempt to describe dust properties and their impacts on stellar observations in the presence of radiative torque disruption from massive RSGs. Therefore, our model composes of some limitations, which are emphasized in this Section.

Firstly, our study adopts a simplified assumption on the spherical geometry of the Betelgeuse envelope produced by radiation pressure on dust grains (see Figure \ref{fig:Betel_model}). This differs from the asymmetric structures with dense clumps in recent optical-NIR observations (e.g., \citealt{Kervella2011}; \citealt{Ohnaka2014}; \citealt{Montarges2021}), which can be a result of mass ejection by large scale convection in the stellar surface (\citealt{Freytag2002}; \citealt{Ren2020}; \citealt{Humphreys2022}). As extensively studied by \cite{Giang2021}, a high concentration of gas inside clumps can enhance the rotational damping by gas collision with a shorter timescale (see Equation \ref{eq:tau_gas}) and reduce the RAT-D efficiency. This can probe the occurrence of large grains of $a > 0.1\,\rm\mu m$ found in the dusty clump of Betelgeuse and other RSGs (\citealt{Scicluna2015}; \citealt{Cannon2021}; \citealt{Montarges2021}).

Secondly, we consider the stellar emission as blackbody radiation for the sake of simplicity. The model disregard the contribution of a complex stellar atmosphere at optical-UV wavelengths, as we described in Section \ref{fig:Fitting observation} for the SED fitting. These factors can improve the model with best-fit parameters and interpret deeper dust properties with various environmental conditions of the evolved star envelopes.

Despite these limitations, we believe that the consideration of the RAT-D mechanism is essential in modeling radiative feedback on circumstellar dust, and can provide a proper insight to understand grain formation and evolution and their effects on AGB/RSG observations.

\section{Summary}
\label{section: Summary}
This paper focuses on modeling the effects of rotational disruption of circumstellar dust by radiative torques and exploring its impact on photometric and spectroscopic observations of RSG and AGB stars. The main findings of our study are summarized as follows.

\begin{enumerate}
  \item Assuming that large grains can form in the dust formation zone, we find that the RAT-D mechanism can destroy the large grains and enhance smaller grains of $a_{\rm disr} < a_{\rm max}$. The mechanism can result in the modification of GSD with the envelope distance, depending on both grain (i.e., sizes, internal structures and rotational properties) and environmental (i.e., dust-to-gas mass ratio).
  
  \item With the GSD determined by RAT-D, we model the dust extinction along the sightline toward Betelgeuse and background stars. The steepness of the extinction curve is related to grain properties and the projected distance of the LOS in the CSE.
  
  \item We calculate the variation of the total-to-selective extinction ratio $R_{\rm V}$. The change in $R_{\rm V}$ correlates with the change in the slope of the observed extinction due to the modification of GSD constrained by RAT-D.
  
  \item We study the circumstellar dust reddening effects of Betelgeuse. The observed stellar spectrum exceeds in UV-NIR regions compared with the case without dust disruption. Grain and environmental properties modify the reddening level.
  
  \item The effects of the RAT-D mechanism can be applied in studying the extinction and reddening of AGB/RSG stars, especially for interpreting the NUV excess in the SEDs of RSGs. Thus, the RAT-D mechanism should be considered in the studies of the dusty envelopes around RSG and AGB stars.
  
  \item The effects of the RAT-D mechanism potentially contribute to probing the spectral flux of Betelgeuse observed by \textcite{Fay1973}. We find that, radiation-driven winds can be generated by both small grains at high-J attractors and large grains at low-J attractors with different dust-gas drift velocities.
  
  \item The properties of gaseous atmosphere and clumpy structures in the Betelgeuse envelope possibly affect the dust reddening and disruption effects. Detailed calculations with more realistic characteristics of RSG envelopes are required to investigate precisely the impacts of radiative feedback and constrain proper circumstellar dust properties through stellar observations.
\end{enumerate}

\textit{Acknowledgement:} We thank the anonymous referee for helpful comments that improved the impact and the presentation of this paper. T.H. acknowledges the support by the National Research Foundation of Korea (NRF) grant (2019R1A2C1087045) funded by the Korean government (MSIT). This research was done in the framework of the Vietnam Astrophysics Research NETwork (VARNET), which aims to foster collaborative research in Astrophysics among Vietnamese researchers and students living in Vietnam and abroad.

\appendix 

\vspace{-0.5cm}
\section{Extinction per column density $A_{\lambda}/N_{\H}$}
\begin{figure*}
    \centering
    \includegraphics[width = 0.48\textwidth]{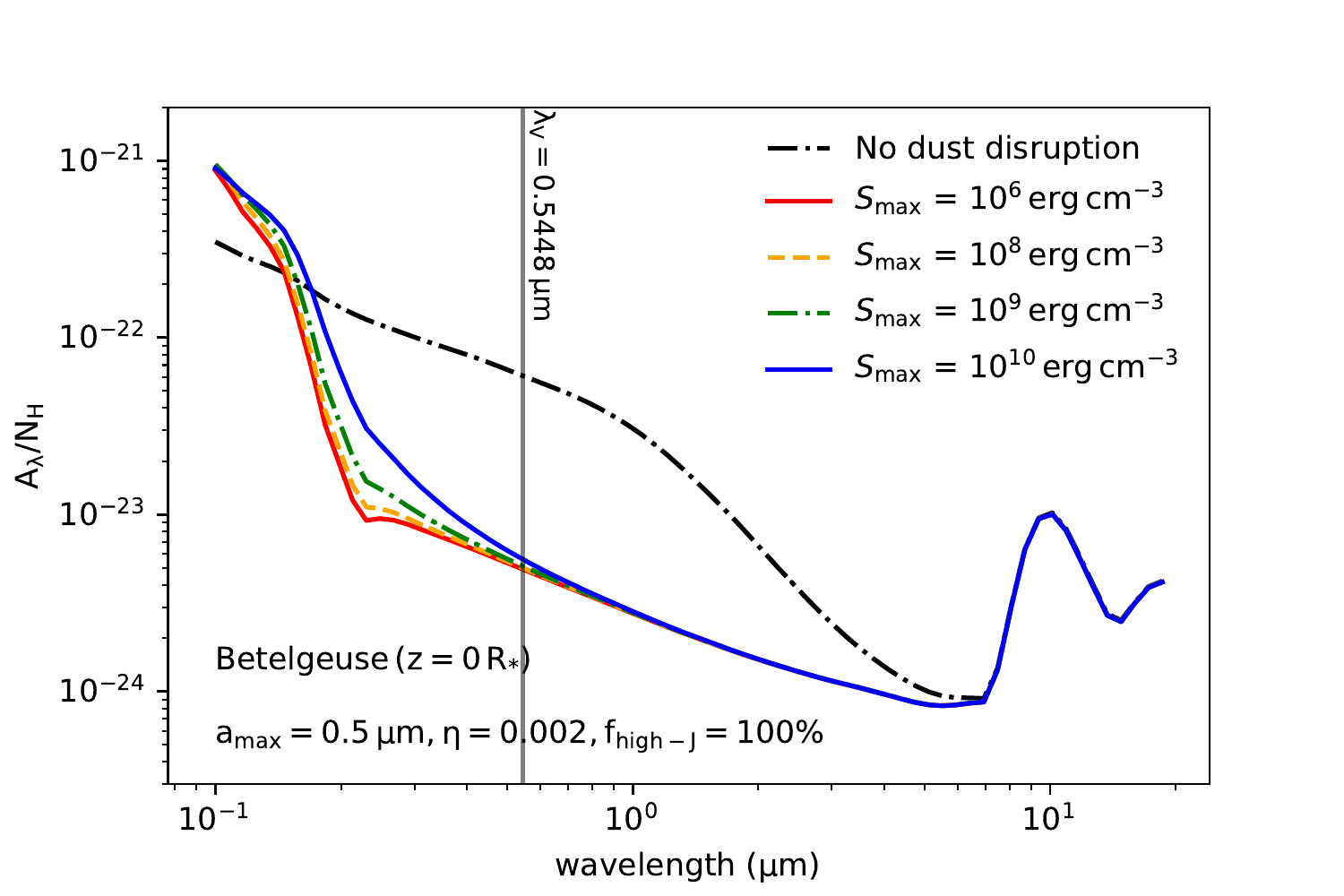}
    \includegraphics[width = 0.48\textwidth]{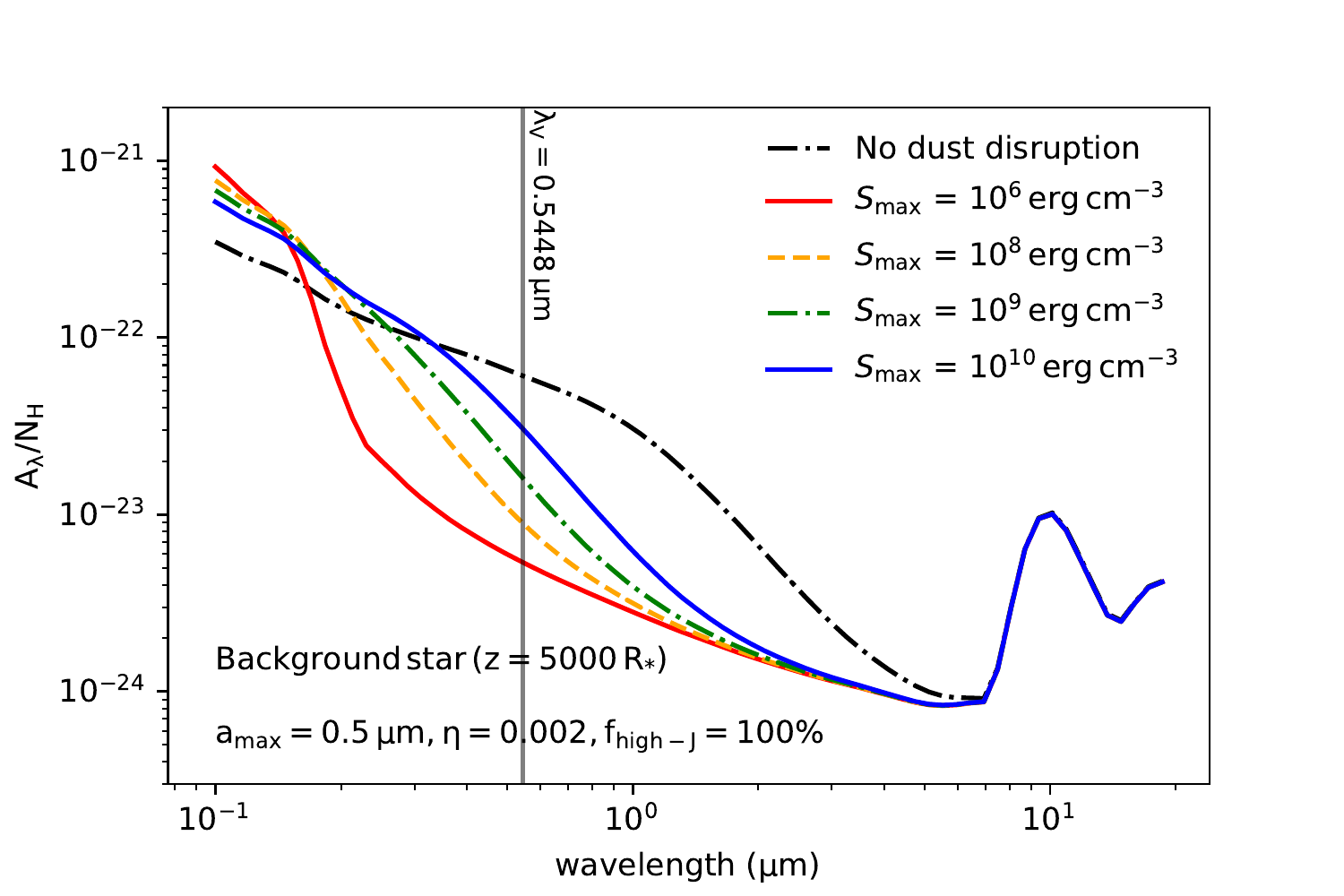}
    \caption{Extinction curve $A_{\lambda}/N_{\H}$ caused by dust grains along a LOS at $z = 0\,\rm R_{*}$ (through Betelgeuse, left panel) and $z = 500\,\rm R_{*}$ (through a background star, right panel) at different values of $S_{\rm max}$ with a fixed $a_{\rm max}=0.5\,\mu$m and $f_{\rm high-J} = 100\%$. Vertical line represents the visual extinction $A_{\rm V}$ at $\lambda = 0.5448\,\mu$m}
    \label{fig:Extinction_Smax_NH}
    \includegraphics[width = 0.48\textwidth]{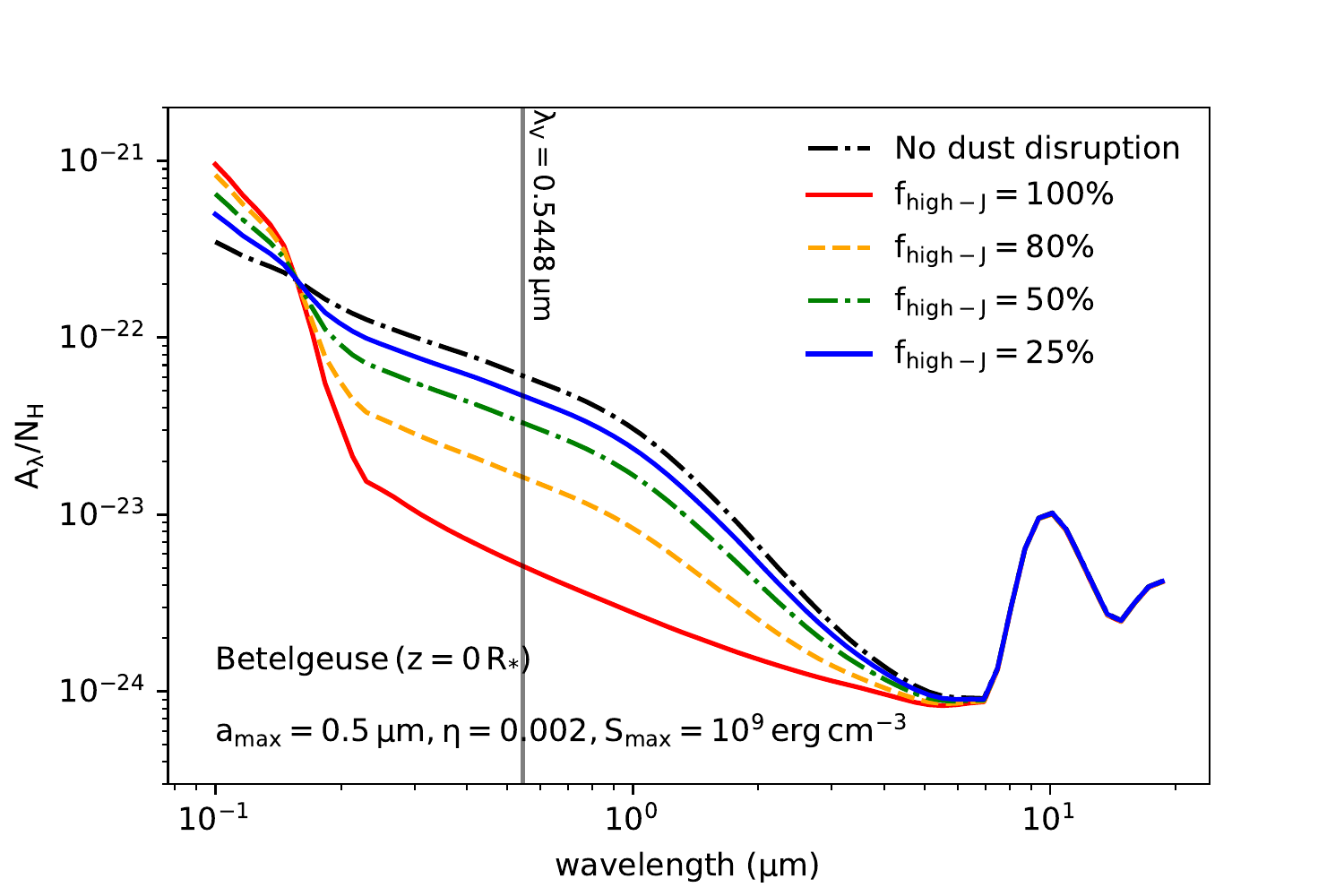}
    \includegraphics[width = 0.48\textwidth]{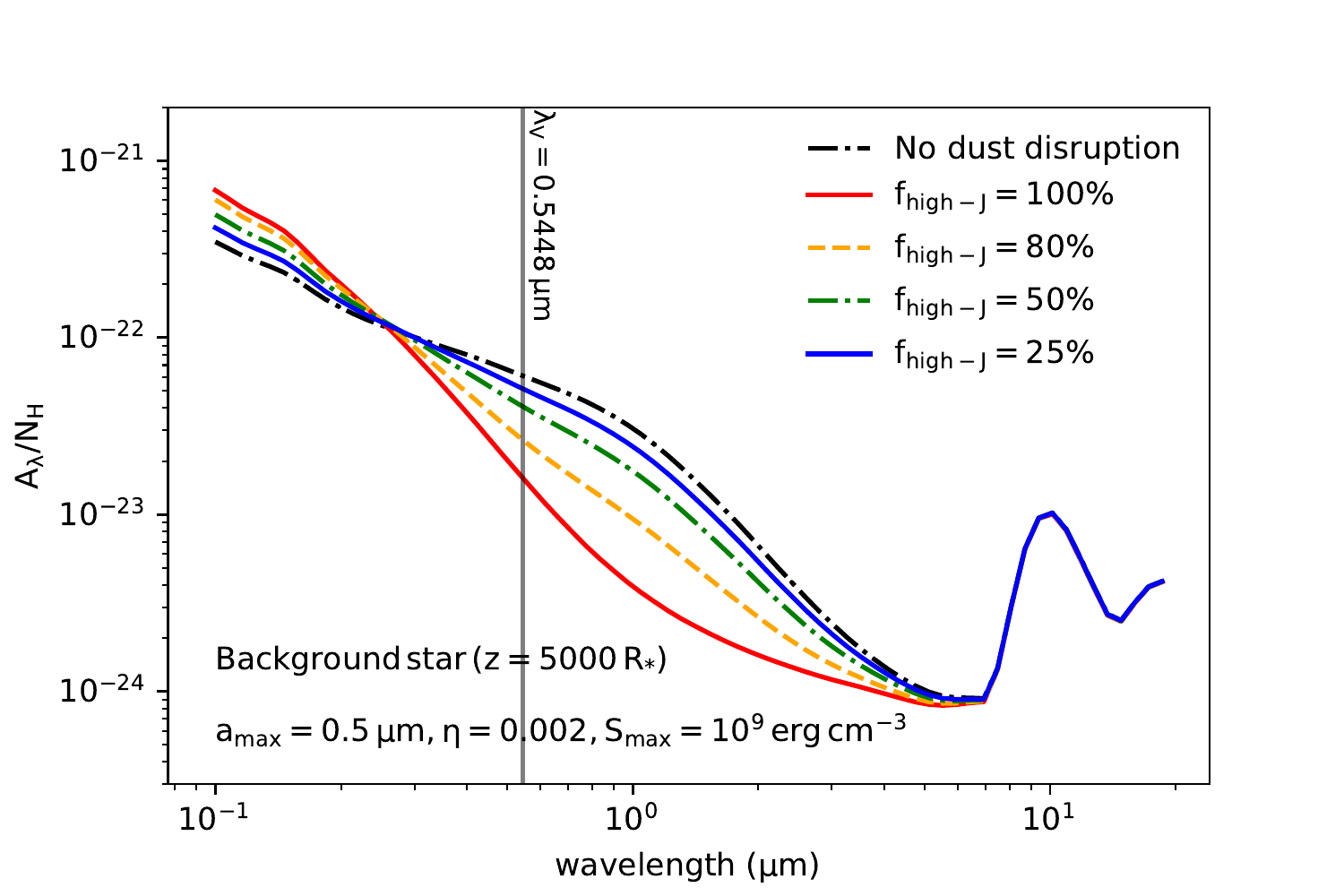}
    \caption{Same as Figure \ref{fig:Extinction_Smax_NH} but for various values of $f_{\rm high-J}$ from $25\%$ to $100\%$. Assuming a fixed $S_{\rm max} = 10^{9}\,\rm erg\,cm^{-3}$ and $\eta = 0.002$.}
    \label{fig:Extinction_fraction_NH}
    \includegraphics[width = 0.48\textwidth]{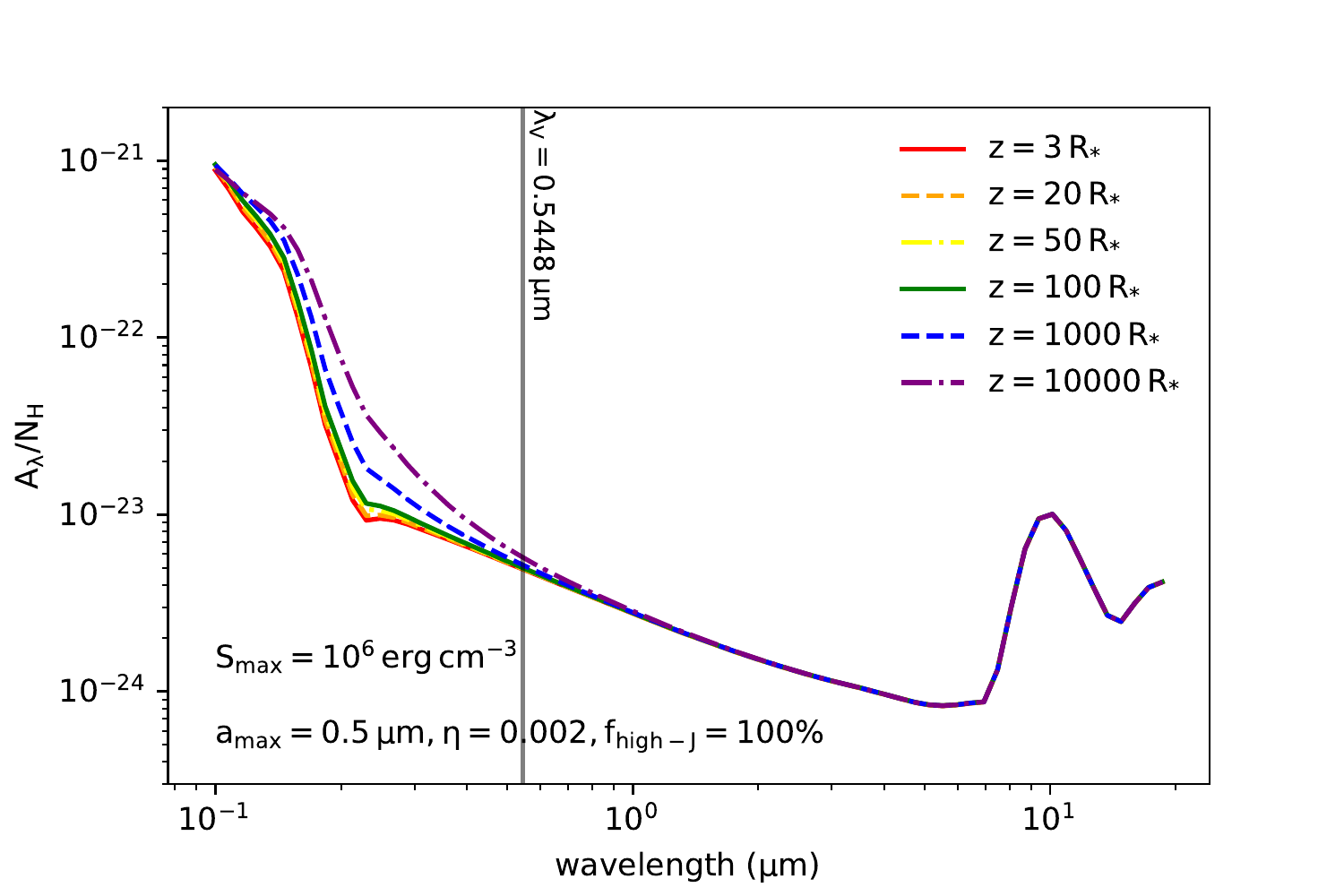}
    \includegraphics[width = 0.48\textwidth]{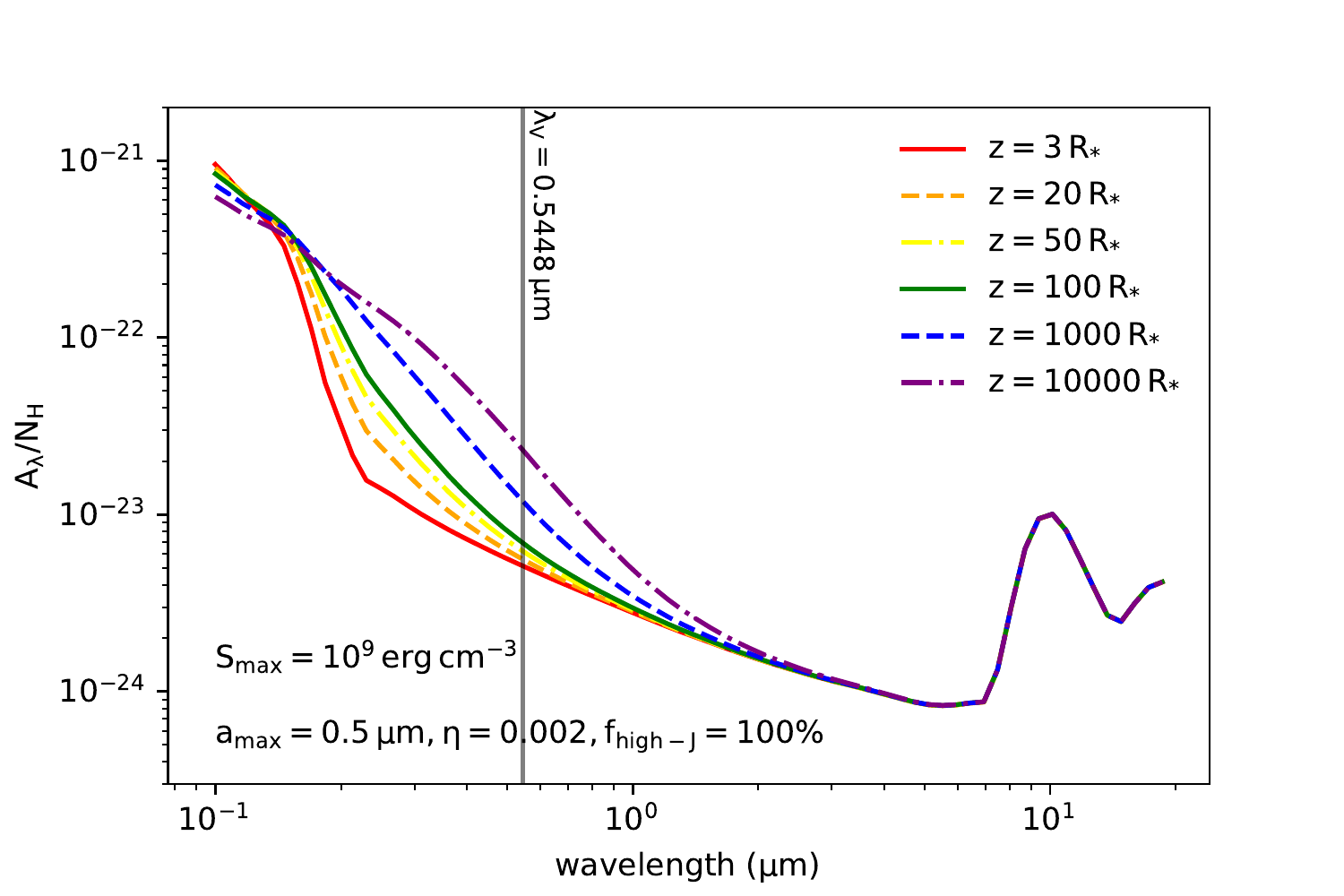}
    \caption{The extinction curve $A_{\lambda}/N_{\H}$ for dust grains at $S_{\rm max} = 10^{6}\,\rm erg\,cm^{-3}$ (left panel) and $S_{\rm max} = 10^{9}\,\rm erg\,cm^{-3}$ (right panel) located at different positions $z$ of the parallel-LOS.}
    \label{fig:Extinction_z_NH}
\end{figure*}

Figure \ref{fig:Extinction_Smax_NH} shows the normalized extinction curve $A_{\lambda}/N_{\rm H}$, where $N_{\rm H} = \int n_{\rm H} ds$ is the hydrogen column density along the LOS with two cases: $z = 0$ (from Betelgeuse) and $z \neq 0$ (from a background star). One can see that the visual extinction $A_{\rm V}$ at $\lambda = 0.5448$ increases with increasing $S_{\rm max}$, which was used for modifying the observed extinction curve $A_{\lambda}/A_{\rm V}$ in Section \ref{section: Extinction curve results}. Similar to Figure \ref{fig:Extinction_Smax_Av}, the figure shows an increase in optical-IR extinction and decrease in FUV extinction for grains with higher $S_{\rm max}$, and the extinction in the right panel exhibits flatter slope than in the left panel. 

Figure \ref{fig:Extinction_fraction_NH} indicates the change in extinction curve $A_{\lambda}/N_{\H}$ with different values of the fraction $f_{\rm high-J}$ of grains at high-J attractors between two cases of the LOS $z = 0$ and $z \neq 0$ (considering a fixed $S_{\max}$). Same as in Figure \ref{fig:Extinction_fraction_Av}, the increased contribution of disrupted grain sizes (i.e., higher $f_{\rm high-J}$) leads to a higher far-UV extinction and a lower optical-IR extinction. The value of $A_{\rm V}$ increases with decreasing $f_{\rm high-J}$.

Figure \ref{fig:Extinction_z_NH} reveals an increase in extinction curve $A_{\lambda}/N_{\rm H}$ at higher positions $z$ of a parallel-LOS due to the survival of larger grains by RAT-D, similar to in Figure \ref{fig:Extinction_z_Av}. The visual extinction $A_{\rm V}$ is higher for larger $z$.

\section{The UBVRI photometric system}
Table \ref{tab:UBVRI} illustrates photometric bands from ultraviolet to infrared in the UBVRI system, which were calibrated and corrected by \cite{Bessel1990}. The marked colors and bandwidths $\Delta \lambda$ were used for setting rectangular boundaries in Figure \ref{fig:Radiation_amax}, \ref{fig:Radiation_Smax} and \ref{fig:Radiation_fraction}.

\begin{table}[h!]
    \centering
    \begin{tabular}{c c c c}
         \toprule
         Photometric bands & Effective wavelength $\lambda_{\rm eff}$ ($\rm\mu m$) & Bandwidth $\Delta \lambda$ ($\rm\mu m$) & Marked color  \\
         \midrule
         U & $0.3659$ & $0.065$ & Purple \\
         B & $0.4363$ & $0.089$ & Blue \\
         V & $0.5448$ & $0.084$ & Green \\
         R & $0.6407$ & $0.158$ & Red \\
         I & $0.7982$ & $0.154$ & Orange \\
         \bottomrule
    \end{tabular}
    \caption{The UBVRI photometric system (\citealt{Bessel1990} and \citealt{Bessell2005}).}
    \label{tab:UBVRI}
\end{table}

\bibliographystyle{aasjournal}

\begin{thebibliography}{}
\expandafter\ifx\csname natexlab\endcsname\relax\def\natexlab#1{#1}\fi
\providecommand{\url}[1]{\href{#1}{#1}}
\providecommand{\dodoi}[1]{doi:~\href{http://doi.org/#1}{\nolinkurl{#1}}}
\providecommand{\doeprint}[1]{\href{http://ascl.net/#1}{\nolinkurl{http://ascl.net/#1}}}
\providecommand{\doarXiv}[1]{\href{https://arxiv.org/abs/#1}{\nolinkurl{https://arxiv.org/abs/#1}}}

\bibitem[{{Antia} {et~al.}(1984){Antia}, {Chitre}, \& {Narasimha}}]{Antia1984}
{Antia}, H.~M., {Chitre}, S.~M., \& {Narasimha}, D. 1984, \apj, 282, 574,
  \dodoi{10.1086/162236}

\bibitem[{{Bennett}(2010)}]{Bennett2010}
{Bennett}, P.~D. 2010, in Astronomical Society of the Pacific Conference
  Series, Vol. 425, Hot and Cool: Bridging Gaps in Massive Star Evolution, ed.
  C.~{Leitherer}, P.~D. {Bennett}, P.~W. {Morris}, \& J.~T. {Van Loon}, 181.
\newblock \doarXiv{1004.1853}

\bibitem[{{Bessell}(1990)}]{Bessel1990}
{Bessell}, M.~S. 1990, \pasp, 102, 1181, \dodoi{10.1086/132749}

\bibitem[{{Bessell}(2005)}]{Bessell2005}
---. 2005, \araa, 43, 293, \dodoi{10.1146/annurev.astro.41.082801.100251}

\bibitem[{{Biscaro} \& {Cherchneff}(2016)}]{Biscaro2016}
{Biscaro}, C., \& {Cherchneff}, I. 2016, \aap, 589, A132,
  \dodoi{10.1051/0004-6361/201527769}

\bibitem[{{Cannon} {et~al.}(2021){Cannon}, {Montarg{\`e}s}, {de Koter},
  {Decin}, {Min}, {Lagadec}, {Kervella}, {Sundqvist}, \& {Sana}}]{Cannon2021}
{Cannon}, E., {Montarg{\`e}s}, M., {de Koter}, A., {et~al.} 2021, \mnras, 502,
  369, \dodoi{10.1093/mnras/stab018}

\bibitem[{{Cardelli} {et~al.}(1989){Cardelli}, {Clayton}, \&
  {Mathis}}]{Cardelli1989}
{Cardelli}, J.~A., {Clayton}, G.~C., \& {Mathis}, J.~S. 1989, \apj, 345, 245,
  \dodoi{10.1086/167900}

\bibitem[{{Carpenter} {et~al.}(1994){Carpenter}, {Robinson}, {Wahlgren},
  {Linsky}, \& {Brown}}]{Carpenter1994}
{Carpenter}, K.~G., {Robinson}, R.~D., {Wahlgren}, G.~M., {Linsky}, J.~L., \&
  {Brown}, A. 1994, \apj, 428, 329, \dodoi{10.1086/174244}

\bibitem[{{Chau Giang} {et~al.}(2021){Chau Giang}, {Hoang}, {Tram}, {Dieu},
  {Diep}, {Phuong}, {Van Tuan}, \& {Le Gia Bao}}]{Giang2021}
{Chau Giang}, N., {Hoang}, T., {Tram}, L.~N., {et~al.} 2021, arXiv e-prints,
  arXiv:2111.11800

\bibitem[{{Cherchneff}(2013)}]{Cherchneff2013}
{Cherchneff}, I. 2013, in EAS Publications Series, Vol.~60, EAS Publications
  Series, ed. P.~{Kervella}, T.~{Le Bertre}, \& G.~{Perrin}, 175--184,
  \dodoi{10.1051/eas/1360020}

\bibitem[{{De Beck} {et~al.}(2019){De Beck}, {Boyer}, {Bujarrabal}, {Decin},
  {Fonfr{\'\i}a}, {Groenewegen}, {H{\"o}fner}, {Jones}, {Kami{\'n}ski},
  {Maercker}, {Marigo}, {Matsuura}, {Meixner}, {Quintana Lacaci Mart{\'\i}nez},
  {Scicluna}, {Szczerba}, {Velilla Prieto}, {Vlemmings}, \&
  {Wiedner}}]{DeBeck2019}
{De Beck}, E., {Boyer}, M.~L., {Bujarrabal}, V., {et~al.} 2019, \baas, 51, 374

\bibitem[{{Decin} {et~al.}(2013){Decin}, {Cox}, {Royer}, {van Marle}, \&
  {Vandenbussche}}]{Decin2013}
{Decin}, L., {Cox}, N.~L.~J., {Royer}, P., {van Marle}, A.~J., \&
  {Vandenbussche}, B. 2013, in EAS Publications Series, Vol.~60, EAS
  Publications Series, ed. P.~{Kervella}, T.~{Le Bertre}, \& G.~{Perrin},
  227--234, \dodoi{10.1051/eas/1360026}

\bibitem[{{Dolan} {et~al.}(2016){Dolan}, {Mathews}, {Lam}, {Quynh Lan},
  {Herczeg}, \& {Dearborn}}]{Dolan2016}
{Dolan}, M.~M., {Mathews}, G.~J., {Lam}, D.~D., {et~al.} 2016, \apj, 819, 7,
  \dodoi{10.3847/0004-637X/819/1/7}

\bibitem[{{Dominik} {et~al.}(1989){Dominik}, {Gail}, \&
  {Sedlmayr}}]{Dominik1989}
{Dominik}, C., {Gail}, H.~P., \& {Sedlmayr}, E. 1989, \aap, 223, 227

\bibitem[{{Draine}(2011)}]{DraineBook2011}
{Draine}, B.~T. 2011, {Physics of the Interstellar and Intergalactic Medium}

\bibitem[{{Draine} \& {Flatau}(2012)}]{Draine2012}
{Draine}, B.~T., \& {Flatau}, P.~J. 2012, arXiv e-prints, arXiv:1202.3424.
\newblock \doarXiv{1202.3424}

\bibitem[{{Draine} \& {Lee}(1984)}]{Draine1984}
{Draine}, B.~T., \& {Lee}, H.~M. 1984, \apj, 285, 89, \dodoi{10.1086/162480}

\bibitem[{{Draine} \& {Salpeter}(1979)}]{Draine1979}
{Draine}, B.~T., \& {Salpeter}, E.~E. 1979, \apj, 231, 77,
  \dodoi{10.1086/157165}

\bibitem[{{Fadeyev}(2012)}]{Fadeyev2012}
{Fadeyev}, Y.~A. 2012, Astronomy Letters, 38, 260,
  \dodoi{10.1134/S1063773712040032}

\bibitem[{{Fay} \& {Johnson}(1973)}]{Fay1973}
{Fay}, T.~D., \& {Johnson}, H.~R. 1973, \apj, 181, 851, \dodoi{10.1086/152097}

\bibitem[{{Fonfr{\'\i}a} {et~al.}(2020){Fonfr{\'\i}a}, {Montiel}, {Cernicharo},
  {DeWitt}, \& {Richter}}]{Fonfria2020}
{Fonfr{\'\i}a}, J.~P., {Montiel}, E.~J., {Cernicharo}, J., {DeWitt}, C.~N., \&
  {Richter}, M.~J. 2020, \aap, 643, L15, \dodoi{10.1051/0004-6361/202039547}

\bibitem[{{Fonfr{\'\i}a} {et~al.}(2021){Fonfr{\'\i}a}, {Montiel}, {Cernicharo},
  {DeWitt}, {Richter}, {Lacy}, {Greathouse}, {Santander-Garc{\'\i}a},
  {Ag{\'u}ndez}, \& {Massalkhi}}]{Fonfria2021}
{Fonfr{\'\i}a}, J.~P., {Montiel}, E.~J., {Cernicharo}, J., {et~al.} 2021, \aap,
  651, A8, \dodoi{10.1051/0004-6361/202040082}

\bibitem[{{Freytag} {et~al.}(2002){Freytag}, {Steffen}, \&
  {Dorch}}]{Freytag2002}
{Freytag}, B., {Steffen}, M., \& {Dorch}, B. 2002, Astronomische Nachrichten,
  323, 213

\bibitem[{{Giang} \& {Hoang}(2021)}]{Giang2020}
{Giang}, N.~C., \& {Hoang}, T. 2021, \apj, 922, 47,
  \dodoi{10.3847/1538-4357/ac1116}

\bibitem[{{Giang} {et~al.}(2020){Giang}, {Hoang}, \& {Tram}}]{GiangSuper2020}
{Giang}, N.~C., {Hoang}, T., \& {Tram}, L.~N. 2020, \apj, 888, 93,
  \dodoi{10.3847/1538-4357/ab5d37}

\bibitem[{{Gilliland} \& {Dupree}(1996)}]{Gillard1996}
{Gilliland}, R.~L., \& {Dupree}, A.~K. 1996, \apjl, 463, L29,
  \dodoi{10.1086/310043}

\bibitem[{{Glassgold} \& {Huggins}(1986)}]{Glassgold1986}
{Glassgold}, A.~E., \& {Huggins}, P.~J. 1986, \apj, 306, 605,
  \dodoi{10.1086/164370}

\bibitem[{{Goldberg}(1984)}]{Goldberg1984}
{Goldberg}, L. 1984, \pasp, 96, 366, \dodoi{10.1086/131347}

\bibitem[{{Greenberg} \& {Li}(1996)}]{Greenberg1996}
{Greenberg}, J.~M., \& {Li}, A. 1996, \aap, 309, 258

\bibitem[{{Groenewegen} {et~al.}(2009){Groenewegen}, {Sloan}, {Soszy{\'n}ski},
  \& {Petersen}}]{Groenewegen2009}
{Groenewegen}, M.~A.~T., {Sloan}, G.~C., {Soszy{\'n}ski}, I., \& {Petersen},
  E.~A. 2009, \aap, 506, 1277, \dodoi{10.1051/0004-6361/200912678}

\bibitem[{{Groh} {et~al.}(2013){Groh}, {Meynet}, {Georgy}, \&
  {Ekstr{\"o}m}}]{Groh2013}
{Groh}, J.~H., {Meynet}, G., {Georgy}, C., \& {Ekstr{\"o}m}, S. 2013, \aap,
  558, A131, \dodoi{10.1051/0004-6361/201321906}

\bibitem[{{Guo} \& {Li}(2002)}]{Gou2002}
{Guo}, J.~H., \& {Li}, Y. 2002, \apj, 565, 559, \dodoi{10.1086/324295}

\bibitem[{{Harper} {et~al.}(2001){Harper}, {Brown}, \& {Lim}}]{Harper2001}
{Harper}, G.~M., {Brown}, A., \& {Lim}, J. 2001, \apj, 551, 1073,
  \dodoi{10.1086/320215}

\bibitem[{{Hartmann} \& {Avrett}(1984)}]{Hartmann1984}
{Hartmann}, L., \& {Avrett}, E.~H. 1984, \apj, 284, 238, \dodoi{10.1086/162402}

\bibitem[{{Heger} {et~al.}(1997){Heger}, {Jeannin}, {Langer}, \&
  {Baraffe}}]{Heger1997}
{Heger}, A., {Jeannin}, L., {Langer}, N., \& {Baraffe}, I. 1997, \aap, 327,
  224.
\newblock \doarXiv{astro-ph/9705097}

\bibitem[{{Herranen} {et~al.}(2021){Herranen}, {Lazarian}, \&
  {Hoang}}]{Herranen2021}
{Herranen}, J., {Lazarian}, A., \& {Hoang}, T. 2021, \apj, 913, 63,
  \dodoi{10.3847/1538-4357/abf096}

\bibitem[{{Hirashita} \& {Voshchinnikov}(2014)}]{Hirashita2014}
{Hirashita}, H., \& {Voshchinnikov}, N.~V. 2014, \mnras, 437, 1636,
  \dodoi{10.1093/mnras/stt1997}

\bibitem[{{Hoang} {et~al.}(2013){Hoang}, {Lazarian}, \& {Martin}}]{Hoang2013}
{Hoang}, T., {Lazarian}, A., \& {Martin}, P.~G. 2013, \apj, 779, 152,
  \dodoi{10.1088/0004-637X/779/2/152}

\bibitem[{{Hoang} {et~al.}(2019){Hoang}, {Tram}, {Lee}, \& {Ahn}}]{Hoang2019}
{Hoang}, T., {Tram}, L.~N., {Lee}, H., \& {Ahn}, S.-H. 2019, Nature Astronomy,
  3, 766, \dodoi{10.1038/s41550-019-0763-6}

\bibitem[{{Hoang} {et~al.}(2021){Hoang}, {Tram}, {Lee}, {Diep}, \&
  {Ngoc}}]{Hoang2021}
{Hoang}, T., {Tram}, L.~N., {Lee}, H., {Diep}, P.~N., \& {Ngoc}, N.~B. 2021,
  \apj, 908, 218, \dodoi{10.3847/1538-4357/abd54f}

\bibitem[{{H{\"o}fner}(2008)}]{Hofner2008}
{H{\"o}fner}, S. 2008, \aap, 491, L1, \dodoi{10.1051/0004-6361:200810641}

\bibitem[{{H{\"o}fner}(2009)}]{Hofner2009}
{H{\"o}fner}, S. 2009, in Astronomical Society of the Pacific Conference
  Series, Vol. 414, Cosmic Dust - Near and Far, ed. T.~{Henning},
  E.~{Gr{\"u}n}, \& J.~{Steinacker}, 3.
\newblock \doarXiv{0903.5280}

\bibitem[{{H{\"o}fner} {et~al.}(2016){H{\"o}fner}, {Bladh}, {Aringer}, \&
  {Ahuja}}]{Hofner2016}
{H{\"o}fner}, S., {Bladh}, S., {Aringer}, B., \& {Ahuja}, R. 2016, \aap, 594,
  A108, \dodoi{10.1051/0004-6361/201628424}

\bibitem[{{H{\"o}fner} \& {Olofsson}(2018)}]{Hofner2018}
{H{\"o}fner}, S., \& {Olofsson}, H. 2018, \aapr, 26, 1,
  \dodoi{10.1007/s00159-017-0106-5}

\bibitem[{{Humphreys}(1978)}]{Humphreys1978}
{Humphreys}, R.~M. 1978, \apjs, 38, 309, \dodoi{10.1086/190559}

\bibitem[{{Humphreys} \& {Jones}(2022)}]{Humphreys2022}
{Humphreys}, R.~M., \& {Jones}, T.~J. 2022, \aj, 163, 103,
  \dodoi{10.3847/1538-3881/ac46ff}

\bibitem[{{Kervella} {et~al.}(2011){Kervella}, {Perrin}, {Chiavassa},
  {Ridgway}, {Cami}, {Haubois}, \& {Verhoelst}}]{Kervella2011}
{Kervella}, P., {Perrin}, G., {Chiavassa}, A., {et~al.} 2011, \aap, 531, A117,
  \dodoi{10.1051/0004-6361/201116962}

\bibitem[{{Kervella} {et~al.}(2016){Kervella}, {Lagadec}, {Montarg{\`e}s},
  {Ridgway}, {Chiavassa}, {Haubois}, {Schmid}, {Langlois}, {Gallenne}, \&
  {Perrin}}]{Kervella2016}
{Kervella}, P., {Lagadec}, E., {Montarg{\`e}s}, M., {et~al.} 2016, \aap, 585,
  A28, \dodoi{10.1051/0004-6361/201527134}

\bibitem[{{Kilpatrick} \& {Foley}(2018)}]{Kilpatrick2018}
{Kilpatrick}, C.~D., \& {Foley}, R.~J. 2018, \mnras, 481, 2536,
  \dodoi{10.1093/mnras/sty2435}

\bibitem[{{Kiss} {et~al.}(2006){Kiss}, {Szab{\'o}}, \& {Bedding}}]{Kiss2006}
{Kiss}, L.~L., {Szab{\'o}}, G.~M., \& {Bedding}, T.~R. 2006, \mnras, 372, 1721,
  \dodoi{10.1111/j.1365-2966.2006.10973.x}

\bibitem[{{Kochanek} {et~al.}(2012){Kochanek}, {Khan}, \& {Dai}}]{Kochanek2012}
{Kochanek}, C.~S., {Khan}, R., \& {Dai}, X. 2012, \apj, 759, 20,
  \dodoi{10.1088/0004-637X/759/1/20}

\bibitem[{{Laor} \& {Draine}(1993)}]{Laor1993}
{Laor}, A., \& {Draine}, B.~T. 1993, \apj, 402, 441, \dodoi{10.1086/172149}

\bibitem[{{Lazarian} \& {Hoang}(2007)}]{Lazarian2007}
{Lazarian}, A., \& {Hoang}, T. 2007, \mnras, 378, 910,
  \dodoi{10.1111/j.1365-2966.2007.11817.x}

\bibitem[{{Lazarian} \& {Hoang}(2021)}]{Lazarian2021}
---. 2021, \apj, 908, 12, \dodoi{10.3847/1538-4357/abd02c}

\bibitem[{{Lee}(1970)}]{Lee1970}
{Lee}, T.~A. 1970, \apj, 162, 217, \dodoi{10.1086/150648}

\bibitem[{{Levesque} {et~al.}(2005){Levesque}, {Massey}, {Olsen}, {Plez},
  {Josselin}, {Maeder}, \& {Meynet}}]{Levesque2005}
{Levesque}, E.~M., {Massey}, P., {Olsen}, K.~A.~G., {et~al.} 2005, \apj, 628,
  973, \dodoi{10.1086/430901}

\bibitem[{{Levesque} {et~al.}(2006){Levesque}, {Massey}, {Olsen}, {Plez},
  {Meynet}, \& {Maeder}}]{Levesque2006}
---. 2006, \apj, 645, 1102, \dodoi{10.1086/504417}

\bibitem[{{Lim} {et~al.}(1998){Lim}, {Carilli}, {White}, {Beasley}, \&
  {Marson}}]{Lim1998}
{Lim}, J., {Carilli}, C.~L., {White}, S.~M., {Beasley}, A.~J., \& {Marson},
  R.~G. 1998, \nat, 392, 575, \dodoi{10.1038/33352}

\bibitem[{{Lobel}(2002)}]{Lobel2002}
{Lobel}, A. 2002, in Cambridge Workshop on Cool Stars, Stellar Systems, and the
  Sun, Vol.~12, Cambridge Workshop on Cool Stars, Stellar Systems, and the Sun,
  1.
\newblock \doarXiv{astro-ph/0211506}

\bibitem[{{Lovegrove} \& {Woosley}(2013)}]{Lovegrove2013}
{Lovegrove}, E., \& {Woosley}, S.~E. 2013, \apj, 769, 109,
  \dodoi{10.1088/0004-637X/769/2/109}

\bibitem[{{Massey} {et~al.}(2005){Massey}, {Plez}, {Levesque}, {Olsen},
  {Clayton}, \& {Josselin}}]{Massey2005}
{Massey}, P., {Plez}, B., {Levesque}, E.~M., {et~al.} 2005, \apj, 634, 1286,
  \dodoi{10.1086/497065}

\bibitem[{{Massey} {et~al.}(2009){Massey}, {Silva}, {Levesque}, {Plez},
  {Olsen}, {Clayton}, {Meynet}, \& {Maeder}}]{Massey2009}
{Massey}, P., {Silva}, D.~R., {Levesque}, E.~M., {et~al.} 2009, \apj, 703, 420,
  \dodoi{10.1088/0004-637X/703/1/420}

\bibitem[{{Mathis} {et~al.}(1983){Mathis}, {Mezger}, \& {Panagia}}]{Mathis1983}
{Mathis}, J.~S., {Mezger}, P.~G., \& {Panagia}, N. 1983, \aap, 500, 259

\bibitem[{{Mathis} {et~al.}(1977){Mathis}, {Rumpl}, \&
  {Nordsieck}}]{Mathis1977}
{Mathis}, J.~S., {Rumpl}, W., \& {Nordsieck}, K.~H. 1977, \apj, 217, 425,
  \dodoi{10.1086/155591}

\bibitem[{{Meynet} \& {Maeder}(2003)}]{Meynet2003}
{Meynet}, G., \& {Maeder}, A. 2003, \aap, 404, 975,
  \dodoi{10.1051/0004-6361:20030512}

\bibitem[{Montarg{\`e}s {et~al.}(2021)Montarg{\`e}s, Cannon, Lagadec, de~Koter,
  Kervella, Sanchez-Bermudez, Paladini, Cantalloube, Decin, Scicluna,
  Kravchenko, Dupree, Ridgway, Wittkowski, Anugu, Norris, Rau, Perrin,
  Chiavassa, Kraus, Monnier, Millour, Le~Bouquin, Haubois, Lopez, Stee, \&
  Danchi}]{Montarges2021}
Montarg{\`e}s, M., Cannon, E., Lagadec, E., {et~al.} 2021, Nature, 594, 365,
  \dodoi{10.1038/s41586-021-03546-8}

\bibitem[{{Nataf} {et~al.}(2016){Nataf}, {Gonzalez}, {Casagrande}, {Zasowski},
  {Wegg}, {Wolf}, {Kunder}, {Alonso-Garcia}, {Minniti}, {Rejkuba}, {Saito},
  {Valenti}, {Zoccali}, {Poleski}, {Pietrzy{\'n}ski}, {Skowron},
  {Soszy{\'n}ski}, {Szyma{\'n}ski}, {Udalski}, {Ulaczyk}, \&
  {Wyrzykowski}}]{Nataf2016}
{Nataf}, D.~M., {Gonzalez}, O.~A., {Casagrande}, L., {et~al.} 2016, \mnras,
  456, 2692, \dodoi{10.1093/mnras/stv2843}

\bibitem[{{Noriega-Crespo} {et~al.}(1997){Noriega-Crespo}, {van Buren}, {Cao},
  \& {Dgani}}]{Noriega1997}
{Noriega-Crespo}, A., {van Buren}, D., {Cao}, Y., \& {Dgani}, R. 1997, \aj,
  114, 837, \dodoi{10.1086/118517}

\bibitem[{{O'Gorman} {et~al.}(2015){O'Gorman}, {Harper}, {Brown}, {Guinan},
  {Richards}, {Vlemmings}, \& {Wasatonic}}]{Gorman2015}
{O'Gorman}, E., {Harper}, G.~M., {Brown}, A., {et~al.} 2015, \aap, 580, A101,
  \dodoi{10.1051/0004-6361/201526136}

\bibitem[{{Ohnaka}(2014)}]{Ohnaka2014}
{Ohnaka}, K. 2014, \aap, 568, A17, \dodoi{10.1051/0004-6361/201423893}

\bibitem[{{Pijpers} \& {Hearn}(1989)}]{Pijpers1989}
{Pijpers}, F.~P., \& {Hearn}, A.~G. 1989, \aap, 209, 198

\bibitem[{{Ren} \& {Jiang}(2020)}]{Ren2020}
{Ren}, Y., \& {Jiang}, B.-W. 2020, \apj, 898, 24,
  \dodoi{10.3847/1538-4357/ab9c17}

\bibitem[{{Rodgers} \& {Glassgold}(1991)}]{Rodger1991}
{Rodgers}, B., \& {Glassgold}, A.~E. 1991, \apj, 382, 606,
  \dodoi{10.1086/170748}

\bibitem[{{Schwarzschild}(1975)}]{Schwarzchild1975}
{Schwarzschild}, M. 1975, \apj, 195, 137, \dodoi{10.1086/153313}

\bibitem[{{Scicluna} {et~al.}(2015){Scicluna}, {Siebenmorgen}, {Wesson},
  {Blommaert}, {Kasper}, {Voshchinnikov}, \& {Wolf}}]{Scicluna2015}
{Scicluna}, P., {Siebenmorgen}, R., {Wesson}, R., {et~al.} 2015, \aap, 584,
  L10, \dodoi{10.1051/0004-6361/201527563}

\bibitem[{{Seab} \& {Snow}(1989)}]{Seab1989}
{Seab}, C.~G., \& {Snow}, T.~P. 1989, \apj, 347, 479, \dodoi{10.1086/168136}

\bibitem[{{Smartt} {et~al.}(2009){Smartt}, {Eldridge}, {Crockett}, \&
  {Maund}}]{Smartt2009}
{Smartt}, S.~J., {Eldridge}, J.~J., {Crockett}, R.~M., \& {Maund}, J.~R. 2009,
  \mnras, 395, 1409, \dodoi{10.1111/j.1365-2966.2009.14506.x}

\bibitem[{{Tielens}(1983)}]{Tielens1983}
{Tielens}, A.~G.~G.~M. 1983, \apj, 271, 702, \dodoi{10.1086/161237}

\bibitem[{{Tielens} {et~al.}(2005){Tielens}, {Waters}, \&
  {Bernatowicz}}]{Tielens2005}
{Tielens}, A.~G.~G.~M., {Waters}, L.~B.~F.~M., \& {Bernatowicz}, T.~J. 2005, in
  Astronomical Society of the Pacific Conference Series, Vol. 341, Chondrites
  and the Protoplanetary Disk, ed. A.~N. {Krot}, E.~R.~D. {Scott}, \&
  B.~{Reipurth}, 605

\bibitem[{{Tielens} {et~al.}(1998){Tielens}, {Waters}, {Molster}, \&
  {Justtanont}}]{Tielens1998}
{Tielens}, A.~G.~G.~M., {Waters}, L.~B.~F.~M., {Molster}, F.~J., \&
  {Justtanont}, K. 1998, \apss, 255, 415, \dodoi{10.1023/A:1001585120472}

\bibitem[{{Tram} {et~al.}(2020){Tram}, {Hoang}, {Soam}, {Lesaffre}, \&
  {Reach}}]{Tram2020}
{Tram}, L.~N., {Hoang}, T., {Soam}, A., {Lesaffre}, P., \& {Reach}, W.~T. 2020,
  \apj, 893, 138, \dodoi{10.3847/1538-4357/ab7b5e}

\bibitem[{{Tsuji}(1976)}]{Tsuji1976}
{Tsuji}, T. 1976, \pasj, 28, 567

\bibitem[{{Uitenbroek} {et~al.}(1998){Uitenbroek}, {Dupree}, \&
  {Gilliland}}]{Uitenbroek1998}
{Uitenbroek}, H., {Dupree}, A.~K., \& {Gilliland}, R.~L. 1998, \aj, 116, 2501,
  \dodoi{10.1086/300596}

\bibitem[{{Van de Sande} {et~al.}(2019){Van de Sande}, {Walsh}, {Mangan}, \&
  {Decin}}]{Sande2019}
{Van de Sande}, M., {Walsh}, C., {Mangan}, T.~P., \& {Decin}, L. 2019, \mnras,
  490, 2023, \dodoi{10.1093/mnras/stz2702}

\bibitem[{{van Loon} {et~al.}(2005){van Loon}, {Cioni}, {Zijlstra}, \&
  {Loup}}]{vanLoon2005}
{van Loon}, J.~T., {Cioni}, M. R.~L., {Zijlstra}, A.~A., \& {Loup}, C. 2005,
  \aap, 438, 273, \dodoi{10.1051/0004-6361:20042555}

\bibitem[{{Verhoelst} {et~al.}(2009){Verhoelst}, {van der Zypen}, {Hony},
  {Decin}, {Cami}, \& {Eriksson}}]{Velhoelst2009}
{Verhoelst}, T., {van der Zypen}, N., {Hony}, S., {et~al.} 2009, \aap, 498,
  127, \dodoi{10.1051/0004-6361/20079063}

\bibitem[{{Verhoelst} {et~al.}(2006){Verhoelst}, {Decin}, {van Malderen},
  {Hony}, {Cami}, {Eriksson}, {Perrin}, {Deroo}, {Vandenbussche}, \&
  {Waters}}]{Velhoelst2006}
{Verhoelst}, T., {Decin}, L., {van Malderen}, R., {et~al.} 2006, \aap, 447,
  311, \dodoi{10.1051/0004-6361:20053359}

\bibitem[{{Walmswell} \& {Eldridge}(2012)}]{Walmswell2012}
{Walmswell}, J.~J., \& {Eldridge}, J.~J. 2012, \mnras, 419, 2054,
  \dodoi{10.1111/j.1365-2966.2011.19860.x}

\bibitem[{{Woosley} \& {Heger}(2012)}]{Woosley2012}
{Woosley}, S.~E., \& {Heger}, A. 2012, \apj, 752, 32,
  \dodoi{10.1088/0004-637X/752/1/32}

\bibitem[{{Yoon} \& {Cantiello}(2010)}]{Yoon2010}
{Yoon}, S.-C., \& {Cantiello}, M. 2010, \apjl, 717, L62,
  \dodoi{10.1088/2041-8205/717/1/L62}

\end{thebibliography}

\end{document}